\documentclass[aps,prb,twocolumn,amsmath,amssymb,superscriptaddress,floatfix]{revtex4-1}
\usepackage{amsmath}
\usepackage{amssymb}
\usepackage{amsfonts}
\usepackage{listings}
\usepackage{enumerate}
\usepackage{latexsym}
\usepackage{bm}
\usepackage{graphicx}
\usepackage{subfigure}
\usepackage{braket}
\usepackage[pdftex,colorlinks=false]{hyperref}
\usepackage{xcolor}
\usepackage{verbatim}

\usepackage{times}

\newcommand{\beq}{\begin{equation}}
\newcommand{\eneq}{\end{equation}}
\newcommand{\bs}[1]{\boldsymbol{#1}}

\usepackage{xr}

\newcommand{\beginsupplement}{%
        \setcounter{table}{0}
        \renewcommand{\thetable}{S\arabic{table}}%
        \setcounter{figure}{0}
        \renewcommand{\thefigure}{S\arabic{figure}}%
     }

\def\mH{\mathcal{H}}

\newcounter{matriz}
\newenvironment{matriz}{\refstepcounter{matriz}\equation}{\tag{S\thematriz}\endequation}

\def\be{\begin{equation}}
\def\ee{\end{equation}}
\def\ba{\begin{eqnarray}}
\def\ea{\end{eqnarray}}

\def\R{{\rm Re}}
\def\Z{\mathbb{Z}}
\def\C{\mathbb{C}}

\def\Tr{{\rm Tr}}

\def\t{\tau}

\renewcommand{\vec}{\bs}

\def\beq{\begin{equation}}
\def\eeq{\end{equation}}
\def\barray{\begin{eqnarray}}
\def\earray{\end{eqnarray}}

\font\upright=cmu10 scaled\magstep1
\def\stroke{\vrule height8pt width0.4pt depth-0.1pt}

\def\Zmath{\mathbb{Z}}
\def\Qmath{\vcenter{\hbox{\upright\rlap{\rlap{Q}\kern
                   3.8pt\stroke}\phantom{Q}}}}
\def\Nmath{\vcenter{\hbox{\upright\rlap{I}\kern 1.7pt N}}}
\def\Cmath{\vcenter{\hbox{\upright\rlap{\rlap{C}\kern
                   3.8pt\stroke}\phantom{C}}}}
\def\Rmath{\vcenter{\hbox{\upright\rlap{I}\kern 1.7pt R}}}
\def\Z{\ifmmode\Zmath\else$\Zmath$\fi}
\def\Q{\ifmmode\Qmath\else$\Qmath$\fi}
\def\N{\ifmmode\Nmath\else$\Nmath$\fi}
\def\C{\ifmmode\Cmath\else$\Cmath$\fi}
\def\R{\ifmmode\Rmath\else$\Rmath$\fi}

\input{epsf}

\newcounter{defcounter}
\usepackage{fancyhdr}
\usepackage{textcomp}

\begin{document}

\tolerance 10000

\newcommand{\cbl}[1]{\color{blue} #1 \color{black}}

\newcommand{\vk}{{\bf k}}

\widowpenalty10000
\clubpenalty10000

\title{Higher-Order Topological Insulators}

\author{
Frank~Schindler}
\address{
 Department of Physics, University of Zurich, Winterthurerstrasse 190, 8057 Zurich, Switzerland
}

\author{
Ashley~M.~Cook}
\address{
 Department of Physics, University of Zurich, Winterthurerstrasse 190, 8057 Zurich, Switzerland
}

\author{
Maia~G.~Vergniory}
\address{
 Donostia International Physics Center, P. Manuel de Lardizabal 4, 20018 Donostia-San Sebastian, Spain
}
\address{
 Department of Applied Physics II, Faculty of Science and Technology,
University of the Basque Country UPV/EHU, Apartado 644, 48080 Bilbao, Spain
}

\author{
Zhijun~Wang}
\address{Department of Physics, Princeton University, Princeton, New Jersey 08544, USA}

\author{
Stuart~S.~P.~Parkin}
\address{
 Max Planck Institute of Microstructure Physics, Weinberg 2, 06120 Halle, Germany
}

\author{
B.~Andrei~Bernevig}
\address{Department of Physics, Princeton University, Princeton, New Jersey 08544, USA}
\address{
 Donostia International Physics Center, P. Manuel de Lardizabal 4, 20018 Donostia-San Sebastian, Spain
}\address{Laboratoire Pierre Aigrain, Ecole Normale Sup\'erieure - PSL Research University, CNRS, Universit\'e Pierre et Marie Curie - Sorbonne Universit\'es,
Universit\'e Paris Diderot - Sorbonne Paris Cit\'e, 24 rue Lhomond, 75231 Paris Cedex 05, France}

\author{
Titus~Neupert}
\address{
 Department of Physics, University of Zurich, Winterthurerstrasse 190, 8057 Zurich, Switzerland
}

\begin{abstract}

Three-dimensional topological (crystalline) insulators are materials with an insulating bulk, but conducting surface states which are topologically protected by time-reversal (or spatial) symmetries.
Here, we extend the notion of three-dimensional topological insulators to systems that host \emph{no gapless surface states}, but exhibit topologically protected \emph{gapless hinge states}. Their topological character is protected by spatio-temporal symmetries, of which we present two cases: 
(1) Chiral higher-order topological insulators protected by the combination of time-reversal and a four-fold rotation symmetry. Their hinge states are chiral modes 
and the bulk topology is $\mathbb{Z}_2$-classified.
(2) Helical higher-order topological insulators protected by time-reversal and mirror symmetries. Their hinge states come in Kramers pairs and the bulk topology is $\mathbb{Z}$-classified. 
We provide the topological invariants for both cases. Furthermore we show that SnTe as well as surface-modified Bi$_2$TeI, BiSe, and BiTe are helical higher-order topological insulators and propose a realistic experimental setup to detect the hinge states.
\end{abstract}

\date{\today}

\maketitle
\thispagestyle{fancy}

The bulk-boundary correspondence is often taken as a defining property of topological insulators (TIs)~\cite{Kane05a
,Kane07
,FuKane2007
}:
if a $d$-dimensional system with given symmetry is insulating in the bulk, but supports gapless boundary excitations which cannot be removed by local boundary perturbations without breaking the symmetry, the system is called a topological insulator.
The electric multipole insulators of Ref.~\onlinecite{Benalcazar16} generalize this bulk-boundary correspondence: in two and three dimensions, these insulators exhibit no edge or surface states, respectively, but feature gapless, topological corner excitations corresponding to quantized higher electric multipole moments. Here, we introduce a new class of three-dimensional (3D) topological phases to which the usual form of the bulk-boundary correspondence also does not apply. The topology of the bulk protects gapless states on the \emph{hinges}, while the surfaces are \emph{gapped}.
Both systems, with gapless corner and hinge states, respectively, can be subsumed under the notion of \emph{higher-order topological insulators} (HOTI): An $n$-th order TI has protected gapless modes at a boundary of the system of co-dimension $n$. Following this terminology, we introduce second-order 3D TIs in this work, while Ref.~\onlinecite{Benalcazar16} has introduced second-order two-dimensional (2D) TIs and third-order 3D TIs. The important aspect of 3D HOTIs is that they exhibit protected hinge states with (spectral) flow between the valence and conduction bands, whereas the corner states have no spectral flow.

\begin{figure}[t]
\begin{center}
\includegraphics[width=0.48 \textwidth]{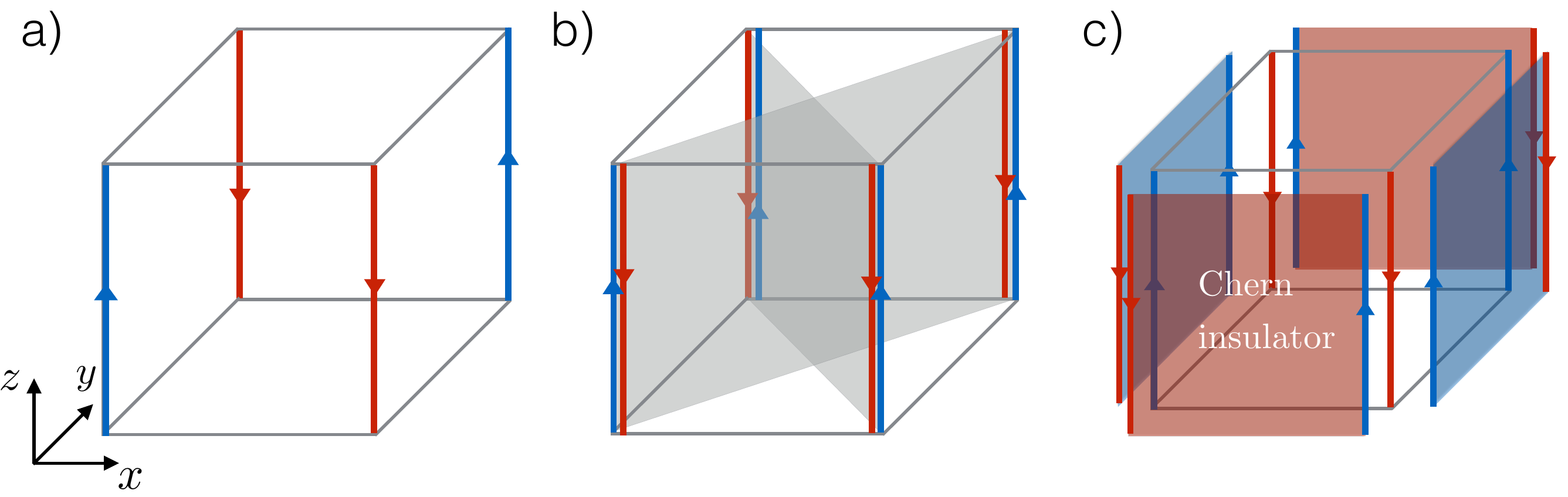}
\caption{
Topologically protected hinge excitations of second-order 3D TIs.
(a) Time-reversal breaking model with chiral hinge currents running along the corners of a $\hat{C}_4^z$-preserving bulk termination, were periodic boundary conditions in $z$-direction are assumed. 
(b) Time-reversal invariant model with anti-propagating Kramers pairs of hinge modes. Highlighted in gray are the planes invariant under the mirror symmetries $\hat{M}_{xy}$ and $\hat{M}_{x\bar{y}}$ that protect the hinge states.
(c) By supplementing each surface of the chiral HOTI in (a) with a Chern insulator with Hall conductivity $\sigma_{xy}=\pm e^2/h$, the number of chiral hinge modes can be changed by 2. 
The Hall conductivities of the additional Chern insulator layers alternate (blue for $+e^2/h$, red for $-e^2/h$) to comply with the $\hat{C}_4^z\hat{T}$ symmetry. The topology is therefore $\mathbb{Z}_2$ classified.
}
\label{fig: physical picture}
\end{center}
\end{figure}

The topological properties of HOTIs are protected by symmetries that involve spatial transformations, possibly augmented by time reversal.
They thus generalize topological crystalline insulators~\cite{Fu11, Hsieh12
}, which have been encompassed in a recent exhaustive classification of topological insulators in Ref.~\onlinecite{Bradlyn17}.
In the current paper, we propose two cases: 
(1) \emph{chiral} HOTIs with hinge modes that propagate unidirectionally, akin to the edge states of a 2D quantum Hall effect~\cite{Klitzing80
}, or Chern insulator~\cite{Haldane88
}. We show that chiral HOTIs may be protected by the product $\hat{C}_4 \hat{T}$ of time reversal $\hat{T}$ and a $\hat{C}_4$ rotation symmetry. The existence of these hinge modes -- but not the direction in which they propagate -- is determined by the topology of the 3D bulk. By a $\hat{C}_4\hat{T}$-respecting surface manipulation, the direction of all hinge modes can be reversed, but they cannot be removed. This constitutes a bulk $\mathbb{Z}_2$ topological classification. We also show that chiral HOTIs may have a bulk $\mathbb{Z}$ topological classification protected by mirror symmetries which leave the hinges invariant when time reversal symmetry $\hat{T}$ is broken.
(2) \emph{helical} HOTIs with Kramers pairs of counter-propagating hinge modes, akin to the edge states of a 2D quantum spin Hall effect~\cite{Kane05a
, Qi2006, Bernevig2006a, Bernevig2006b
}. We show that helical HOTIs may occur when a system is invariant under time reversal $\hat{T}$ and a $\hat{C}_4$ rotation symmetry. We further show that helical HOTIs can also be protected by $\hat{T}$ and mirror symmetries which leave the hinges invariant. Any integer number of Kramers pairs is topologically protected against symmetry preserving surface manipulations, yielding a $\mathbb{Z}$ classification.

For both cases, we show the topological bulk-surface-hinge correspondence, provide concrete lattice-model realizations, and provide expressions for the bulk topological invariants. The latter are given by the magneto-electric polarizability and mirror Chern numbers~\cite{Teo08,Hsieh12}, for chiral and helical HOTIs, respectively. For the case where a chiral HOTI also respects the product of inversion times time-reversal symmetry $\hat{I}\hat{T}$, we formulate a simplified topological index akin to the Fu-Kane formula for inversion symmetric TIs.~\cite{Kane07} Finally, on the basis of tight-binding and ab-initio calculations, we propose SnTe as a material realization for helical HOTIs. We also propose an explicit experimental setup to cleanly create hinge states in a topological SnTe coaxial cable. Chiral HOTIs, in contrast, may arise in 3D TI materials that exhibit noncollinear antiferromagnetic order at low temperatures.

Our work is complemented by two related articles:
Ref.~\onlinecite{Brouwer17} provides a general classification of second-order phases with reflection symmetry for all ten Altland-Zirnbauer symmetry classes, Ref.~\onlinecite{BABHughesBenalcazar17} establishes a physical interpretation of the topological invariants of higher-order phases in terms of electric multipole moments.


\begin{center}
\textbf{
Chiral Higher-Order Topological Insulator}
\end{center}

We first give an intuitive argument for the topological nature of a chiral 3D HOTI.
We consider a hypothetical but realizable electronic structure where gapless degrees of freedom are only found on the hinge. 
For concreteness, let us consider a system with a square cross-section, periodic boundary conditions in $z$-direction and $\hat{C}_4^z\hat{T}$ symmetry that has a single chiral mode at each hinge, as sketched in Fig.~\ref{fig: physical picture}~(a).
For these modes to be a feature associated with the 3D bulk topology of the system, they should be protected against any $\hat{C}^z_4\hat{T}$ preserving \emph{surface} or \emph{hinge} perturbation of the system. The minimal relevant surface perturbation of that kind is the addition of an integer quantum Hall (or Chern insulator) layer with Hall conductivity $\sigma_{xy}=e^2/h$ and $\sigma_{xy}=-e^2/h$ on the (100) surfaces and the (010) surfaces, respectively, which respects $\hat{C}^z_4\hat{T}$. As seen from Fig.~\ref{fig: physical picture}~(c), 
this adds to each hinge \emph{two} chiral hinge channels. Repeating this procedure, we can change--via a pure surface manipulation--the number of chiral channels on each hinge by any even number. Hence, only the $\mathbb{Z}_2$ \emph{parity} of hinge channels can be a topological property protected by the system's 3D bulk.

A concrete model of this phase is defined via the four-band Bloch Hamiltonian
\begin{equation} \label{eq: H}
\begin{aligned}
\mathcal{H}_{\mathrm{c}}(\vec{k}) = &\Bigl(M+t\sum_i \cos k_i\Bigr) \, \tau_z \sigma_0 +\Delta_1\sum_i \sin k_i \, \tau_x \sigma_i \\&+\Delta_2 (\cos k_x - \cos k_y) \, \tau_y \sigma_0,
\end{aligned}
\end{equation}
where $\sigma_i$ and $\tau_i$, $i=x,y,z$, are the three Pauli matrices acting on spin and orbital degree of freedoms, respectively (see the Methods Section for a real space representation of the model).
For $\Delta_2=0$, $\mathcal{H}_{\mathrm{c}}(\vec{k})$ represents the well-known 3D TI if $1<|M|<3$.
Time-reversal symmetry is represented by $T \mathcal{H}_{\mathrm{c}}(\vec{k}) T^{-1}=\mathcal{H}_{\mathrm{c}}(-\vec{k}),$ with $T \equiv \tau_0 \sigma_y \mathit{K}$, where $\mathit{K}$ denotes complex conjugation. For $\Delta_2=0$, Hamiltonian~\eqref{eq: H} has a $\hat{C}^z_4$ rotation symmetry $C_4^z  \mathcal{H}_{\mathrm{c}}(\vec{k}) \left(C_4^z \right)^{-1}=\mathcal{H}_{\mathrm{c}}(D_{\hat{C}^{z}_4}\vec{k})$, where $C_4^z \equiv \tau_0 e^{-\mathrm{i} \frac{\pi}{4} \sigma_z}$  and $D_{\hat{C}^{z}_4}\vec{k}=(-k_y,k_x,k_z)$.

\begin{figure}[t]
\begin{center}
\includegraphics[width=0.48 \textwidth]{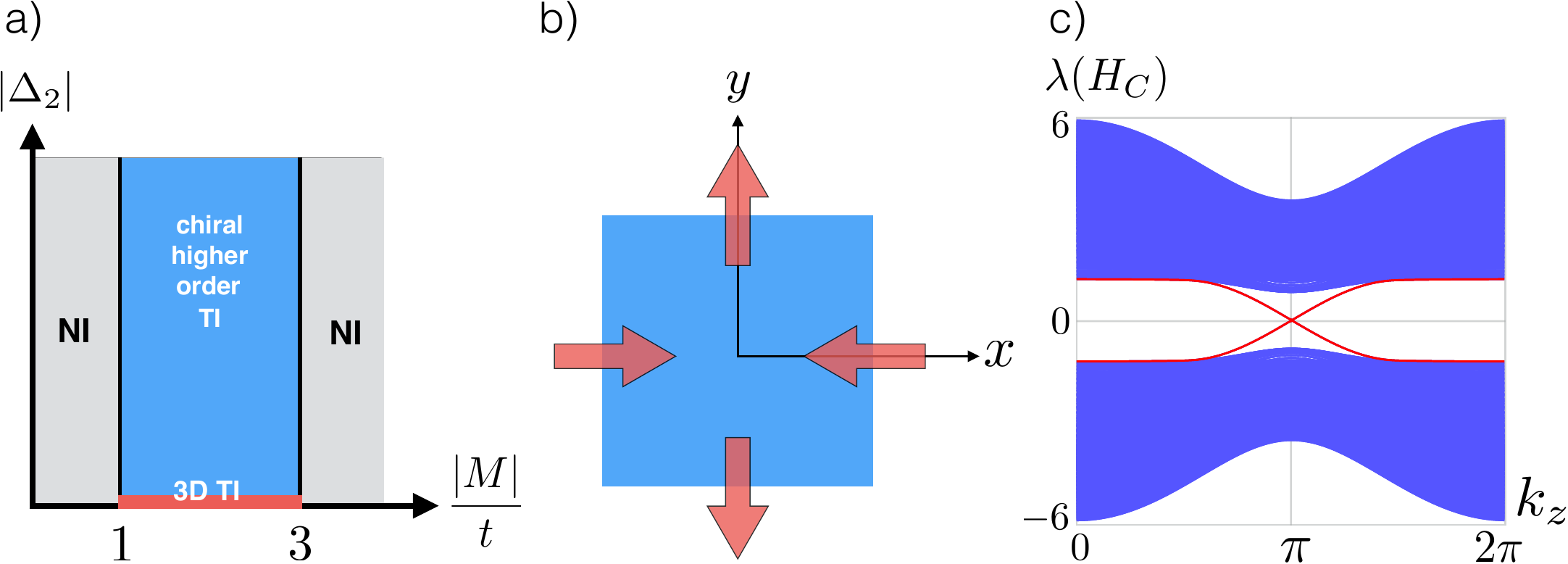}
\caption{
(a) Schematic phase diagram for model~\eqref{eq: H}, where NI stands for normal insulator.
(b) A unit cell of noncollinear magnetic order with $\hat{C}_4^z \hat{T}$ symmetry.
(c) Energy spectrum of model~\eqref{eq: H} with chiral hinge currents (red) in the geometry of Fig.~\ref{fig: physical picture}~(a). For a slab geometry, where the bulk is terminated in just one direction of space, there are in general no gapless modes. 
}
\label{fig: phsdiagtrb}
\end{center}
\end{figure}

The term proportional to $\Delta_2$ breaks both $\hat{T}$ and $\hat{C}_4^z$ individually, but respects the anti-unitary combination $\hat{C}^z_4 \hat{T}$, which means that 
\begin{equation} \label{eq: C4ZTRep}
\begin{gathered}
\left(C_4^z T\right) \mathcal{H}_{\mathrm{c}}(\vec{k}) \left(C_4^z T\right)^{-1}=\mathcal{H}_{\mathrm{c}}(D_{\hat{C}^{z}_4 \hat{T}}\vec{k}),\\
D_{\hat{C}^{z}_4 \hat{T}}\vec{k} = (k_y, -k_x, -k_z)
\end{gathered}
\end{equation}
is a symmetry of the Hamiltonian also when $\Delta_2\neq0$. 
Since $[\hat{C}_4^z,\hat{T}]=0$, we have $(\hat{C}_4^z \hat{T})^4 = -1$, independent of the choice of representation.

The phase diagram of Hamiltonian~\eqref{eq: H} is shown in Fig.~\ref{fig: phsdiagtrb}~(a). 
For $1<|M/t|<3$ and $\Delta_1,\Delta_2\neq 0$ the system is a chiral 3D HOTI.
The spectrum in the case of open boundary conditions in $x$ and $y$ directions is presented in Fig.~\ref{fig: phsdiagtrb}~(c), where the chiral hinge modes (each 2-fold degenerate) are seen to traverse the bulk gap. 
Physically, the term multiplied by $\Delta_2$ corresponds to orbital currents that break TRS oppositely in the $x$ and $y$-directions.
When infinitesimally small, its main effect is thus to open gaps with alternating signs for the surface Dirac electrons of the 3D TI on the (100) and (010) surfaces.
The four hinges are then domain walls at which the Dirac mass changes sign. It is well known~\cite{
Sitte12,FanZhang13} that such a domain wall on the surface of a 3D TI binds a gapless chiral mode, which in the case at hand is reinterpreted as the hinge mode of the HOTI.
 [Another physical mechanism that breaks time-reversal symmetry and preserves $\hat{C}^z_4 \hat{T}$ would be $(\pi,\pi,0)$ noncollinear antiferromagnetic order with a unit cell as shown in Fig.~\ref{fig: phsdiagtrb}~(b).] Note that even with finite $\Delta_2$, the (001) surface of the model remains gapless, since its Dirac cone is protected by the $\hat{C}_4^z \hat{T}$ symmetry which leaves the surface invariant and enforces a Kramers-like degeneracy discussed in the Supplementary Information. The gapless nature of the (001) surface in the geometry of Fig.~\ref{fig: physical picture} is also required by current conservation, because the chiral hinge currents cannot terminate in a gapped region of the sample. A current-conserving geometry with gapped surfaces is given in the Supplementary Information, Sec.~\ref{app: isotropicmodel}.

We turn to the bulk topological invariant that describes the $\mathbb{Z}_2$ topology. The topological invariant of 3D TIs is the theta angle, or Chern Simons invariant $\theta$ (see the Methods Section for its definition) which is quantized by time-reversal symmetry to be $\theta=0,\pi\,\mathrm{mod}\,2\pi$, with $\theta=\pi$ being the nontrivial case\cite{Qi08}. In fact, the very same quantity $\theta$ is the topological invariant of chiral HOTIs. What changes is that its quantization to values $0,\pi$ is not enforced by $\hat{T}$ but by $\hat{C}_4 \hat{T}$ symmetry in this case. 
$\theta$ attains a new meaning in the second-order picture: it  uniquely characterizes a different symmetry-protected topological phase which exhibits $\hat{T}$ breaking, but $\hat{C}_4\hat{T}$ preserving hinge currents instead of $\hat{T}$ preserving gapless surface excitations. In the Supplementary Information, Sec.~\ref{sec: CS invairant}, we show the quantization of $\theta$ enforced by $\hat{C}_4 \hat{T}$ symmetry and explicitly evaluate $\theta=\pi$ for the model~\eqref{eq: H}. We furthermore note that for a nontrivial $\theta$ in the presence of $\hat{C}_4 \hat{T}$ symmetry to uniquely characterize the presence of gapless hinge excitations, the bulk \emph{and} the surfaces of the material which adjoin the hinge are required to be insulating. This constitutes the \emph{bulk-surface-hinge} correspondence of chiral HOTIs.

The explicit evaluation of $\theta$ is impractical for ab-initio computations in generic insulators. This motivates the discussion of alternative forms of the topological invariant. 
The Pfaffian invariant~\cite{Kane05a
} used to define first-order 3D TIs rests on the group relation $\hat{T}^2=-1$, it fails in our case where $(\hat{C}_4 \hat{T})^4 = -1$. We may instead use a non-Abelian Wilson loop characterization of the topology, as presented in the Supplementary Information.~\cite{Yu11,Alexandradinata14}
There, we also provide two further topological characterizations, one based on so-called nested Wilson loop\cite{Benalcazar16} and entanglement spectra~\cite{Li08,Peschel03,Fidkowski10-2
}, and one applicable to systems that are in addition invariant under the product $\hat{I}\hat{T}$ of inversion symmetry $\hat{I}$ and $\hat{T}$.\cite{FuKane2007}

\begin{center}
\textbf{
Helical Higher-Order Topological Insulator}
\end{center}

Helical higher-order TIs feature Kramers pairs of counter-propagating hinge modes. They are protected by time-reversal symmetry and a spatial symmetry. For concreteness, let us consider a system with a square (or rhombic) cross-section, periodic boundary conditions in $z$-direction, and two mirror symmetries $\hat{M}_{xy}$ and $\hat{M}_{x\bar{y}}$ that leave, respectively, the $x=-y$ and the $x=y$ planes invariant, and with it a pair of hinges each [sketched in Fig.~\ref{fig: physical picture}~(b)].
We consider a hypothetical but realizable electronic structure where gapless degrees of freedom are only found on the hinge. At a given hinge, for instance one that is invariant under $\hat{M}_{xy}$, we can choose all hinge modes as eigenstates of $\hat{M}_{xy}$. We denote the number of modes that propagate parallel, R,  (antiparallel, L) to the $z$ direction and have $\hat{M}_{xy}$ eigenvalue $\mathrm{i}\lambda$, $\lambda=\pm 1$, by $N_{\mathrm{R},\lambda}$ ($N_{\mathrm{L},\lambda}$). We argue that the net number of helical hinge pairs $n\equiv N_{\mathrm{R},+}-N_{\mathrm{L},+}$ (which by time-reversal symmetry is equal to $N_{\mathrm{L},-}-N_{\mathrm{R},-}$) is topologically protected. In particular, $n$ cannot be changed by any \emph{surface or hinge} manipulation that respects both $\hat{T}$ and $\hat{M}_{xy}$. First note that if both $N_{\mathrm{R},+}$ and $N_{\mathrm{L},+}$ are nonzero (assuming from now on that $N_{\mathrm{R},+} > N_{\mathrm{R},-}$), we can always hybridize $N_{\mathrm{L},+}$ right-moving modes with all  $N_{\mathrm{L},+}$ left-moving modes within the $\lambda=+$ subspace without breaking any symmetry. Therefore, only the difference $n$ is well defined and corresponds to the number of remaining pairs of modes. 

\begin{figure}[t]
\begin{center}
\includegraphics[width= 0.48\textwidth]{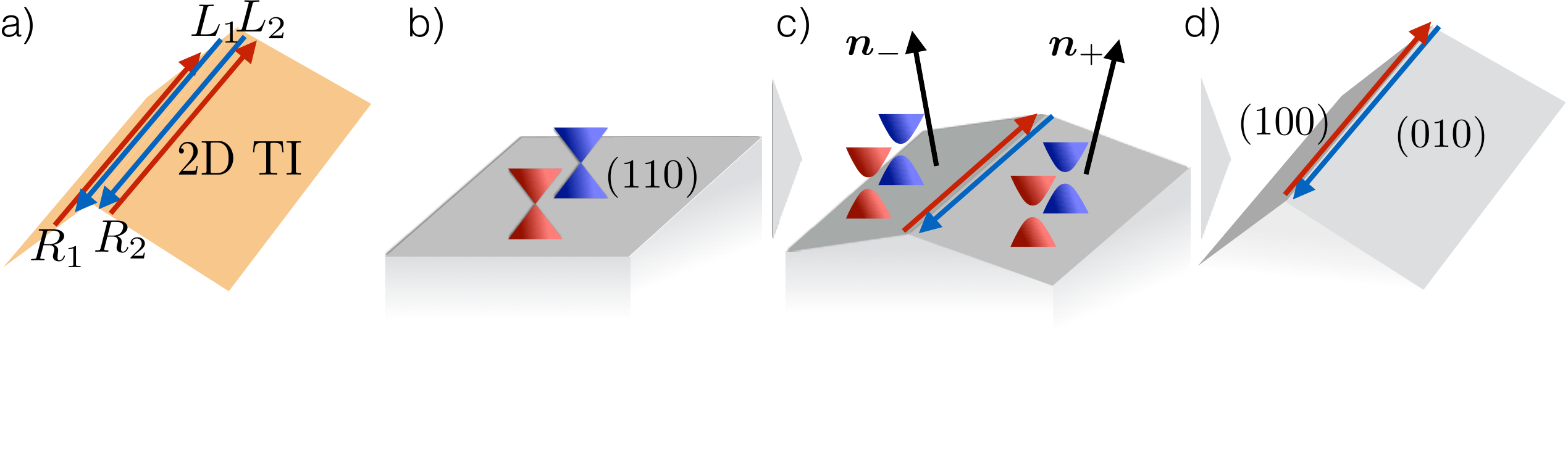}
\caption{
Bulk-surface-hinge correspondence of helical HOTIs.
(a) Additional hinge modes obtained by decorating the surfaces with 2D time-reversal symmetric TIs in a mirror-symmetric fashion. They can always be combined in `bonding' and `anti-bonding' pairs $\{R_1+R_2,L_1+L_2\}$ and $\{R_1-R_2,L_1-L_2\}$, with mirror eigenvalues $+\mathrm{i}$ and $-\mathrm{i}$, respectively. Therefore, they do not change the net mirror chirality of the hinge.
(b) Mirror-symmetry protected Dirac cones on a (110) surface. (c) Slightly tilting the surface normal out of the mirror plane gaps the Dirac cones and forms a Kramers pair of domain wall states between two surfaces with opposite tilting. The mirror eigenvalues of the hinge modes are tied to those of the Dirac cones, which in turn are related to a bulk topological invariant, the mirror Chern number $C_{\mathrm{m}}$.
(d) Further deforming the surface to the (100) and (010) orientation in a mirror-symmetry preserving manner does not change this correspondence. 
}
\label{fig: bulk-boundary}
\end{center}
\end{figure}

The argument for their topological protection proceeds similar to the chiral HOTI case by considering a minimal symmetry-preserving surface perturbation. It consists of a layer of a 2D time-reversal symmetric TI and its mirror-conjugated partner added to surfaces that border the hinge under consideration. Each of the TIs contributes a single Kramers pair of boundary modes to the hinge so that $(N_{\mathrm{L},-}+N_{\mathrm{L},+})$ and $(N_{\mathrm{R},-}+N_{\mathrm{R},+})$ each increase by 2 [see Fig.~\ref{fig: bulk-boundary}~(a)]. Since mirror symmetry maps the right-moving modes of the two Kramers pairs onto one another (and the same for the two left-moving modes) we can form a `bonding' and `anti-bonding' superposition with mirror eigenvalues $+\mathrm{i}$ and $-\mathrm{i}$ out of each pair. Thus each of $N_{\mathrm{L},+}$, $N_{\mathrm{L},-}$, $N_{\mathrm{R},+}$, and $N_{\mathrm{R},-}$ increase by 1 due to this minimal surface manipulation. This leaves $n$ invariant, suggesting a $\mathbb{Z}$ classification of the helical HOTI for each pair of mirror-invariant hinges. The case depicted in Fig.~\ref{fig: physical picture}~(b) with two mirror symmetries is then $\mathbb{Z}\times\mathbb{Z}$ classified. A more rigorous version of this argument can be found in the Supplementary Information.

The topological invariant for the $\mathbb{Z}\times\mathbb{Z}$ classification of the helical HOTI is the set of mirror Chern numbers~\cite{Teo08,Hsieh12} $C_{\mathrm{m}}/2$ on the 
$\hat{M}_{xy}$ and $\hat{M}_{x\bar{y}}$ mirror planes (see the Methods Section for the definition of $C_{\mathrm{m}}$).
First observe that if $C_{\mathrm{m}}$ were odd, the system would be a strong 3D TI: the $\hat{M}_{xy}$ mirror planes in momentum space include all time-reversal invariant momenta in the (110) surface Brillouin zone. Thus if $C_{\mathrm{m}}$ is odd, there is an odd number of Dirac cones on the (110) surface and time-reversal symmetry implies that such a system is a strong 3D TI. As the surfaces of a strong 3D TI cannot be gapped out with a time-reversal symmetric perturbation, we cannot construct a helical HOTI from it. We conclude that $C_{\mathrm{m}}$ is even for all systems of interest to us.

We now discuss the correspondence between the bulk topological invariant $C_{\mathrm{m}}/2$ and the existence of Kramers paired hinge modes. For this, we first consider the electronic structure of the (110) surface, which is invariant under $\hat{M}_{xy}$, and then that of a pair of surfaces with a normal $\bs{n}_\pm=(1\pm\alpha,1 \mp\alpha,0)$ for small $\alpha$, which are mapped into each other under $\hat{M}_{xy}$ and form a hinge at their interface [see Fig.~\ref{fig: bulk-boundary}~(b)--(d)]. 

A nonzero bulk mirror Chern number $C_{\mathrm{m}}$ with respect to the $\hat{M}_{xy}$ symmetry enforces the existence of gapless Dirac cones on the (110) surface. These Dirac cones are pinned to the mirror invariant lines $k_1=0,\pi$ in the surface Brillouin zone of the (110) surface, where $k_1$ is the momentum along the direction with unit vector $\hat{\bs{e}}_1=(\hat{\bs{e}}_x-\hat{\bs{e}}_y)/\sqrt{2}$.
If we consider the electronic structure along these lines in momentum space, see Fig.~\ref{fig: bulk-boundary}~(b), each Dirac cone 
has an effective Hamiltonian $\mathcal{H}_{\mathrm{D}}=v_1\sigma_z (k_1-k_1^{(0)})+v_z\sigma_x (k_z-k_z^{(0)})$ when expanded around a Dirac point at $(k_1,k_z)=(k_1^{(0)},k_z^{(0)})$ for $k_1^{(0)}=0$ or $k_1^{(0)}=\pi$. The mirror symmetry is represented by $M_{xy}=\mathrm{i}\sigma_x$, preventing mass terms of the form $m\sigma_y$ from appearing. 
The sign of $v_z$ is tied to the $\hat{M}_{xy}$ eigenvalue $(\mathrm{i}\,\mathrm{sgn}\,v_z)$ of the eigenstate with positive group velocity in $z$ direction (at $k_1-k_1^{(0)}=0$).
Denoting the total number of Dirac cones with $v_z>0$ ($v_z<0$) by $n_{+}$ ($n_{-}$), the bulk-boundary correspondence of a topological crystalline insulator~\cite{Fu11} implies
\begin{equation}
C_{\mathrm{m}}=n_{+}-n_{-}.
\label{eq: bulk-boundary TCI}
\end{equation}

Consider now a pair of surfaces with slightly tilted normals $\bs{n}_+$ and $\bs{n}_-$, which are not invariant under the mirror symmetry but map into each other. Mass terms are allowed and the Hamiltonians on the surfaces with normal $\bs{n}_\pm$ read
\begin{equation}
\mathcal{H}_{\mathrm{D},\pm}=
v_1\sigma_z (k_1-k_1^{(0)}\pm\kappa \alpha)
+v_z\sigma_x (k_z-k_z^{(0)})
\pm m \alpha \sigma_y
\label{eq: low-energy massive surface Dirac}
\end{equation}
to linear order in $\alpha$, $(k_1-k_1^{(0)})$ and $(k_z-k_z^{(0)})$ with $m$ and $\kappa$ real parameters. The two surfaces with normals $\bs{n}_+$ and $\bs{n}_-$ meet in a hinge [see Fig.~\ref{fig: bulk-boundary}~(b)]. Equation~\eqref{eq: low-energy massive surface Dirac} describes a Dirac fermion with a mass of opposite sign on the two surfaces. The hinge therefore forms a domain wall in the Dirac mass, from which a single chiral channel connecting valence and conduction bands arises~\cite{Jackiw76}.
As we show in the Supplementary Information, this domain wall either binds a R moving mode with $M_{xy}$ mirror eigenvalue $\mathrm{i}\lambda=\mathrm{i}\,\mathrm{sgn}(v_z)$ or a L moving mode with mirror eigenvalue $-\mathrm{i}\,\mathrm{sgn}(v_z)$. The equality $n_{\mathrm{sgn}(v_z)}=N_{\mathrm{R},\mathrm{sgn}(v_z)}+N_{\mathrm{L},-\mathrm{sgn}(v_z)}$ follows, which connects the number of hinge modes $N_{\mathrm{L}/\mathrm{R},\pm}$ we had introduced before to the mirror-graded numbers of Dirac cones on the (110) surface $n_\pm$. From Eq.~\eqref{eq: bulk-boundary TCI} we obtain
\begin{equation}
C_{\mathrm{m}}=(N_{\mathrm{R},+}-N_{\mathrm{R},-}+N_{\mathrm{L},-}-N_{\mathrm{L},+}) \equiv 2n,
\label{eq: bulk-hinge helical}
\end{equation}
relating the 3D bulk invariant $C_{\mathrm{m}}$ to the number of protected helical hinge pairs $n$ of the HOTI. Notice that by time-reversal symmetry $N_{\mathrm{R},+}-N_{\mathrm{R},-}=N_{\mathrm{L},-}-N_{\mathrm{L},+}$, so that $n$ in Eq.~\eqref{eq: bulk-hinge helical} is integer. ($C_{\mathrm{m}}$ is even as forementioned.)

Note that the above deformation of the surfaces can be extended to nonperturbative angles $\alpha$, until for example the (100) and (010) surface orientations are reached. The surfaces on either side of the hinge may undergo gap-closing transitions as $\alpha$ is increased, but as we argued at the beginning of the section, surface transitions of this kind may not change the net number of helical hinge states with a given mirror eigenvalue, if they occur in a mirror-symmetric way.

We remark that an equation similar to Eq.~\eqref{eq: bulk-hinge helical} also holds in the absence of time-reversal symmetry for each mirror subspace.
Then the Chern number in each mirror subspace is an independent topological invariant, which gives rise to a $\mathbb{Z}\times\mathbb{Z}$ classification on each hinge (as opposed to $\mathbb{Z}$ with time-reversal symmetry). This case corresponds to chiral HOTIs protected by mirror symmetries instead of the $\hat{C}_4\hat{T}$ symmetry employed in Eq.~\eqref{eq: C4ZTRep}. Conversely, we show in the Supplementary Information that a helical HOTI protected by $\hat{C}_4$ \emph{and } $\hat{T}$ exists and has a $\mathbb{Z}_2$ classification.

\begin{figure*}[t]
\begin{center}
\includegraphics[width=0.90 \textwidth]{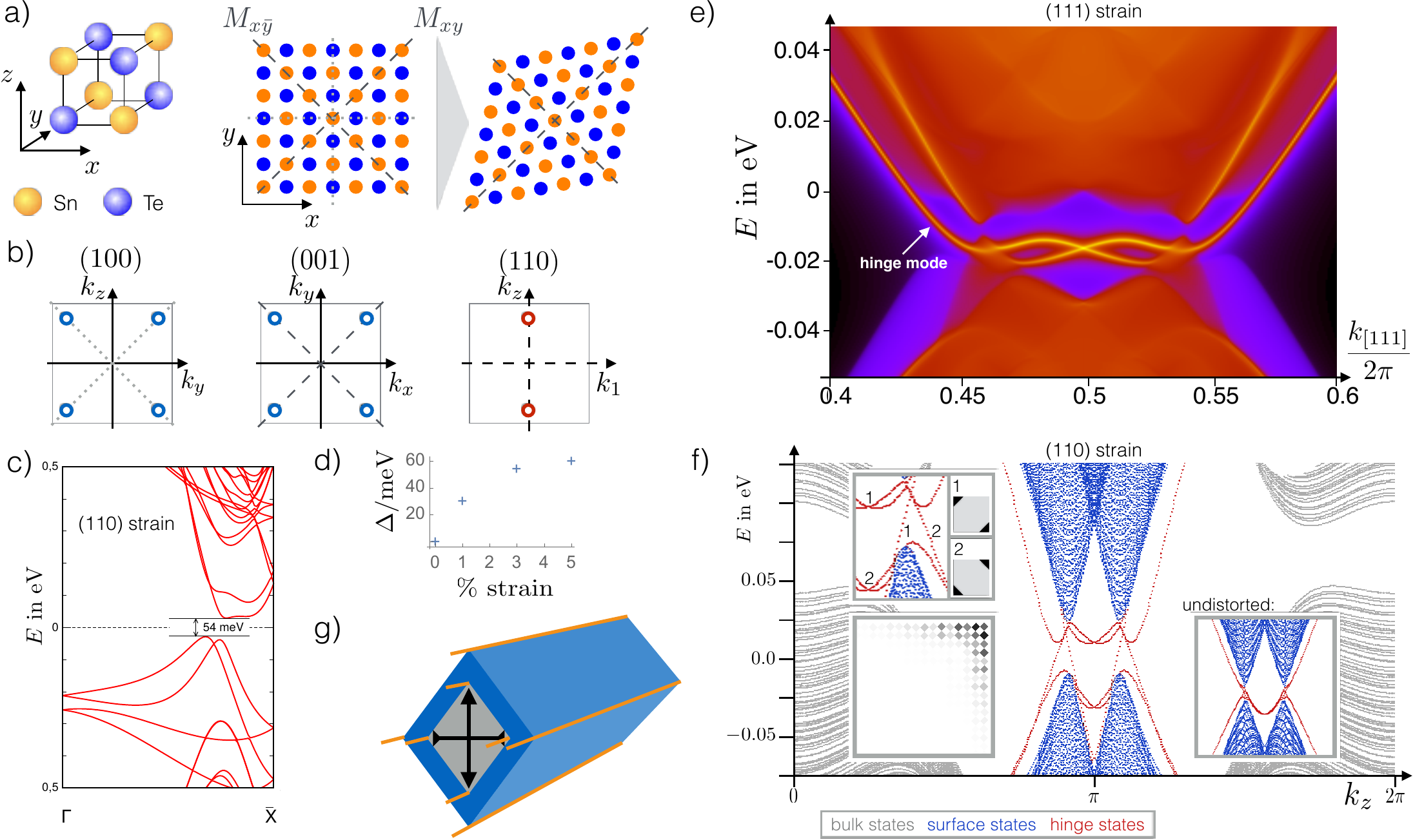}
\caption{
Helical HOTI emerging from the topological crystalline insulator SnTe.
(a)~Rocksalt lattice structure of SnTe. Uniaxial strain along the (110) direction breaks the mirror symmetries represented by dotted lines, but preserves the ones represented by dashed lines.
(b)~Circles indicate the location of Dirac cones in the surface Brillouin zone of pristine SnTe for various surface terminations. Those crossed by dotted mirror symmetries are gapped in SnTe with uniaxial strain while the others are retained.
The two red Dirac cones are enforced by a mirror Chern number $C_{\mathrm{m}}=2$, corresponding to one helical pair of hinge modes. $k_1$ is the momentum along the direction with unit vector $\hat{\bs{e}}_1=(\hat{\bs{e}}_x-\hat{\bs{e}}_y)/\sqrt{2}$.
(c) DFT band structure of a slab of SnTe with open boundary conditions in the (100) direction under 3\% strain in the (110) direction.
(d) DFT calculation of the gap $\Delta$ that develops on the (100) surface of SnTe under (110) uniaxial strain.
(e) DFT-based Wannier tight-binding calculation of SnTe with the (111) ferroelectric displacement in a semi-infinite geometry in which the (0\=11) surface and the (\=101) surface meet at a hinge that is parallel to the (111) direction. A single Kramers pair of hinge states is visible. This distortion beaks all mirror symmetries except those with normal (0\=11), (\=101), (1\=10), which retain their mirror Chern number 2 for a sufficiently small distortion. The (0\=11) and (\=101) surfaces considered here are both not invariant under these mirror symmetries, but the hinge formed between them is invariant under the mirror symmetry with normal (1\=10), supporting topological hinge states.
(f) Low-energy finite size spectrum of SnTe with uniaxial (110) strain obtained using a tight-binding model (see Supplementary Information) for open boundary conditions in the $x$ and $y$ directions (with $L_x=L_y=111$ atoms) and periodic boundary conditions in the $z$ direction.
States localized in the bulk, on the (100)/(010) surfaces, and on the hinges are color coded. 
Near $k_z=\pi$ four Kramers pairs of hinge modes, one localized on each hinge, are found.
Upper left inset: localization of the gapless modes. Lower left inset: spatial structure of one such mode near a hinge.
(Only a small portion of the lattice near the hinge is shown).
Right inset: electronic structure of undistorted SnTe in the same geometry, showing two `flat band' hinge modes in addition to the gapless surface Dirac cones.
(g) Topological coaxial cable geometry to realize (110) uniaxial displacement. A Si or SiO substrate (gray) is etched to have a rhombohedral cross-section and then coated with SnTe (blue) yielding Kramers pairs of hinge modes (orange).
}
\label{fig: TRS TI}
\end{center}
\end{figure*}

\begin{center}
\textbf{
Material Candidates and Experimental Setup}
\end{center}

We propose that SnTe realizes a helical HOTI.
In its cubic rocksalt structure, SnTe is known to be a topological crystalline insulator~\cite{Fu11,Hsieh12
}.
This crystal structure has mirror symmetries $\hat{M}_{xy}$  [acting as $(x,y,z)\to(y,x,z)$] as well as its partners under cubic symmetry,  $\hat{M}_{x\bar{y}}$, $\hat{M}_{xz}$, $\hat{M}_{x\bar{z}}$, $\hat{M}_{yz}$, $\hat{M}_{y\bar{z}}$. (Further spatial symmetries irrelevant to the discussion are not mentioned.) 
The bulk electronic structure of SnTe is insulating and topologically characterized by a mirror Chern number $C_{\mathrm{m}}=2$ with respect to the mirror symmetries on the mirror planes which include the $\Gamma$ point in momentum space. All other mirror planes have $C_{\mathrm{m}}=0$. 
As a result, cubic SnTe has mirror-symmetry protected Dirac cones on specific surfaces.
We consider the geometry of Fig.~\ref{fig: physical picture}~(b) with open boundary conditions in the $x$ and $y$ direction, and periodic boundary conditions in the $z$ direction. The $\hat{M}_{xz}$, $\hat{M}_{x\bar{z}}$ and $\hat{M}_{yz}$, $\hat{M}_{y\bar{z}}$ symmetries along with their mirror Chern numbers protect either four Dirac cones at generic surface momenta, or two at the surface Brillouin zone Kramers points on the (100) as well as the (010) surfaces [see Fig.~\ref{fig: TRS TI}~(b)]. In the case at hand, the former possibility is realized.
We now discuss two distortions of the crystal structure that turn SnTe into a HOTI.

\emph{(i)} At about $98$~K, SnTe undergoes a structural distortion into a low-temperature rhombohedral phase 
via a relative displacement of the two sublattices along the (111) direction.~\cite{Iizumi75
,Liang17} This breaks the mirror symmetries $\hat{M}_{x\bar{z}}$, $\hat{M}_{y\bar{z}}$, and $\hat{M}_{x\bar{y}}$, but preserves $\hat{M}_{xz}$, $\hat{M}_{yz}$, and $\hat{M}_{xy}$. On the (100) surface in the geometry in Fig.~\ref{fig: physical picture}~(b), for instance, the two Dirac cones protected by $\hat{M}_{y\bar{z}}$ can thus be gapped out, while the two Dirac cones protected by $\hat{M}_{yz}$ remain [and similarly for the (010) surface]. 
Therefore, the (100) and (010) surfaces remain gapless and the geometry of Fig.~\ref{fig: physical picture}~(b) cannot be used to expose the HOTI nature of SnTe with (111) uniaxial displacement. 
For that reason, we instead consider the (\=101) and (0\=11) surfaces, which are not invariant under any mirror symmetry of SnTe with (111) uniaxial displacement.
The spectral function focused on the hinge weight of semi-infinite geometry with a single hinge formed between the (\=101) and (0\=11) surfaces is shown in Fig.~\ref{fig: TRS TI}~(e). This tight-binding calculation, based on density functional theory (DFT)-derived Wannier functions (see Methods Section), demonstrates the existence of this single Kramers pair of states on the two hinges invariant under $\hat{M}_{xy}$, in line with the prediction of Eq.~\eqref{eq: bulk-hinge helical} for $C_{\mathrm{m}}=2$.

\emph{(ii)} If uniaxial strain along the (110) direction is applied to SnTe, $\hat{M}_{xz}$, $\hat{M}_{x\bar{z}}$, $\hat{M}_{yz}$, and $\hat{M}_{y\bar{z}}$ symmetries are broken, but $\hat{M}_{xy}$ and $\hat{M}_{x\bar{y}}$ are preserved.
This gaps the (100) and (010) surfaces in the geometry in Fig.~\ref{fig: physical picture}~(b) completely. We calculated the surface states by using a slab geometry along the (100)
direction with DFT. Due to the smallness of the band gap induced by
strain, we needed to achieve a negligible interaction between
the surface states from both sides of the slab. To reduce the
overlap between top and bottom surface states, we considered
a slab of 45 layers, 1 nm vacuum thickness and artificially localized
the states on one of the surfaces, and adding one layer of hydrogen on one of the surfaces. The evolution of the surface gap size with strain is shown in Fig.~\ref{fig: TRS TI}~(d) (see Supplementary Information for more details). Figure~\ref{fig: TRS TI}~(f) is the spectrum of a tight-binding calculation~\cite{Hsieh12} with (110) strain, demonstrating that there exists one Kramers pair of hinge modes on all four hinges in the geometry of Fig.~\ref{fig: physical picture}~(b).

We propose to physically realize the (110) uniaxial strain in SnTe with a \emph{topological coaxial cable} geometry, which would enable the use of its protected hinge states as quasi one-dimensional dissipationless conduction channels [see Fig.~\ref{fig: TRS TI}~(f)]. Starting point is an insulating nanowire substrate made from Si or SiO, with a slightly rhombohedral cross-section imprinted by anisotropic etching. SnTe is grown in layers on the surfaces by using molecular beam epitaxy, with a thickness of about ten layers. SnTe will experience the uniaxial strain to gap out its surfaces and protect the helical HOTI phase. The hinge states can be studied by scanning tunneling microscopy and transport experiments with contacts applied through electronic-beam lithography. Note that in the process of growth, regions with step edges are likely to form on the surfaces and should be avoided in measurements, as they may carry their own gapless modes~\cite{Sessi2016}. Alternatively, we propose to use a superconducting substrate to study proximity-induced superconductivity on the helical hinge states.

In addition to the topological crystalline insulator SnTe, we propose weak TIs with nonvanishing mirror Chern number as possible avenues to realize helical HOTIs. We computed the relevant mirror Chern numbers for the weak TIs Bi$_2$TeI\cite{Rusinov16}, BiSe\cite{Kunjalata17}, and BiTe\cite{Eschbach2017}, which all turn out to be $2$. These materials are therefore dual topological insulators, in the sense that they carry nontrivial weak \emph{and} crystalline topological invariants. 
Their surface Dirac cones are protected by a nontrivial weak index, i.e., by time reversal together with translation symmetry. To gap them, it is necessary to break at least one of these symmetries, which is possible by inducing magnetic or charge density wave order.

\begin{center}
\textbf{
Summary}
\end{center}

We have introduced 3D HOTIs, which have gapped surfaces, but gapless hinge modes, as intrinsically 3D topological phases of matter. Both time-reversal symmetry breaking and time-reversal symmetric systems were explored, which support hinge states akin to those of the integer quantum Hall effect and 2D time-reversal symmetric TIs, respectively. The former may be realized in magnetically ordered topological insulators; we propose the naturally occurring rhombohedral or a uniaxially distorted phase of SnTe as a material realization for the latter. Despite their global topological characterization based on spatial symmetries, the hinge states are as robust against local perturbations as quantum (spin) Hall edge modes. The concepts introduced here can be extended to define novel topological superconductors with chiral and helical Majorana modes at their hinges and may further be transferred to strongly interacting, possibly topologically ordered, states of matter and to mechanical\cite{Huber18}, electrical\cite{Imhof17} and photonic analogues of Bloch Hamiltonians.

\newpage 
\begin{center}
\textbf{
Acknowledgments}
\end{center}

FS and TN acknowledge support from the Swiss National Science Foundation (grant number: 200021\_169061) and from the European Union's Horizon 2020 research and innovation program (ERC-StG-Neupert-757867-PARATOP). AMC wishes to thank the Aspen Center for Physics, which is supported by National Science Foundation grant PHY-1066293, for hosting during some stages of this work. MGV was supported by FIS2016- 75862-P  national projects of the Spanish MINECO. BAB acknowledges support for the analytic work Department of Energy de-sc0016239, Simons Investigator Award, the Packard Foundation, and the Schmidt Fund for Innovative Research. The computational part of the Princeton work was performed under NSF EAGER grant DMR -- 1643312, ONR - N00014-14-1-0330, ARO MURI W911NF-12-1-0461, NSF-MRSEC DMR-1420541. BAB also wishes to thank Ecole Normale Superieure, UPMC Paris, and Donostia International Physics Center for their generous sabbatical hosting during some of the stages of this work. \textbf{Competing interests:} the authors declare that they have no competing interests. \textbf{Author contributions:} FS, AMC, AB and TN worked out the theoretical results presented here, MGV and ZW performed first-principles calculations, and SSP contributed the experimental proposal for SnTe nanowires. \textbf{Data availability:} all data needed to evaluate the conclusions in the paper are present in the paper and/or the Supplementary Materials. Additional data available from authors upon reasonable request.

\clearpage
\let\oldaddcontentsline\addcontentsline
\renewcommand{\addcontentsline}[3]{}

\bibliography{Ref-Lib}

\begin{thebibliography}{40}%
\makeatletter
\providecommand \@ifxundefined [1]{%
 \@ifx{#1\undefined}
}%
\providecommand \@ifnum [1]{%
 \ifnum #1\expandafter \@firstoftwo
 \else \expandafter \@secondoftwo
 \fi
}%
\providecommand \@ifx [1]{%
 \ifx #1\expandafter \@firstoftwo
 \else \expandafter \@secondoftwo
 \fi
}%
\providecommand \natexlab [1]{#1}%
\providecommand \enquote  [1]{``#1''}%
\providecommand \bibnamefont  [1]{#1}%
\providecommand \bibfnamefont [1]{#1}%
\providecommand \citenamefont [1]{#1}%
\providecommand \href@noop [0]{\@secondoftwo}%
\providecommand \href [0]{\begingroup \@sanitize@url \@href}%
\providecommand \@href[1]{\@@startlink{#1}\@@href}%
\providecommand \@@href[1]{\endgroup#1\@@endlink}%
\providecommand \@sanitize@url [0]{\catcode `\\12\catcode `\$12\catcode
  `\&12\catcode `\#12\catcode `\^12\catcode `\_12\catcode `\%12\relax}%
\providecommand \@@startlink[1]{}%
\providecommand \@@endlink[0]{}%
\providecommand \url  [0]{\begingroup\@sanitize@url \@url }%
\providecommand \@url [1]{\endgroup\@href {#1}{\urlprefix }}%
\providecommand \urlprefix  [0]{URL }%
\providecommand \Eprint [0]{\href }%
\providecommand \doibase [0]{http://dx.doi.org/}%
\providecommand \selectlanguage [0]{\@gobble}%
\providecommand \bibinfo  [0]{\@secondoftwo}%
\providecommand \bibfield  [0]{\@secondoftwo}%
\providecommand \translation [1]{[#1]}%
\providecommand \BibitemOpen [0]{}%
\providecommand \bibitemStop [0]{}%
\providecommand \bibitemNoStop [0]{.\EOS\space}%
\providecommand \EOS [0]{\spacefactor3000\relax}%
\providecommand \BibitemShut  [1]{\csname bibitem#1\endcsname}%
\let\auto@bib@innerbib\@empty
\bibitem [{\citenamefont {Kane}\ and\ \citenamefont {Mele}(2005)}]{Kane05a}%
  \BibitemOpen
  \bibfield  {author} {\bibinfo {author} {\bibfnamefont {C.~L.}\ \bibnamefont
  {Kane}}\ and\ \bibinfo {author} {\bibfnamefont {E.~J.}\ \bibnamefont
  {Mele}},\ }\href {\doibase 10.1103/PhysRevLett.95.146802} {\bibfield
  {journal} {\bibinfo  {journal} {Phys. Rev. Lett.}\ }\textbf {\bibinfo
  {volume} {95}},\ \bibinfo {pages} {146802} (\bibinfo {year}
  {2005})}\BibitemShut {NoStop}%
\bibitem [{\citenamefont {Fu}\ \emph {et~al.}(2007)\citenamefont {Fu},
  \citenamefont {Kane},\ and\ \citenamefont {Mele}}]{Kane07}%
  \BibitemOpen
  \bibfield  {author} {\bibinfo {author} {\bibfnamefont {L.}~\bibnamefont
  {Fu}}, \bibinfo {author} {\bibfnamefont {C.~L.}\ \bibnamefont {Kane}}, \ and\
  \bibinfo {author} {\bibfnamefont {E.~J.}\ \bibnamefont {Mele}},\ }\href
  {\doibase 10.1103/PhysRevLett.98.106803} {\bibfield  {journal} {\bibinfo
  {journal} {Phys. Rev. Lett.}\ }\textbf {\bibinfo {volume} {98}},\ \bibinfo
  {pages} {106803} (\bibinfo {year} {2007})}\BibitemShut {NoStop}%
\bibitem [{\citenamefont {Fu}\ and\ \citenamefont {Kane}(2007)}]{FuKane2007}%
  \BibitemOpen
  \bibfield  {author} {\bibinfo {author} {\bibfnamefont {L.}~\bibnamefont
  {Fu}}\ and\ \bibinfo {author} {\bibfnamefont {C.~L.}\ \bibnamefont {Kane}},\
  }\href {\doibase 10.1103/PhysRevB.76.045302} {\bibfield  {journal} {\bibinfo
  {journal} {Phys. Rev. B}\ }\textbf {\bibinfo {volume} {76}},\ \bibinfo
  {pages} {045302} (\bibinfo {year} {2007})}\BibitemShut {NoStop}%
\bibitem [{\citenamefont {Benalcazar}\ \emph
  {et~al.}(2017{\natexlab{a}})\citenamefont {Benalcazar}, \citenamefont
  {Bernevig},\ and\ \citenamefont {Hughes}}]{Benalcazar16}%
  \BibitemOpen
  \bibfield  {author} {\bibinfo {author} {\bibfnamefont {W.~A.}\ \bibnamefont
  {Benalcazar}}, \bibinfo {author} {\bibfnamefont {B.~A.}\ \bibnamefont
  {Bernevig}}, \ and\ \bibinfo {author} {\bibfnamefont {T.~L.}\ \bibnamefont
  {Hughes}},\ }\href {\doibase 10.1126/science.aah6442} {\bibfield  {journal}
  {\bibinfo  {journal} {Science}\ }\textbf {\bibinfo {volume} {357}},\ \bibinfo
  {pages} {61} (\bibinfo {year} {2017}{\natexlab{a}})},\ \Eprint
  {http://arxiv.org/abs/http://science.sciencemag.org/content/357/6346/61.full.pdf}
  {http://science.sciencemag.org/content/357/6346/61.full.pdf} \BibitemShut
  {NoStop}%
\bibitem [{\citenamefont {Fu}(2011)}]{Fu11}%
  \BibitemOpen
  \bibfield  {author} {\bibinfo {author} {\bibfnamefont {L.}~\bibnamefont
  {Fu}},\ }\href {\doibase 10.1103/PhysRevLett.106.106802} {\bibfield
  {journal} {\bibinfo  {journal} {Phys. Rev. Lett.}\ }\textbf {\bibinfo
  {volume} {106}},\ \bibinfo {pages} {106802} (\bibinfo {year}
  {2011})}\BibitemShut {NoStop}%
\bibitem [{\citenamefont {Hsieh}\ \emph {et~al.}(2012)\citenamefont {Hsieh},
  \citenamefont {Lin}, \citenamefont {Liu}, \citenamefont {Duan}, \citenamefont
  {Bansil},\ and\ \citenamefont {Fu}}]{Hsieh12}%
  \BibitemOpen
  \bibfield  {author} {\bibinfo {author} {\bibfnamefont {T.~H.}\ \bibnamefont
  {Hsieh}}, \bibinfo {author} {\bibfnamefont {H.}~\bibnamefont {Lin}}, \bibinfo
  {author} {\bibfnamefont {J.}~\bibnamefont {Liu}}, \bibinfo {author}
  {\bibfnamefont {W.}~\bibnamefont {Duan}}, \bibinfo {author} {\bibfnamefont
  {A.}~\bibnamefont {Bansil}}, \ and\ \bibinfo {author} {\bibfnamefont
  {L.}~\bibnamefont {Fu}},\ }\href {http://dx.doi.org/10.1038/ncomms1969}
  {\bibfield  {journal} {\bibinfo  {journal} {Nat Commun}\ }\textbf {\bibinfo
  {volume} {3}},\ \bibinfo {pages} {982} (\bibinfo {year} {2012})}\BibitemShut
  {NoStop}%
\bibitem [{\citenamefont {Bradlyn}\ \emph {et~al.}(2017)\citenamefont
  {Bradlyn}, \citenamefont {Elcoro}, \citenamefont {Cano}, \citenamefont
  {Vergniory}, \citenamefont {Wang}, \citenamefont {Felser}, \citenamefont
  {Aroyo},\ and\ \citenamefont {Bernevig}}]{Bradlyn17}%
  \BibitemOpen
  \bibfield  {author} {\bibinfo {author} {\bibfnamefont {B.}~\bibnamefont
  {Bradlyn}}, \bibinfo {author} {\bibfnamefont {L.}~\bibnamefont {Elcoro}},
  \bibinfo {author} {\bibfnamefont {J.}~\bibnamefont {Cano}}, \bibinfo {author}
  {\bibfnamefont {M.~G.}\ \bibnamefont {Vergniory}}, \bibinfo {author}
  {\bibfnamefont {Z.}~\bibnamefont {Wang}}, \bibinfo {author} {\bibfnamefont
  {C.}~\bibnamefont {Felser}}, \bibinfo {author} {\bibfnamefont {M.~I.}\
  \bibnamefont {Aroyo}}, \ and\ \bibinfo {author} {\bibfnamefont {B.~A.}\
  \bibnamefont {Bernevig}},\ }\href {http://dx.doi.org/10.1038/nature23268}
  {\bibfield  {journal} {\bibinfo  {journal} {Nature}\ }\textbf {\bibinfo
  {volume} {547}},\ \bibinfo {pages} {298} (\bibinfo {year}
  {2017})}\BibitemShut {NoStop}%
\bibitem [{\citenamefont {Klitzing}\ \emph {et~al.}(1980)\citenamefont
  {Klitzing}, \citenamefont {Dorda},\ and\ \citenamefont
  {Pepper}}]{Klitzing80}%
  \BibitemOpen
  \bibfield  {author} {\bibinfo {author} {\bibfnamefont {K.~v.}\ \bibnamefont
  {Klitzing}}, \bibinfo {author} {\bibfnamefont {G.}~\bibnamefont {Dorda}}, \
  and\ \bibinfo {author} {\bibfnamefont {M.}~\bibnamefont {Pepper}},\ }\href
  {\doibase 10.1103/PhysRevLett.45.494} {\bibfield  {journal} {\bibinfo
  {journal} {Phys. Rev. Lett.}\ }\textbf {\bibinfo {volume} {45}},\ \bibinfo
  {pages} {494} (\bibinfo {year} {1980})}\BibitemShut {NoStop}%
\bibitem [{\citenamefont {Haldane}(1988)}]{Haldane88}%
  \BibitemOpen
  \bibfield  {author} {\bibinfo {author} {\bibfnamefont {F.~D.~M.}\
  \bibnamefont {Haldane}},\ }\href {\doibase 10.1103/PhysRevLett.61.2015}
  {\bibfield  {journal} {\bibinfo  {journal} {Phys. Rev. Lett.}\ }\textbf
  {\bibinfo {volume} {61}},\ \bibinfo {pages} {2015} (\bibinfo {year}
  {1988})}\BibitemShut {NoStop}%
\bibitem [{\citenamefont {Qi}\ \emph {et~al.}(2006)\citenamefont {Qi},
  \citenamefont {Wu},\ and\ \citenamefont {Zhang}}]{Qi2006}%
  \BibitemOpen
  \bibfield  {author} {\bibinfo {author} {\bibfnamefont {X.-L.}\ \bibnamefont
  {Qi}}, \bibinfo {author} {\bibfnamefont {Y.-S.}\ \bibnamefont {Wu}}, \ and\
  \bibinfo {author} {\bibfnamefont {S.-C.}\ \bibnamefont {Zhang}},\ }\href
  {\doibase 10.1103/PhysRevB.74.085308} {\bibfield  {journal} {\bibinfo
  {journal} {Phys. Rev. B}\ }\textbf {\bibinfo {volume} {74}},\ \bibinfo
  {pages} {085308} (\bibinfo {year} {2006})}\BibitemShut {NoStop}%
\bibitem [{\citenamefont {Bernevig}\ \emph {et~al.}(2006)\citenamefont
  {Bernevig}, \citenamefont {Hughes},\ and\ \citenamefont
  {Zhang}}]{Bernevig2006a}%
  \BibitemOpen
  \bibfield  {author} {\bibinfo {author} {\bibfnamefont {B.~A.}\ \bibnamefont
  {Bernevig}}, \bibinfo {author} {\bibfnamefont {T.~L.}\ \bibnamefont
  {Hughes}}, \ and\ \bibinfo {author} {\bibfnamefont {S.-C.}\ \bibnamefont
  {Zhang}},\ }\href {\doibase 10.1126/science.1133734} {\bibfield  {journal}
  {\bibinfo  {journal} {Science}\ }\textbf {\bibinfo {volume} {314}},\ \bibinfo
  {pages} {1757} (\bibinfo {year} {2006})},\ \Eprint
  {http://arxiv.org/abs/http://science.sciencemag.org/content/314/5806/1757.full.pdf}
  {http://science.sciencemag.org/content/314/5806/1757.full.pdf} \BibitemShut
  {NoStop}%
\bibitem [{\citenamefont {Bernevig}\ and\ \citenamefont
  {Zhang}(2006)}]{Bernevig2006b}%
  \BibitemOpen
  \bibfield  {author} {\bibinfo {author} {\bibfnamefont {B.~A.}\ \bibnamefont
  {Bernevig}}\ and\ \bibinfo {author} {\bibfnamefont {S.-C.}\ \bibnamefont
  {Zhang}},\ }\href {\doibase 10.1103/PhysRevLett.96.106802} {\bibfield
  {journal} {\bibinfo  {journal} {Phys. Rev. Lett.}\ }\textbf {\bibinfo
  {volume} {96}},\ \bibinfo {pages} {106802} (\bibinfo {year}
  {2006})}\BibitemShut {NoStop}%
\bibitem [{\citenamefont {Teo}\ \emph {et~al.}(2008)\citenamefont {Teo},
  \citenamefont {Fu},\ and\ \citenamefont {Kane}}]{Teo08}%
  \BibitemOpen
  \bibfield  {author} {\bibinfo {author} {\bibfnamefont {J.~C.~Y.}\
  \bibnamefont {Teo}}, \bibinfo {author} {\bibfnamefont {L.}~\bibnamefont
  {Fu}}, \ and\ \bibinfo {author} {\bibfnamefont {C.~L.}\ \bibnamefont
  {Kane}},\ }\href {\doibase 10.1103/PhysRevB.78.045426} {\bibfield  {journal}
  {\bibinfo  {journal} {Phys. Rev. B}\ }\textbf {\bibinfo {volume} {78}},\
  \bibinfo {pages} {045426} (\bibinfo {year} {2008})}\BibitemShut {NoStop}%
\bibitem [{\citenamefont {Langbehn}\ \emph {et~al.}(2017)\citenamefont
  {Langbehn}, \citenamefont {Peng}, \citenamefont {Trifunovic}, \citenamefont
  {von Oppen},\ and\ \citenamefont {Brouwer}}]{Brouwer17}%
  \BibitemOpen
  \bibfield  {author} {\bibinfo {author} {\bibfnamefont {J.}~\bibnamefont
  {Langbehn}}, \bibinfo {author} {\bibfnamefont {Y.}~\bibnamefont {Peng}},
  \bibinfo {author} {\bibfnamefont {L.}~\bibnamefont {Trifunovic}}, \bibinfo
  {author} {\bibfnamefont {F.}~\bibnamefont {von Oppen}}, \ and\ \bibinfo
  {author} {\bibfnamefont {P.~W.}\ \bibnamefont {Brouwer}},\ }\href {\doibase
  10.1103/PhysRevLett.119.246401} {\bibfield  {journal} {\bibinfo  {journal}
  {Phys. Rev. Lett.}\ }\textbf {\bibinfo {volume} {119}},\ \bibinfo {pages}
  {246401} (\bibinfo {year} {2017})}\BibitemShut {NoStop}%
\bibitem [{\citenamefont {Benalcazar}\ \emph
  {et~al.}(2017{\natexlab{b}})\citenamefont {Benalcazar}, \citenamefont
  {Bernevig},\ and\ \citenamefont {Hughes}}]{BABHughesBenalcazar17}%
  \BibitemOpen
  \bibfield  {author} {\bibinfo {author} {\bibfnamefont {W.~A.}\ \bibnamefont
  {Benalcazar}}, \bibinfo {author} {\bibfnamefont {B.~A.}\ \bibnamefont
  {Bernevig}}, \ and\ \bibinfo {author} {\bibfnamefont {T.~L.}\ \bibnamefont
  {Hughes}},\ }\href {\doibase 10.1103/PhysRevB.96.245115} {\bibfield
  {journal} {\bibinfo  {journal} {Phys. Rev. B}\ }\textbf {\bibinfo {volume}
  {96}},\ \bibinfo {pages} {245115} (\bibinfo {year}
  {2017}{\natexlab{b}})}\BibitemShut {NoStop}%
\bibitem [{\citenamefont {Sitte}\ \emph {et~al.}(2012)\citenamefont {Sitte},
  \citenamefont {Rosch}, \citenamefont {Altman},\ and\ \citenamefont
  {Fritz}}]{Sitte12}%
  \BibitemOpen
  \bibfield  {author} {\bibinfo {author} {\bibfnamefont {M.}~\bibnamefont
  {Sitte}}, \bibinfo {author} {\bibfnamefont {A.}~\bibnamefont {Rosch}},
  \bibinfo {author} {\bibfnamefont {E.}~\bibnamefont {Altman}}, \ and\ \bibinfo
  {author} {\bibfnamefont {L.}~\bibnamefont {Fritz}},\ }\href {\doibase
  10.1103/PhysRevLett.108.126807} {\bibfield  {journal} {\bibinfo  {journal}
  {Phys. Rev. Lett.}\ }\textbf {\bibinfo {volume} {108}},\ \bibinfo {pages}
  {126807} (\bibinfo {year} {2012})}\BibitemShut {NoStop}%
\bibitem [{\citenamefont {Zhang}\ \emph {et~al.}(2013)\citenamefont {Zhang},
  \citenamefont {Kane},\ and\ \citenamefont {Mele}}]{FanZhang13}%
  \BibitemOpen
  \bibfield  {author} {\bibinfo {author} {\bibfnamefont {F.}~\bibnamefont
  {Zhang}}, \bibinfo {author} {\bibfnamefont {C.~L.}\ \bibnamefont {Kane}}, \
  and\ \bibinfo {author} {\bibfnamefont {E.~J.}\ \bibnamefont {Mele}},\ }\href
  {\doibase 10.1103/PhysRevLett.110.046404} {\bibfield  {journal} {\bibinfo
  {journal} {Phys. Rev. Lett.}\ }\textbf {\bibinfo {volume} {110}},\ \bibinfo
  {pages} {046404} (\bibinfo {year} {2013})}\BibitemShut {NoStop}%
\bibitem [{\citenamefont {Qi}\ \emph {et~al.}(2008)\citenamefont {Qi},
  \citenamefont {Hughes},\ and\ \citenamefont {Zhang}}]{Qi08}%
  \BibitemOpen
  \bibfield  {author} {\bibinfo {author} {\bibfnamefont {X.-L.}\ \bibnamefont
  {Qi}}, \bibinfo {author} {\bibfnamefont {T.~L.}\ \bibnamefont {Hughes}}, \
  and\ \bibinfo {author} {\bibfnamefont {S.-C.}\ \bibnamefont {Zhang}},\ }\href
  {\doibase 10.1103/PhysRevB.78.195424} {\bibfield  {journal} {\bibinfo
  {journal} {Phys. Rev. B}\ }\textbf {\bibinfo {volume} {78}},\ \bibinfo
  {pages} {195424} (\bibinfo {year} {2008})}\BibitemShut {NoStop}%
\bibitem [{\citenamefont {Yu}\ \emph {et~al.}(2011)\citenamefont {Yu},
  \citenamefont {Qi}, \citenamefont {Bernevig}, \citenamefont {Fang},\ and\
  \citenamefont {Dai}}]{Yu11}%
  \BibitemOpen
  \bibfield  {author} {\bibinfo {author} {\bibfnamefont {R.}~\bibnamefont
  {Yu}}, \bibinfo {author} {\bibfnamefont {X.~L.}\ \bibnamefont {Qi}}, \bibinfo
  {author} {\bibfnamefont {A.}~\bibnamefont {Bernevig}}, \bibinfo {author}
  {\bibfnamefont {Z.}~\bibnamefont {Fang}}, \ and\ \bibinfo {author}
  {\bibfnamefont {X.}~\bibnamefont {Dai}},\ }\href {\doibase
  10.1103/PhysRevB.84.075119} {\bibfield  {journal} {\bibinfo  {journal} {Phys.
  Rev. B}\ }\textbf {\bibinfo {volume} {84}},\ \bibinfo {pages} {075119}
  (\bibinfo {year} {2011})}\BibitemShut {NoStop}%
\bibitem [{\citenamefont {Alexandradinata}\ \emph {et~al.}(2014)\citenamefont
  {Alexandradinata}, \citenamefont {Dai},\ and\ \citenamefont
  {Bernevig}}]{Alexandradinata14}%
  \BibitemOpen
  \bibfield  {author} {\bibinfo {author} {\bibfnamefont {A.}~\bibnamefont
  {Alexandradinata}}, \bibinfo {author} {\bibfnamefont {X.}~\bibnamefont
  {Dai}}, \ and\ \bibinfo {author} {\bibfnamefont {B.~A.}\ \bibnamefont
  {Bernevig}},\ }\href {\doibase 10.1103/PhysRevB.89.155114} {\bibfield
  {journal} {\bibinfo  {journal} {Phys. Rev. B}\ }\textbf {\bibinfo {volume}
  {89}},\ \bibinfo {pages} {155114} (\bibinfo {year} {2014})}\BibitemShut
  {NoStop}%
\bibitem [{\citenamefont {Li}\ and\ \citenamefont {Haldane}(2008)}]{Li08}%
  \BibitemOpen
  \bibfield  {author} {\bibinfo {author} {\bibfnamefont {H.}~\bibnamefont
  {Li}}\ and\ \bibinfo {author} {\bibfnamefont {F.~D.~M.}\ \bibnamefont
  {Haldane}},\ }\href {\doibase 10.1103/PhysRevLett.101.010504} {\bibfield
  {journal} {\bibinfo  {journal} {Phys. Rev. Lett.}\ }\textbf {\bibinfo
  {volume} {101}},\ \bibinfo {pages} {010504} (\bibinfo {year}
  {2008})}\BibitemShut {NoStop}%
\bibitem [{\citenamefont {Peschel}(2003)}]{Peschel03}%
  \BibitemOpen
  \bibfield  {author} {\bibinfo {author} {\bibfnamefont {I.}~\bibnamefont
  {Peschel}},\ }\href {http://stacks.iop.org/0305-4470/36/i=14/a=101}
  {\bibfield  {journal} {\bibinfo  {journal} {Journal of Physics A:
  Mathematical and General}\ }\textbf {\bibinfo {volume} {36}},\ \bibinfo
  {pages} {L205} (\bibinfo {year} {2003})}\BibitemShut {NoStop}%
\bibitem [{\citenamefont {Fidkowski}(2010)}]{Fidkowski10-2}%
  \BibitemOpen
  \bibfield  {author} {\bibinfo {author} {\bibfnamefont {L.}~\bibnamefont
  {Fidkowski}},\ }\href {\doibase 10.1103/PhysRevLett.104.130502} {\bibfield
  {journal} {\bibinfo  {journal} {Phys. Rev. Lett.}\ }\textbf {\bibinfo
  {volume} {104}},\ \bibinfo {pages} {130502} (\bibinfo {year}
  {2010})}\BibitemShut {NoStop}%
\bibitem [{\citenamefont {Jackiw}\ and\ \citenamefont
  {Rebbi}(1976)}]{Jackiw76}%
  \BibitemOpen
  \bibfield  {author} {\bibinfo {author} {\bibfnamefont {R.}~\bibnamefont
  {Jackiw}}\ and\ \bibinfo {author} {\bibfnamefont {C.}~\bibnamefont {Rebbi}},\
  }\href {\doibase 10.1103/PhysRevD.13.3398} {\bibfield  {journal} {\bibinfo
  {journal} {Phys. Rev. D}\ }\textbf {\bibinfo {volume} {13}},\ \bibinfo
  {pages} {3398} (\bibinfo {year} {1976})}\BibitemShut {NoStop}%
\bibitem [{\citenamefont {Iizumi}\ \emph {et~al.}(1975)\citenamefont {Iizumi},
  \citenamefont {Hamaguchi}, \citenamefont {Komatsubara},\ and\ \citenamefont
  {Kato}}]{Iizumi75}%
  \BibitemOpen
  \bibfield  {author} {\bibinfo {author} {\bibfnamefont {M.}~\bibnamefont
  {Iizumi}}, \bibinfo {author} {\bibfnamefont {Y.}~\bibnamefont {Hamaguchi}},
  \bibinfo {author} {\bibfnamefont {K.~F.}\ \bibnamefont {Komatsubara}}, \ and\
  \bibinfo {author} {\bibfnamefont {Y.}~\bibnamefont {Kato}},\ }\href {\doibase
  10.1143/JPSJ.38.443} {\bibfield  {journal} {\bibinfo  {journal} {Journal of
  the Physical Society of Japan}\ }\textbf {\bibinfo {volume} {38}},\ \bibinfo
  {pages} {443} (\bibinfo {year} {1975})},\ \Eprint
  {http://arxiv.org/abs/http://dx.doi.org/10.1143/JPSJ.38.443}
  {http://dx.doi.org/10.1143/JPSJ.38.443} \BibitemShut {NoStop}%
\bibitem [{\citenamefont {Liang}\ \emph {et~al.}(2017)\citenamefont {Liang},
  \citenamefont {Kushwaha}, \citenamefont {Kim}, \citenamefont {Gibson},
  \citenamefont {Lin}, \citenamefont {Kioussis}, \citenamefont {Cava},\ and\
  \citenamefont {Ong}}]{Liang17}%
  \BibitemOpen
  \bibfield  {author} {\bibinfo {author} {\bibfnamefont {T.}~\bibnamefont
  {Liang}}, \bibinfo {author} {\bibfnamefont {S.}~\bibnamefont {Kushwaha}},
  \bibinfo {author} {\bibfnamefont {J.}~\bibnamefont {Kim}}, \bibinfo {author}
  {\bibfnamefont {Q.}~\bibnamefont {Gibson}}, \bibinfo {author} {\bibfnamefont
  {J.}~\bibnamefont {Lin}}, \bibinfo {author} {\bibfnamefont {N.}~\bibnamefont
  {Kioussis}}, \bibinfo {author} {\bibfnamefont {R.~J.}\ \bibnamefont {Cava}},
  \ and\ \bibinfo {author} {\bibfnamefont {N.~P.}\ \bibnamefont {Ong}},\ }\href
  {\doibase 10.1126/sciadv.1602510} {\bibfield  {journal} {\bibinfo  {journal}
  {Science Advances}\ }\textbf {\bibinfo {volume} {3}} (\bibinfo {year}
  {2017}),\ 10.1126/sciadv.1602510},\ \Eprint
  {http://arxiv.org/abs/http://advances.sciencemag.org/content/3/5/e1602510.full.pdf}
  {http://advances.sciencemag.org/content/3/5/e1602510.full.pdf} \BibitemShut
  {NoStop}%
\bibitem [{\citenamefont {Sessi}\ \emph {et~al.}(2016)\citenamefont {Sessi},
  \citenamefont {Di~Sante}, \citenamefont {Szczerbakow}, \citenamefont {Glott},
  \citenamefont {Wilfert}, \citenamefont {Schmidt}, \citenamefont {Bathon},
  \citenamefont {Dziawa}, \citenamefont {Greiter}, \citenamefont {Neupert},
  \citenamefont {Sangiovanni}, \citenamefont {Story}, \citenamefont {Thomale},\
  and\ \citenamefont {Bode}}]{Sessi2016}%
  \BibitemOpen
  \bibfield  {author} {\bibinfo {author} {\bibfnamefont {P.}~\bibnamefont
  {Sessi}}, \bibinfo {author} {\bibfnamefont {D.}~\bibnamefont {Di~Sante}},
  \bibinfo {author} {\bibfnamefont {A.}~\bibnamefont {Szczerbakow}}, \bibinfo
  {author} {\bibfnamefont {F.}~\bibnamefont {Glott}}, \bibinfo {author}
  {\bibfnamefont {S.}~\bibnamefont {Wilfert}}, \bibinfo {author} {\bibfnamefont
  {H.}~\bibnamefont {Schmidt}}, \bibinfo {author} {\bibfnamefont
  {T.}~\bibnamefont {Bathon}}, \bibinfo {author} {\bibfnamefont
  {P.}~\bibnamefont {Dziawa}}, \bibinfo {author} {\bibfnamefont
  {M.}~\bibnamefont {Greiter}}, \bibinfo {author} {\bibfnamefont
  {T.}~\bibnamefont {Neupert}}, \bibinfo {author} {\bibfnamefont
  {G.}~\bibnamefont {Sangiovanni}}, \bibinfo {author} {\bibfnamefont
  {T.}~\bibnamefont {Story}}, \bibinfo {author} {\bibfnamefont
  {R.}~\bibnamefont {Thomale}}, \ and\ \bibinfo {author} {\bibfnamefont
  {M.}~\bibnamefont {Bode}},\ }\href {\doibase 10.1126/science.aah6233}
  {\bibfield  {journal} {\bibinfo  {journal} {Science}\ }\textbf {\bibinfo
  {volume} {354}},\ \bibinfo {pages} {1269} (\bibinfo {year} {2016})},\ \Eprint
  {http://arxiv.org/abs/http://science.sciencemag.org/content/354/6317/1269.full.pdf}
  {http://science.sciencemag.org/content/354/6317/1269.full.pdf} \BibitemShut
  {NoStop}%
\bibitem [{\citenamefont {Rusinov}\ \emph {et~al.}(2016)\citenamefont
  {Rusinov}, \citenamefont {Menshchikova}, \citenamefont {Isaeva},
  \citenamefont {Eremeev}, \citenamefont {Koroteev}, \citenamefont {Vergniory},
  \citenamefont {Echenique},\ and\ \citenamefont {Chulkov}}]{Rusinov16}%
  \BibitemOpen
  \bibfield  {author} {\bibinfo {author} {\bibfnamefont {I.~P.}\ \bibnamefont
  {Rusinov}}, \bibinfo {author} {\bibfnamefont {T.~V.}\ \bibnamefont
  {Menshchikova}}, \bibinfo {author} {\bibfnamefont {A.}~\bibnamefont
  {Isaeva}}, \bibinfo {author} {\bibfnamefont {S.~V.}\ \bibnamefont {Eremeev}},
  \bibinfo {author} {\bibfnamefont {Y.~M.}\ \bibnamefont {Koroteev}}, \bibinfo
  {author} {\bibfnamefont {M.~G.}\ \bibnamefont {Vergniory}}, \bibinfo {author}
  {\bibfnamefont {P.~M.}\ \bibnamefont {Echenique}}, \ and\ \bibinfo {author}
  {\bibfnamefont {E.~V.}\ \bibnamefont {Chulkov}},\ }\href
  {http://dx.doi.org/10.1038/srep20734} {\bibfield  {journal} {\bibinfo
  {journal} {Scientific Reports}\ }\textbf {\bibinfo {volume} {6}},\ \bibinfo
  {pages} {20734 EP } (\bibinfo {year} {2016})}\BibitemShut {NoStop}%
\bibitem [{\citenamefont {Majhi}\ \emph {et~al.}(2017)\citenamefont {Majhi},
  \citenamefont {Pal}, \citenamefont {Lohani}, \citenamefont {Banerjee},
  \citenamefont {Mishra}, \citenamefont {Yadav}, \citenamefont {Ganesan},
  \citenamefont {Sekhar}, \citenamefont {Waghmare},\ and\ \citenamefont
  {Kumar}}]{Kunjalata17}%
  \BibitemOpen
  \bibfield  {author} {\bibinfo {author} {\bibfnamefont {K.}~\bibnamefont
  {Majhi}}, \bibinfo {author} {\bibfnamefont {K.}~\bibnamefont {Pal}}, \bibinfo
  {author} {\bibfnamefont {H.}~\bibnamefont {Lohani}}, \bibinfo {author}
  {\bibfnamefont {A.}~\bibnamefont {Banerjee}}, \bibinfo {author}
  {\bibfnamefont {P.}~\bibnamefont {Mishra}}, \bibinfo {author} {\bibfnamefont
  {A.~K.}\ \bibnamefont {Yadav}}, \bibinfo {author} {\bibfnamefont
  {R.}~\bibnamefont {Ganesan}}, \bibinfo {author} {\bibfnamefont {B.~R.}\
  \bibnamefont {Sekhar}}, \bibinfo {author} {\bibfnamefont {U.~V.}\
  \bibnamefont {Waghmare}}, \ and\ \bibinfo {author} {\bibfnamefont {P.~S.~A.}\
  \bibnamefont {Kumar}},\ }\href {\doibase 10.1063/1.4981875} {\bibfield
  {journal} {\bibinfo  {journal} {Applied Physics Letters}\ }\textbf {\bibinfo
  {volume} {110}},\ \bibinfo {pages} {162102} (\bibinfo {year} {2017})},\
  \Eprint {http://arxiv.org/abs/http://dx.doi.org/10.1063/1.4981875}
  {http://dx.doi.org/10.1063/1.4981875} \BibitemShut {NoStop}%
\bibitem [{\citenamefont {Eschbach}\ \emph {et~al.}(2017)\citenamefont
  {Eschbach}, \citenamefont {Lanius}, \citenamefont {Niu}, \citenamefont
  {Mlynczak}, \citenamefont {Gospodaric}, \citenamefont {Kellner},
  \citenamefont {Sch\"uffelgen}, \citenamefont {Gehlmann}, \citenamefont
  {D\"oring}, \citenamefont {Neumann}, \citenamefont {Luysberg}, \citenamefont
  {Mussler}, \citenamefont {Plucinski}, \citenamefont {Morgenstern},
  \citenamefont {Gr\"utzmacher}, \citenamefont {Bihlmayer}, \citenamefont
  {Bl\"ugel},\ and\ \citenamefont {Schneider}}]{Eschbach2017}%
  \BibitemOpen
  \bibfield  {author} {\bibinfo {author} {\bibfnamefont {M.}~\bibnamefont
  {Eschbach}}, \bibinfo {author} {\bibfnamefont {M.}~\bibnamefont {Lanius}},
  \bibinfo {author} {\bibfnamefont {C.}~\bibnamefont {Niu}}, \bibinfo {author}
  {\bibfnamefont {E.}~\bibnamefont {Mlynczak}}, \bibinfo {author}
  {\bibfnamefont {P.}~\bibnamefont {Gospodaric}}, \bibinfo {author}
  {\bibfnamefont {J.}~\bibnamefont {Kellner}}, \bibinfo {author} {\bibfnamefont
  {P.}~\bibnamefont {Sch\"uffelgen}}, \bibinfo {author} {\bibfnamefont
  {M.}~\bibnamefont {Gehlmann}}, \bibinfo {author} {\bibfnamefont
  {S.}~\bibnamefont {D\"oring}}, \bibinfo {author} {\bibfnamefont
  {E.}~\bibnamefont {Neumann}}, \bibinfo {author} {\bibfnamefont
  {M.}~\bibnamefont {Luysberg}}, \bibinfo {author} {\bibfnamefont
  {G.}~\bibnamefont {Mussler}}, \bibinfo {author} {\bibfnamefont
  {L.}~\bibnamefont {Plucinski}}, \bibinfo {author} {\bibfnamefont
  {M.}~\bibnamefont {Morgenstern}}, \bibinfo {author} {\bibfnamefont
  {D.}~\bibnamefont {Gr\"utzmacher}}, \bibinfo {author} {\bibfnamefont
  {G.}~\bibnamefont {Bihlmayer}}, \bibinfo {author} {\bibfnamefont
  {S.}~\bibnamefont {Bl\"ugel}}, \ and\ \bibinfo {author} {\bibfnamefont
  {C.~M.}\ \bibnamefont {Schneider}},\ }\href@noop {} {\bibfield  {journal}
  {\bibinfo  {journal} {Nature Communications}\ }\textbf {\bibinfo {volume}
  {8}},\ \bibinfo {pages} {14976} (\bibinfo {year} {2017})}\BibitemShut
  {NoStop}%
\bibitem [{\citenamefont {Serra-Garcia}\ \emph {et~al.}(2018)\citenamefont
  {Serra-Garcia}, \citenamefont {Peri}, \citenamefont {S{\"u}sstrunk},
  \citenamefont {Bilal}, \citenamefont {Larsen}, \citenamefont {Villanueva},\
  and\ \citenamefont {Huber}}]{Huber18}%
  \BibitemOpen
  \bibfield  {author} {\bibinfo {author} {\bibfnamefont {M.}~\bibnamefont
  {Serra-Garcia}}, \bibinfo {author} {\bibfnamefont {V.}~\bibnamefont {Peri}},
  \bibinfo {author} {\bibfnamefont {R.}~\bibnamefont {S{\"u}sstrunk}}, \bibinfo
  {author} {\bibfnamefont {O.~R.}\ \bibnamefont {Bilal}}, \bibinfo {author}
  {\bibfnamefont {T.}~\bibnamefont {Larsen}}, \bibinfo {author} {\bibfnamefont
  {L.~G.}\ \bibnamefont {Villanueva}}, \ and\ \bibinfo {author} {\bibfnamefont
  {S.~D.}\ \bibnamefont {Huber}},\ }\href
  {http://dx.doi.org/10.1038/nature25156} {\bibfield  {journal} {\bibinfo
  {journal} {Nature}\ ,\ \bibinfo {pages} {EP }} (\bibinfo {year}
  {2018})}\BibitemShut {NoStop}%
\bibitem [{\citenamefont {{Imhof}}\ \emph {et~al.}(2017)\citenamefont
  {{Imhof}}, \citenamefont {{Berger}}, \citenamefont {{Bayer}}, \citenamefont
  {{Brehm}}, \citenamefont {{Molenkamp}}, \citenamefont {{Kiessling}},
  \citenamefont {{Schindler}}, \citenamefont {{Lee}}, \citenamefont
  {{Greiter}}, \citenamefont {{Neupert}},\ and\ \citenamefont
  {{Thomale}}}]{Imhof17}%
  \BibitemOpen
  \bibfield  {author} {\bibinfo {author} {\bibfnamefont {S.}~\bibnamefont
  {{Imhof}}}, \bibinfo {author} {\bibfnamefont {C.}~\bibnamefont {{Berger}}},
  \bibinfo {author} {\bibfnamefont {F.}~\bibnamefont {{Bayer}}}, \bibinfo
  {author} {\bibfnamefont {J.}~\bibnamefont {{Brehm}}}, \bibinfo {author}
  {\bibfnamefont {L.}~\bibnamefont {{Molenkamp}}}, \bibinfo {author}
  {\bibfnamefont {T.}~\bibnamefont {{Kiessling}}}, \bibinfo {author}
  {\bibfnamefont {F.}~\bibnamefont {{Schindler}}}, \bibinfo {author}
  {\bibfnamefont {C.~H.}\ \bibnamefont {{Lee}}}, \bibinfo {author}
  {\bibfnamefont {M.}~\bibnamefont {{Greiter}}}, \bibinfo {author}
  {\bibfnamefont {T.}~\bibnamefont {{Neupert}}}, \ and\ \bibinfo {author}
  {\bibfnamefont {R.}~\bibnamefont {{Thomale}}},\ }\href@noop {} {\bibfield
  {journal} {\bibinfo  {journal} {ArXiv e-prints}\ } (\bibinfo {year}
  {2017})},\ \Eprint {http://arxiv.org/abs/1708.03647} {arXiv:1708.03647
  [cond-mat.mes-hall]} \BibitemShut {NoStop}%
\bibitem [{\citenamefont {Kresse}\ and\ \citenamefont
  {Furthmueller}(1996)}]{vasp2}%
  \BibitemOpen
  \bibfield  {author} {\bibinfo {author} {\bibfnamefont {G.}~\bibnamefont
  {Kresse}}\ and\ \bibinfo {author} {\bibfnamefont {J.}~\bibnamefont
  {Furthmueller}},\ }\href {\doibase
  http://dx.doi.org/10.1016/0927-0256(96)00008-0} {\bibfield  {journal}
  {\bibinfo  {journal} {Computational Materials Science}\ }\textbf {\bibinfo
  {volume} {6}},\ \bibinfo {pages} {15 } (\bibinfo {year} {1996})}\BibitemShut
  {NoStop}%
\bibitem [{\citenamefont {Perdew}\ \emph {et~al.}(1996)\citenamefont {Perdew},
  \citenamefont {Burke},\ and\ \citenamefont {Ernzerhof}}]{PBE-1996}%
  \BibitemOpen
  \bibfield  {author} {\bibinfo {author} {\bibfnamefont {J.~P.}\ \bibnamefont
  {Perdew}}, \bibinfo {author} {\bibfnamefont {K.}~\bibnamefont {Burke}}, \
  and\ \bibinfo {author} {\bibfnamefont {M.}~\bibnamefont {Ernzerhof}},\ }\href
  {\doibase 10.1103/PhysRevLett.77.3865} {\bibfield  {journal} {\bibinfo
  {journal} {Phys. Rev. Lett.}\ }\textbf {\bibinfo {volume} {77}},\ \bibinfo
  {pages} {3865} (\bibinfo {year} {1996})}\BibitemShut {NoStop}%
\bibitem [{\citenamefont {Plekhanov}\ \emph {et~al.}(2014)\citenamefont
  {Plekhanov}, \citenamefont {Barone}, \citenamefont {Di~Sante},\ and\
  \citenamefont {Picozzi}}]{Plekhanov14}%
  \BibitemOpen
  \bibfield  {author} {\bibinfo {author} {\bibfnamefont {E.}~\bibnamefont
  {Plekhanov}}, \bibinfo {author} {\bibfnamefont {P.}~\bibnamefont {Barone}},
  \bibinfo {author} {\bibfnamefont {D.}~\bibnamefont {Di~Sante}}, \ and\
  \bibinfo {author} {\bibfnamefont {S.}~\bibnamefont {Picozzi}},\ }\href
  {\doibase 10.1103/PhysRevB.90.161108} {\bibfield  {journal} {\bibinfo
  {journal} {Phys. Rev. B}\ }\textbf {\bibinfo {volume} {90}},\ \bibinfo
  {pages} {161108} (\bibinfo {year} {2014})}\BibitemShut {NoStop}%
\bibitem [{\citenamefont {Hanke}\ \emph {et~al.}(2017)\citenamefont {Hanke},
  \citenamefont {Freimuth}, \citenamefont {Bl{\"u}gel},\ and\ \citenamefont
  {Mokrousov}}]{Hanke16}%
  \BibitemOpen
  \bibfield  {author} {\bibinfo {author} {\bibfnamefont {J.-P.}\ \bibnamefont
  {Hanke}}, \bibinfo {author} {\bibfnamefont {F.}~\bibnamefont {Freimuth}},
  \bibinfo {author} {\bibfnamefont {S.}~\bibnamefont {Bl{\"u}gel}}, \ and\
  \bibinfo {author} {\bibfnamefont {Y.}~\bibnamefont {Mokrousov}},\ }\href
  {http://dx.doi.org/10.1038/srep41078} {\bibfield  {journal} {\bibinfo
  {journal} {Scientific Reports}\ }\textbf {\bibinfo {volume} {7}},\ \bibinfo
  {pages} {41078 EP } (\bibinfo {year} {2017})}\BibitemShut {NoStop}%
\bibitem [{\citenamefont {Fulga}\ \emph {et~al.}(2016)\citenamefont {Fulga},
  \citenamefont {Avraham}, \citenamefont {Beidenkopf},\ and\ \citenamefont
  {Stern}}]{fulga2016}%
  \BibitemOpen
  \bibfield  {author} {\bibinfo {author} {\bibfnamefont {I.~C.}\ \bibnamefont
  {Fulga}}, \bibinfo {author} {\bibfnamefont {N.}~\bibnamefont {Avraham}},
  \bibinfo {author} {\bibfnamefont {H.}~\bibnamefont {Beidenkopf}}, \ and\
  \bibinfo {author} {\bibfnamefont {A.}~\bibnamefont {Stern}},\ }\href
  {\doibase 10.1103/PhysRevB.94.125405} {\bibfield  {journal} {\bibinfo
  {journal} {Phys. Rev. B}\ }\textbf {\bibinfo {volume} {94}},\ \bibinfo
  {pages} {125405} (\bibinfo {year} {2016})}\BibitemShut {NoStop}%
\bibitem [{\citenamefont {Liu}\ \emph {et~al.}(2013)\citenamefont {Liu},
  \citenamefont {Duan},\ and\ \citenamefont {Fu}}]{Liu13}%
  \BibitemOpen
  \bibfield  {author} {\bibinfo {author} {\bibfnamefont {J.}~\bibnamefont
  {Liu}}, \bibinfo {author} {\bibfnamefont {W.}~\bibnamefont {Duan}}, \ and\
  \bibinfo {author} {\bibfnamefont {L.}~\bibnamefont {Fu}},\ }\href {\doibase
  10.1103/PhysRevB.88.241303} {\bibfield  {journal} {\bibinfo  {journal} {Phys.
  Rev. B}\ }\textbf {\bibinfo {volume} {88}},\ \bibinfo {pages} {241303}
  (\bibinfo {year} {2013})}\BibitemShut {NoStop}%
\bibitem [{\citenamefont {Wang}\ \emph {et~al.}(2013)\citenamefont {Wang},
  \citenamefont {Tsai}, \citenamefont {Lin}, \citenamefont {Xu}, \citenamefont
  {Neupane}, \citenamefont {Hasan},\ and\ \citenamefont {Bansil}}]{Wang13}%
  \BibitemOpen
  \bibfield  {author} {\bibinfo {author} {\bibfnamefont {Y.~J.}\ \bibnamefont
  {Wang}}, \bibinfo {author} {\bibfnamefont {W.-F.}\ \bibnamefont {Tsai}},
  \bibinfo {author} {\bibfnamefont {H.}~\bibnamefont {Lin}}, \bibinfo {author}
  {\bibfnamefont {S.-Y.}\ \bibnamefont {Xu}}, \bibinfo {author} {\bibfnamefont
  {M.}~\bibnamefont {Neupane}}, \bibinfo {author} {\bibfnamefont {M.~Z.}\
  \bibnamefont {Hasan}}, \ and\ \bibinfo {author} {\bibfnamefont
  {A.}~\bibnamefont {Bansil}},\ }\href {\doibase 10.1103/PhysRevB.87.235317}
  {\bibfield  {journal} {\bibinfo  {journal} {Phys. Rev. B}\ }\textbf {\bibinfo
  {volume} {87}},\ \bibinfo {pages} {235317} (\bibinfo {year}
  {2013})}\BibitemShut {NoStop}%
\bibitem [{\citenamefont {Jackiw}\ and\ \citenamefont
  {Rossi}(1981)}]{Jackiw81}%
  \BibitemOpen
  \bibfield  {author} {\bibinfo {author} {\bibfnamefont {R.}~\bibnamefont
  {Jackiw}}\ and\ \bibinfo {author} {\bibfnamefont {P.}~\bibnamefont {Rossi}},\
  }\href {\doibase http://dx.doi.org/10.1016/0550-3213(81)90044-4} {\bibfield
  {journal} {\bibinfo  {journal} {Nuclear Physics B}\ }\textbf {\bibinfo
  {volume} {190}},\ \bibinfo {pages} {681 } (\bibinfo {year}
  {1981})}\BibitemShut {NoStop}%
\end{thebibliography}%

\let\addcontentsline\oldaddcontentsline

\clearpage
\begin{center}
\textbf{
METHODS}
\end{center}

\textbf{First-principle calculations.}
We employed density functional theory (DFT) as implemented in the Vienna Ab Initio Simulation Package (VASP)\cite{
vasp2
}. The exchange correlation term is described according to the Perdew-Burke-Ernzerhof (PBE) prescription together with
projected augmented-wave pseudopotentials\cite{PBE-1996
}. For the autoconsistent calculations we used a $12 \times 12 \times 12$ $\bs{k}$-points mesh for the bulk and $7 \times 7 \times 1$ for the slab calculations. 

For the electronic structure of SnTe with (110) distortion, the kinetic energy cut off was set to 400 eV.
We calculated the surface states by using a slab geometry along the (100) direction. Due to the smallness of the band gap induced by strain, we needed to achieve
a negligible interaction between the surface states from both sides of the slab (to avoid a spurious gap opened by the creation of bonding and anti-bonding states from the top and bottom surface states). To reduce the overlap between top and bottom surface states, we considered a slab of 45 layers, 1~nm vacuum thickness and  artificially localized the states on one of the surfaces. The latter was done by adding one layer of hydrogen to one of the surfaces.

To obtain the electronic structure of bulk SnTe with (111) ferroelectric distortion, the cutoff energy for wave-function expansion was set to 500~eV.
We use the parameter $\lambda$ introduced in Ref.~\onlinecite{Plekhanov14} to parameterize a path linearly connecting the cubic structure (space group Fm3m) to the rhombohedral structure (space group R3m). Our calculations are focused on the $\lambda=0.1$ structure.  
Then, to obtain the hinge electronic structure, we first constructed the maximally-localized Wannier functions (WFs) from the bulk ab-initio calculations. These WFs were used in a Green's function calculation for a system finite in $a$ direction, semi-infinite in $b$ direction and periodic in $c$ direction ($a,~b,~c$ are the conventional lattice vectors in space group R3m). The hinge state spectrum is obtain by projecting on the atoms at the corner, which preserve the mirror symmetry $\hat{M}_{xy}$.

\textbf{Chiral higher-order TI tight-binding model.} We consider a model on a simple cubic lattice spanned by the basis vectors $\hat{\bs{e}}_i$, $i=x,y,z$, with two orbitals $d_{x^2-y^2}$ (denoted $\alpha=0$ below), and $f_{z(x^2-y^2)}$ ($\alpha=1$) on each site, which is populated by spin 1/2 electrons. It is defined by the tight-binding Hamiltonian
\begin{equation}
\begin{aligned}
H_{\mathrm{c}} = &\hphantom{+} \frac{M}{2} \sum_{\bs{r},\alpha} (-1)^\alpha \, c^\dagger_{\bs{r},\alpha} c_{\bs{r},\alpha} \\
&+ \frac{t}{2} \sum_{\bs{r},\alpha} \sum_{i=x,y,z}
 (-1)^\alpha \, c^\dagger_{\bs{r}+\bs{\hat{e}}_i,\alpha} c_{\bs{r},\alpha} 
\\
&+ \frac{\Delta_1}{2} \sum_{\bs{r},\alpha} \sum_{i=x,y,z} \, 
c^\dagger_{\bs{r}+\bs{\hat{e}}_i,\alpha+1} \, \sigma_i \, c_{\bs{r},\alpha}
 \\
&- \frac{\Delta_2}{2 \mathrm{i}} \sum_{\bs{r},\alpha} \sum_{i=x,y,z} (-1)^{\alpha} \, n_i \, c^\dagger_{\bs{r}+\bs{\hat{e}}_i,\alpha + 1} c_{\bs{r},\alpha}
+\mathrm{h.c.},
\end{aligned}
\end{equation}
where $\alpha$ is defined modulo $2$, $\hat{\bs{n}} = (1,-1,0)$, and $c^\dagger_{\bs{r},\alpha}=(c^\dagger_{\bs{r},\alpha,\uparrow},c^\dagger_{\bs{r},\alpha,\downarrow})$ creates a spinor in orbital $\alpha$ at lattice site $\bs{r}$. We denote by $\sigma_0$ and $\sigma_i$, $i=x,y,z$, respectively, the $2\times 2$ identity matrix and the three Pauli matrices acting on the spin 1/2 degree of freedom.

\textbf{Chern-Simons topological invariant.} The invariant for chiral HOTIs with $\hat{C}_4 \hat{T}$ symmetry is given by
\begin{equation}
\theta
=\frac{1}{4\pi}
\int \mathrm{d}^3\bs{k} \, \epsilon_{abc} 
\mathrm{tr}\left[
\mathcal{A}_a\partial_b\mathcal{A}_c + \mathrm{i}\frac23 \mathcal{A}_a\mathcal{A}_b\mathcal{A}_c
\right],
\label{eq: topo alpha}
\end{equation}
written in terms of the Berry gauge field $\mathcal{A}_{a;n,n'}=-\mathrm{i}\braket{u_n|\partial_a|u_{n'}}$, where $\ket{u_{n}}$ are the Bloch eigenstates of the Bloch Hamiltonian, and $n,n'$ are running over the occupied bands of the insulator. $\partial_a$ is the partial derivative with respect to the momentum component $k_a$, $a=x,z,y$. The trace is performed with respect to band indices.

\textbf{Mirror Chern number.} The topological invariant of a 3D helical HOTI is the mirror Chern number $C_{\mathrm{m}}$. Since for a spinful system, a mirror symmetry $\hat{M}$ satisfies $\hat{M}^2 = -1$, its representation $M$ has eigenvalues $\pm \mathrm{i}$. Given a surface $\Sigma$ in the Brillouin zone which is left invariant under the action of $\hat{M}$, the eigenstates $\ket{u_{n}}$ of the Bloch Hamiltonian on $\Sigma$ can be decomposed into two groups, $\{\ket{u_{l}^+}\}$ and $\{\ket{u_{l'}^-}\}$, with mirror eigenvalue $\pm \mathrm{i}$, respectively. Time-reversal maps one mirror eigenspace into the other; if time-reversal symmetry is present, the two mirror eigenspaces are of the same dimension. We can define the Chern number in each mirror subspace as
\begin{equation}
C_{\pm} = \frac{1}{2\pi} \int_\Sigma dk_x dk_y \mathcal{F}^\pm_{xy}(\bs{k}).
\end{equation}
Here
\begin{equation}
\mathcal{F}^\pm_{ab}(\bs{k})
=
\partial_a\mathcal{A}^+_b(\bs{k})-\partial_b\mathcal{A}^+_a(\bs{k})
+\mathrm{i}\left[\mathcal{A}^+_a(\bs{k}),\mathcal{A}^+_b(\bs{k})\right]
\end{equation}
is the non-Abelian Berry curvature field in the $\pm\mathrm{i}$ mirror subspace, with $\mathcal{A}^\pm_{a;l,l'}=-\mathrm{i}\braket{u^\pm_l|\partial_a|u^\pm_{l'}}$, and matrix multiplication is implied in the expressions. 
Note that in time-reversal symmetric systems $C_{+}=-C_{-}$ and we define the mirror Chern number 
\begin{equation}
C_{\mathrm{m}}\equiv(C_{+}-C_{-})/2.
\end{equation}

\newpage
\begin{center}
\textbf{\large Supplementary Material: Higher-Order Topological Insulators}
\end{center}

\beginsupplement
\setcounter{page}{1}

\tableofcontents

%
%
%
%

\hspace{2cm}
\section{Chern-Simons topological invariant}
\label{sec: CS invairant}
\subsection{Quantization from electromagnetic response}
We first argue that the topological invariant for chiral HOTIs, the Chern-Simons (CS) form, is quantized for systems with $\hat{C}^z_4\hat{T}$ symmetry to evaluate to either $0$ or $\pi$, just as in the case of $\hat{T}$ symmetry alone. The CS form is given by
\begin{matriz}
\theta=\frac{1}{4\pi}
\int\mathrm{d}^3k
\epsilon_{abc} \mathrm{tr}\left[
\mathcal{A}_a\partial_b\mathcal{A}_c+\mathrm{i}\frac23 \mathcal{A}_a\mathcal{A}_b\mathcal{A}_c
\right],
\label{eq: theta invariant}
\end{matriz}
written in terms of the non-Abelian Berry gauge field $\mathcal{A}_{a;n,n'}=-\mathrm{i}
\braket{u_n|\partial_a|u_{n'}}
$, where $\ket{u_{n}}$ are the Bloch eigenstates of the chiral HOTI Hamiltonian and the indices $n,n'$ run over its filled bands.

To derive the quantization from the electromagnetic response, we consider the effective action quantifying the response to an external electromagnetic $U(1)$ field $A_\mu$ of a system with non-vanishing $\theta$. The effective action is given by a contribution to the path integral in the form of the axion term
\begin{matriz}
\mathrm{exp}\left[\mathrm{i}\frac{\theta}{8\pi^2 }\int\mathrm{d}^3 x\mathrm{d}t\,
\epsilon_{\mu\nu\sigma\tau}\partial_\mu A_\nu\partial_\sigma A_\tau\right].
\end{matriz}
Observe that
\begin{matriz}
\label{eq: cstermapp}
\frac{\mathrm{i}}{8 \pi^2} \int\mathrm{d}^3 x\mathrm{d}t\epsilon_{\mu\nu\sigma\tau} \, \partial_\mu A_\nu\partial_\sigma A_\tau
\end{matriz}
changes sign under a  $\hat{C}^z_4 \hat{T}$ transformation due to the anti-unitary nature of $\hat{T}$ which takes $\mathrm{i} \rightarrow -\mathrm{i}$ (the axion term is otherwise rotationally invariant and therefore unaffected by $\hat{C}_4^z$). Furthermore, Eq.~\eqref{eq: cstermapp} is a topological invariant, the second Chern number, which is quantized to integer values. Thus, for the theory to be invariant under $\hat{T}$ or $\hat{C}^z_4 \hat{T}$, $\theta=0,\pi \ \mathrm{mod} \ 2\pi$ is required.
We conclude that $\theta$ serves as a topological invariant in $\hat{C}^z_4 \hat{T}$ symmetric TIs, exactly as in $\hat{T}$ symmetric ones.

\subsection{Explicit proof of the quantization of the Chern-Simons invariant}
\label{sec: higher-order TCS}
Here we adapt the explicit proof given in Ref.~\onlinecite{Qi08} 
for the case of helical 3D HOTIs to show that the CS invariant~\eqref{eq: theta invariant} is quantized in the same way by $\hat{C}^z_4\hat{T}$ symmetry as it is by $\hat{T}$ or $\hat{I}$ alone.

\begin{widetext}
The CS invariant~$\theta$ is proportional to the time-reversal invariant polarization
\begin{matriz}
P_3=
\frac{1}{16\pi^2}
\int\mathrm{d}^3 k
\epsilon_{abc}
\mathrm{tr}
\left[
\mathcal{F}_{ab}(\bs{k})\mathcal{A}_c(\bs{k})
-\frac{\mathrm{i}}{3}\left[\mathcal{A}_a(\bs{k}),\mathcal{A}_b(\bs{k})\right]\mathcal{A}_c(\bs{k})
\right].
\end{matriz}
Here
\begin{matriz}
\mathcal{F}_{ab}(\bs{k})
=
\partial_a\mathcal{A}_b(\bs{k})-\partial_b\mathcal{A}_a(\bs{k})
+\mathrm{i}\left[\mathcal{A}_a(\bs{k}),\mathcal{A}_b(\bs{k})\right]
\end{matriz}
is the non-Abelian Berry curvature field and matrix multiplication is implied in the expressions. 

Consider now a general Bloch Hamiltonian $\mathcal{H}(\bs{k})$ that is $\hat{C}^z_4\hat{T}$ invariant, i.e., 
\begin{matriz}
\left(C_4^z T\right) \mathcal{H}(\vec{k}) \left(C_4^z T\right)^{-1}=\mathcal{H}(D_{\hat{C}^{z}_4 \hat{T}}\vec{k}),
\end{matriz}
where
$D_{\hat{C}^{z}_4\hat{T}} \vec{k} = (k_y, -k_x, -k_z)$ and $\left(C_4^z T\right)^4=-1$.
Due to this symmetry, the eigenstates of $\mathcal{H}(\vec{k})$ at $\bs{k}$ and $D_{\hat{C}^{z}_4\hat{T}} \vec{k}$ must be related by a gauge transformation. Explicitly, we can write for any eigenstate $|\bs{k},n\rangle$ of $\mathcal{H}(\vec{k})$ with eigenvalue $\varepsilon_n(\bs{k})$
\begin{matriz}
\begin{split}
 \mathcal{H}(D_{\hat{C}^{z}_4 \hat{T}}\vec{k})\left(C_4^z T\right)|\bs{k},n\rangle
 =&\,\left(C_4^z T\right) \mathcal{H}(\vec{k})|\bs{k},n\rangle
 \\
 =&\,\varepsilon_n(\bs{k})\left(C_4^z T\right)|\bs{k},n\rangle.
\end{split}
\end{matriz}
Thus, $\left(C_4^z T\right)|\bs{k},n\rangle$ is an eigenstate of $\mathcal{H}(D_{\hat{C}^{z}_4 \hat{T}}\vec{k})$ with the same energy. We can thus expand
\begin{matriz}
\left(C_4^z T\right)|\bs{k},n\rangle
=\sum_{m}
B_{n,m}(\bs{k})|D_{\hat{C}^{z}_4 \hat{T}}\bs{k},m\rangle,
\end{matriz}
where $B_{n,m}(\bs{k})$ are the matrix elements of a unitary transformation acting on the space of occupied bands. 
We can factor out the complex conjugation $\mathcal{K}$ from $C_4^z T$ as $C_4^z T=C_4^z T'\mathcal{K}$, thereby defining the unitary operator $T'$ [which can either satisfy $T'(T')^*=-1$ or $T'(T')^*=+1$, as long as $(C_4^z T)^4=-1$ holds], and obtain
\begin{matriz}
|\bs{k},n\rangle
=\sum_{m}
\left[B_{n,m}(\bs{k})\left(C_4^z T'\right)^{-1}|D_{\hat{C}^{z}_4 \hat{T}}\bs{k},m\rangle\right]^*.
\end{matriz}

We can then rewrite the Berry connection
\begin{matriz}
\begin{split}
\mathcal{A}_{a;n,n'}(\bs{k})=&\,
-\mathrm{i}\langle \bs{k},n|\partial_a|\bs{k},n'\rangle
\\
=&\,
-\mathrm{i}
\left\{\sum_{m,m'}B^*_{n,m}(\bs{k})\langle D_{\hat{C}^{z}_4 \hat{T}} \bs{k},m|\partial_a\left[B_{n',m'}(\bs{k})|D_{\hat{C}^{z}_4 \hat{T}} \bs{k},m'\rangle\right]\right\}^*
\\
=&\, - J_{ab}
\sum_{m,m'} B_{n,m}(\bs{k}) \mathcal{A}^\mathsf{T}_{b;m,m'}(D_{\hat{C}^{z}_4 \hat{T}} \bs{k}) \left[B^\dagger(\bs{k})\right]_{m',n'}
-\mathrm{i}\sum_m B_{n,m}(\bs{k})\partial_a \left[B^\dagger(\bs{k})\right]_{m,n'}
,
\end{split}
\end{matriz}
where $J_{ab} = \partial (D_{\hat{C}^{z}_4 \hat{T}} \bs{k})^b/\partial k^a$, confirming that the connections are related by a non-Abelian gauge transformation.
One verifies that the Berry field satisfies
\begin{matriz}
\mathcal{F}_{ab}(\bs{k})= - J_{ad} J_{be} \, B(\bs{k}) \mathcal{F}^\mathsf{T}_{de}(D_{\hat{C}^{z}_4 \hat{T}}\bs{k})B^\dagger(\bs{k}).
\end{matriz}
We use these transformations to re-express 
\begin{matriz}
\begin{split}
P_3
=
&\,\frac{1}{16\pi^2}
\int\mathrm{d}^3 k
\epsilon_{abc}
\mathrm{tr}
\Biggl[
\left(B(\bs{k}) (- J_{cf}) \mathcal{A}^\mathsf{T}_{f}(D_{\hat{C}^{z}_4 \hat{T}} \bs{k}) B^\dagger(\bs{k})-\mathrm{i}B(\bs{k})\partial_c B^\dagger(\bs{k})\right)
\Biggl\{
B(\bs{k}) (- J_{ad} J_{be}) \mathcal{F}^\mathsf{T}_{de}(D_{\hat{C}^{z}_4 \hat{T}}\bs{k})B^\dagger(\bs{k})
\\
&-
\frac{\mathrm{i}}{3}\left[
\left(B(\bs{k}) (- J_{ad}) \mathcal{A}^\mathsf{T}_{d}(D_{\hat{C}^{z}_4 \hat{T}} \bs{k}) B^\dagger(\bs{k})-\mathrm{i}B(\bs{k})\partial_a B^\dagger(\bs{k})\right),
\left(B(\bs{k}) (- J_{be}) \mathcal{A}^\mathsf{T}_{e}(D_{\hat{C}^{z}_4 \hat{T}} \bs{k}) B^\dagger(\bs{k})-\mathrm{i}B(\bs{k})\partial_b B^\dagger(\bs{k})\right)\right]
\Biggr\}
\Biggr].
\end{split}
\end{matriz}
Expanding this, we obtain
\begin{matriz}
\begin{split}
P_3
=
&\,\frac{1}{16\pi^2}
\int\mathrm{d}^3 k
\epsilon_{abc}
\mathrm{tr}\left[
(- J_{cf}) \mathcal{A}^\mathsf{T}_{f} (D_{\hat{C}^{z}_4 \hat{T}} \bs{k})
\Biggl\{
(- J_{ad} J_{be}) \mathcal{F}^\mathsf{T}_{de} (D_{\hat{C}^{z}_4 \hat{T}} \bs{k})
-
\frac{\mathrm{i}}{3}\left[
 (- J_{ad}) \mathcal{A}^\mathsf{T}_{d} (D_{\hat{C}^{z}_4 \hat{T}} \bs{k}) ,
 (- J_{be}) \mathcal{A}^\mathsf{T}_{e} (D_{\hat{C}^{z}_4 \hat{T}} \bs{k}) \right]
\Biggr\}
\right]
\\
&
-
\frac{\mathrm{i}}{16\pi^2}
\int\mathrm{d}^3 k
\epsilon_{abc}
\mathrm{tr}
\Biggl[
B(\bs{k})\partial_c B(\bs{k})^\dagger
\Biggl\{
B(\bs{k}) (- J_{ad} J_{be}) \mathcal{F}^\mathsf{T}_{de} (D_{\hat{C}^{z}_4 \hat{T}} \bs{k})B(\bs{k})^\dagger \\&-
\frac{\mathrm{i}}{3}\left[
\left(B(\bs{k}) (- J_{ad}) \mathcal{A}^\mathsf{T}_{d} (D_{\hat{C}^{z}_4 \hat{T}} \bs{k}) B(\bs{k})^\dagger-\mathrm{i}B(\bs{k})\partial_a B(\bs{k})^\dagger\right), 
\left(B(\bs{k}) (- J_{be}) \mathcal{A}^\mathsf{T}_{e} (D_{\hat{C}^{z}_4 \hat{T}} \bs{k}) B(\bs{k})^\dagger-\mathrm{i}B(\bs{k})\partial_b B(\bs{k})^\dagger\right)\right]
\Biggr\}
\Biggr]
\\
&
-
\frac{\mathrm{i}}{16\pi^2}\frac{1}{3}
\int\mathrm{d}^3 k
\epsilon_{abc}
\mathrm{tr}
\Biggl[
B(\bs{k}) (- J_{cf}) \mathcal{A}^\mathsf{T}_{f} (D_{\hat{C}^{z}_4 \hat{T}} \bs{k}) B(\bs{k})^\dagger
\Bigl(
-\left[
B(\bs{k})\partial_a B(\bs{k})^\dagger,
B(\bs{k})\partial_b B(\bs{k})^\dagger\right] 
\\&
-\mathrm{i}
\left[
B(\bs{k}) (- J_{ad}) \mathcal{A}^\mathsf{T}_{d} (D_{\hat{C}^{z}_4 \hat{T}} \bs{k}) B(\bs{k})^\dagger,
B(\bs{k})\partial_b B(\bs{k})^\dagger\right]
-\mathrm{i}
\left[
B(\bs{k})\partial_a B(\bs{k})^\dagger,
B(\bs{k}) (- J_{be}) \mathcal{A}^\mathsf{T}_{e} (D_{\hat{C}^{z}_4 \hat{T}} \bs{k}) B(\bs{k})^\dagger\right]
\Bigr)
\Biggr]
\\
=
&\,-P_3
+
\frac{2}{3\times16\pi^2}
\int\mathrm{d}^3 k
\epsilon_{abc}
\mathrm{tr}
\left[
\left(B(\bs{k})\partial_a B(\bs{k})^\dagger\right)
\left(B(\bs{k})\partial_b B(\bs{k})^\dagger\right)
\left(B(\bs{k})\partial_c B(\bs{k})^\dagger\right)\right]
\\
&
-
\frac{\mathrm{i}}{16\pi^2}
\int\mathrm{d}^3 k
\epsilon_{abc}
\mathrm{tr}
\Biggl[
B(\bs{k})\partial_c B(\bs{k})^\dagger
\Biggl\{
B(\bs{k}) (- J_{ad} J_{be}) \mathcal{F}^\mathsf{T}_{de} (D_{\hat{C}^{z}_4 \hat{T}} \bs{k})B(\bs{k})^\dagger
\\&-
\frac{\mathrm{i}}{3}
\Bigl(
\left[
B(\bs{k}) (- J_{ad}) \mathcal{A}^\mathsf{T}_{d} (D_{\hat{C}^{z}_4 \hat{T}} \bs{k}) B(\bs{k})^\dagger,
B(\bs{k}) (- J_{be}) \mathcal{A}^\mathsf{T}_{e} (D_{\hat{C}^{z}_4 \hat{T}} \bs{k}) B(\bs{k})^\dagger \right]
\\&-\mathrm{i}
\left[
B(\bs{k}) (- J_{ad}) \mathcal{A}^\mathsf{T}_{d} (D_{\hat{C}^{z}_4 \hat{T}} \bs{k}) B(\bs{k})^\dagger,
B(\bs{k})\partial_b B(\bs{k})^\dagger\right]
-\mathrm{i}
\left[
B(\bs{k})\partial_a B(\bs{k})^\dagger,B(\bs{k}) (- J_{be}) \mathcal{A}^\mathsf{T}_{e} (D_{\hat{C}^{z}_4 \hat{T}} \bs{k}) B(\bs{k})^\dagger\right]
\Bigr)
\Biggr\}
\Biggr]
\\
&
-
\frac{\mathrm{i}}{16\pi^2}\frac{1}{3}
\int\mathrm{d}^3 k
\epsilon_{abc}
\mathrm{tr}
\Biggl[
B(\bs{k}) (- J_{cf}) \mathcal{A}^\mathsf{T}_{f} (D_{\hat{C}^{z}_4 \hat{T}} \bs{k}) B(\bs{k})^\dagger
\Bigl(
-\left[
B(\bs{k})\partial_a B(\bs{k})^\dagger,
B(\bs{k})\partial_b B(\bs{k})^\dagger\right]
\\&
-\mathrm{i}
\left[
B(\bs{k}) (- J_{ad}) \mathcal{A}^\mathsf{T}_{d} (D_{\hat{C}^{z}_4 \hat{T}} \bs{k}) B(\bs{k})^\dagger,
B(\bs{k})\partial_b B(\bs{k})^\dagger\right]
-\mathrm{i}
\left[
B(\bs{k})\partial_a B(\bs{k})^\dagger,
B(\bs{k}) (- J_{be}) \mathcal{A}^\mathsf{T}_{e} (D_{\hat{C}^{z}_4 \hat{T}} \bs{k}) B(\bs{k})^\dagger\right]
\Bigr)
\Biggr]
\\
=
&\,-P_3
+
\frac{2}{3\times16\pi^2}
\int\mathrm{d}^3 k
\epsilon_{abc}
\mathrm{tr}
\left[
\left(B(\bs{k})\partial_a B(\bs{k})^\dagger\right)
\left(B(\bs{k})\partial_b B(\bs{k})^\dagger\right)
\left(B(\bs{k})\partial_c B(\bs{k})^\dagger\right)\right]
\\
&
-
\frac{\mathrm{i}}{16\pi^2}
\int\mathrm{d}^3 k
\epsilon_{abc}
\mathrm{tr}
\Biggl[
(\partial_c B(\bs{k})^\dagger) B(\bs{k})
\Biggl\{
(- J_{ad} J_{be}) \mathcal{F}^\mathsf{T}_{de} (D_{\hat{C}^{z}_4 \hat{T}} \bs{k})\\&-
\frac{2\mathrm{i}}{3}
\Bigl(
 (- J_{ad}) \mathcal{A}^\mathsf{T}_{d} (D_{\hat{C}^{z}_4 \hat{T}} \bs{k})  (- J_{be}) \mathcal{A}^\mathsf{T}_{e} (D_{\hat{C}^{z}_4 \hat{T}} \bs{k}) 
-\mathrm{i}
 (- J_{ad}) \mathcal{A}^\mathsf{T}_{d} (D_{\hat{C}^{z}_4 \hat{T}} \bs{k}) (\partial_b B(\bs{k})^\dagger) B(\bs{k})
 \\&
-\mathrm{i}
(\partial_a B(\bs{k})^\dagger) B(\bs{k}) (- J_{be}) \mathcal{A}^\mathsf{T}_{e} (D_{\hat{C}^{z}_4 \hat{T}} \bs{k}) 
\Bigr)
\Biggr\}
\Biggr]
\\
&
+
\frac{\mathrm{i}}{16\pi^2}\frac{2}{3}
\int\mathrm{d}^3 k
\epsilon_{abc}
\mathrm{tr}
\Biggl[
 (- J_{cf}) \mathcal{A}^\mathsf{T}_{f} (D_{\hat{C}^{z}_4 \hat{T}} \bs{k}) 
\Bigl\{
(\partial_a B(\bs{k})^\dagger)
B(\bs{k})(\partial_b B(\bs{k})^\dagger)B(\bs{k})
\\&
+\mathrm{i}
 (- J_{ad}) \mathcal{A}^\mathsf{T}_{d} (D_{\hat{C}^{z}_4 \hat{T}} \bs{k}) (\partial_b B(\bs{k})^\dagger) B(\bs{k})
+\mathrm{i}
(\partial_a B(\bs{k})^\dagger)
B(\bs{k}) (- J_{be}) \mathcal{A}^\mathsf{T}_{e} (D_{\hat{C}^{z}_4 \hat{T}} \bs{k}) 
\Bigr\}
\Biggr].
\end{split}
\end{matriz}
We have used $\mathrm{tr}M^\mathsf{T} = \mathrm{tr}M$ for any matrix $M$ and $\epsilon_{abc} J_{ad} J_{be} J_{cf} = \epsilon_{def} (\det{J}) = - \epsilon_{def}$ (the minus sign comes from the time-reversal transformation of $k_z$) in the second line to obtain the crucial result that the first term becomes $-P_3$. 
This, in essence, is what implies the quantization of the invariant. Any other spatial symmetry $\hat{S}$, when combined with TRS $\hat{T}$,
would effect the same quantization if the Jacobian of the combined transformation is $-1$ in momentum space. (This does not, however, imply the existence of a phase with $\theta=\pi$ in each such case. 
In the case of $\hat{C}_4\hat{T}$ we have shown by an explicit example that the topologically nontrivial phase exists, which proves that we have not overlooked any symmetry constraints that would always render the invariant trivial.)
We proceed with the manipulations
\begin{matriz}
\begin{split}
2P_3
=
&\,
+
\frac{2}{3\times16\pi^2}
\int\mathrm{d}^3 k
\epsilon_{abc}
\mathrm{tr}
\left[
\left(B(\bs{k})\partial_a B(\bs{k})^\dagger\right)
\left(B(\bs{k})\partial_b B(\bs{k})^\dagger\right)
\left(B(\bs{k})\partial_c B(\bs{k})^\dagger\right)\right]
\\
&
-
\frac{\mathrm{i}}{16\pi^2}
\int\mathrm{d}^3 k
\epsilon_{abc}
\mathrm{tr}
\Biggl[
(\partial_c B(\bs{k})^\dagger) B(\bs{k})
\Biggl\{
(- J_{ad} J_{be}) \mathcal{F}^\mathsf{T}_{de} (D_{\hat{C}^{z}_4 \hat{T}} \bs{k})-
2
(\partial_a B(\bs{k})^\dagger) B(\bs{k}) (- J_{be}) \mathcal{A}^\mathsf{T}_{e} (D_{\hat{C}^{z}_4 \hat{T}} \bs{k}) 
\\&
-\mathrm{i}2
(- J_{ad}) \mathcal{A}^\mathsf{T}_{d} (D_{\hat{C}^{z}_4 \hat{T}} \bs{k}) (- J_{be}) \mathcal{A}^\mathsf{T}_{e} (D_{\hat{C}^{z}_4 \hat{T}} \bs{k}) 
\Biggr\}
\Biggr]
\\
=
&\,
+
\frac{1}{24\pi^2}
\int\mathrm{d}^3 k
\epsilon_{abc}
\mathrm{tr}
\left[
\left(B(\bs{k})\partial_a B(\bs{k})^\dagger\right)
\left(B(\bs{k})\partial_b B(\bs{k})^\dagger\right)
\left(B(\bs{k})\partial_c B(\bs{k})^\dagger\right)\right]
\\
&
-
\frac{\mathrm{i}}{16\pi^2}
\int\mathrm{d}^3 k
\epsilon_{abc}
\mathrm{tr}
\Biggl[
2(\partial_c B(\bs{k})^\dagger) B(\bs{k})
\partial_a (- J_{be}) \mathcal{A}^\mathsf{T}_{e} (D_{\hat{C}^{z}_4 \hat{T}} \bs{k}) 
+
2
(\partial_c B(\bs{k})^\dagger)
(\partial_a B(\bs{k})) 
(- J_{be}) \mathcal{A}^\mathsf{T}_{e} (D_{\hat{C}^{z}_4 \hat{T}} \bs{k}) 
\Biggr]
\\
=
&\,
+
\frac{1}{24\pi^2}
\int\mathrm{d}^3 k
\epsilon_{abc}
\mathrm{tr}
\left[
\left(B(\bs{k})\partial_a B(\bs{k})^\dagger\right)
\left(B(\bs{k})\partial_b B(\bs{k})^\dagger\right)
\left(B(\bs{k})\partial_c B(\bs{k})^\dagger\right)\right],
\end{split}
\end{matriz}
where partial integration is required to obtain the last line. In the righthand side of the last line, we recognize the winding number of a unitary matrix over the three-torus. This integral is necessarily an integer. For $B(\bs{k}) \in U(1)$ it is in fact always 0, which explains why there is no nontrivial 3D TI phase for spinless electron systems. Indeed these satisfy $\hat{T}^2=1$ and therefore have singly-degenerate bands on which the representation of $B(\bs{k})$ is one-dimensional and hence the winding number vanishes trivially. In our case, however, since $\hat{C}^{z}_4 \hat{T}$ implies a double degeneracy at all $\vec{k} \in \mathcal{I}_{\hat{C}_4^z \hat{T}}$, we have $B(\bs{k}) \in U(2)$ and the integral can evaluate to nontrivial integers. Hence, we find $2P_3\in\mathbb{Z}$, i.e., $P_3$ is quantized to half-integer values by $\hat{C}^{z}_4 \hat{T}$ symmetry. 
This implies a quantization of $\theta$ in units of $\pi$.

\end{widetext}

\subsection{Evaluation of the Chern-Simons invariant}
\label{sec: evaluation of invariant}

Here, we calculate $\theta$ explicitly for the chiral HOTI model from Eq.~\eqref{eq: H} in the main text, which in its topological phase is given by the representative parameter choice $M=2$, 
\begin{matriz} \label{eq: H mod}
\begin{aligned}
\mathcal{H}_{\mathrm{c}}(\vec{k}) = &\left(2+\sum_i \cos k_i\right) \, \tau_z \sigma_0 +\sum_i \sin k_i \, \tau_y \sigma_i \\&+\Delta_2 (\cos k_x - \cos k_y) \, \tau_x \sigma_0,
\end{aligned}
\end{matriz}
containing a $\hat{T}$ breaking term proportional to $\Delta_2$.
This Hamiltonian has a band inversion at $\bs{k}_0 = (\pi, \pi, \pi)$. Expanding around this point with $\bs{k} = \bs{k}_0 + \tilde{\bs{k}}$, we obtain
\begin{matriz}
\mathcal{H}_{\mathrm{c}}(\vec{\tilde{k}}) = - \tau_z \sigma_0 +\sum_i \tilde{k}_i \, \tau_y \sigma_i.
\end{matriz}
We choose a gauge in which the normalized eigentstates of the two occupied bands are (to first order in $\tilde{\bs{k}}$)
\begin{matriz}
\begin{aligned}
u_1(\tilde{\bs{k}}) = &\frac{1}{f(\tilde{\bs{k}})} \Bigl\{\tilde{k}_x-\mathrm{i} \tilde{k}_y,-\tilde{k}_z,0,0\Bigr\}, \\
u_2(\tilde{\bs{k}}) = &\frac{1}{f(\tilde{\bs{k}})} \Bigl\{+\tilde{k}_z,\tilde{k}_x+\mathrm{i} \tilde{k}_y,0,0\Bigr\},
\end{aligned}
\end{matriz}
where we have defined the normalization $f(\tilde{\bs{k}}) = |\bs{\tilde{k}}|$.
Note that for $\tilde{k}_z = 0$, we have $\frac{\tilde{k}_x-i \tilde{k}_y}{f(\tilde{\bs{k}})} = 
\frac{\tilde{k}_x - \mathrm{i}\tilde{k}_y}{\sqrt{\tilde{k}_x^2 + \tilde{k}_y^2}}$. This expression, which occurs in both eigenstates, is multi-valued at $\tilde{\bs{k}} = 0$ [for instance, in the limit $(\tilde{k}_x\rightarrow0, \tilde{k}_y = 0)$ it evaluates to $1$, while for $(\tilde{k}_y\rightarrow0, \tilde{k}_x = 0)$ it evaluates to $\mathrm{i}$]. On the other hand, for an expansion around the other $\hat{C}_4^z \hat{T}$ invariant momenta $\bs{k}_0 \in \{(0, 0, 0), (\pi, \pi, 0), (0, 0, \pi)\}$, we obtain $f(\bs{\tilde{k}}) = \sqrt{4+3 \bs{\tilde{k}}^2}$ and the eigenstates are well defined in the vicinity of these points.

Note that Eq.~\eqref{eq: theta invariant} can be written as the integral of a total derivative which vanishes on the Brillouin zone torus (which has no boundary) as long as all functions are single-valued. Therefore, contributions to the CS form can be thought of as arising from points in momentum space where Bloch states in a given gauge are multi-valued. The resulting form of the Berry gauge field near $\bs{k}_0 = (\pi, \pi, \pi)$, expanded in the basis of Pauli matrices, is given by
\begin{matriz}
\bs{\mathcal{A}}(\tilde{\bs{k}})
=\frac{1}{|\tilde{\bs{k}}|^2}\tilde{\bs{k}}\wedge\
\begin{pmatrix} -\sigma_x\\ \sigma_y\\ \sigma_z\end{pmatrix}+\mathcal{O}(1),
\end{matriz}
in which we recognize the gauge field of an SU(2) monopole. As a result of this monopole, the CS form evaluates to $\theta=\pi$ independent of $\Delta_2$, i.e., even if $\hat{T}$ symmetry is broken as long as $\hat{C}_4^z\hat{T}$ symmetry is preserved. 
\
\section{Topological characterization of chiral higher-order topological insulators with inversion symmetry}
\label{sec: topochar}
Note that the Hamiltonian given in Eq.~(\ref{eq: H}) in the main text is invariant under the combination of time-reversal and inversion symmetry $\hat{I} \hat{T}$, with representations $I = \tau_z \sigma_0$ and $T = \tau_0 \sigma_y \mathit{K}$. This symmetry, which forces the bands to be two-fold degenerate at all $\bs{k}$, is not essential for the topological phase of the model, and may be broken by adding a perturbation $\delta \, \tau_x \sigma_0$ to the Hamiltonian, with $\delta$ a small real parameter. However, when $\hat{I} \hat{T}$ symmetry holds, we may formulate a topological index that simplifies the topological characterization.

To achieve this,
the unitary symmetry $\hat{C}_4 \hat{I}$ with matrix representation
$
S_{mn}(\vec{k}) = \bra{u_m(\vec{k})} C_4 I \ket{u_n(\vec{k})},
$
where $\ket{u_n(\vec{k})}$ span the filled subspace, can be used in the same way inversion symmetry is used in Ref.~\onlinecite{FuKane2007}.
We can study the eigenvalues of $S(\vec{k})$ at the four high symmetry points $\vec{k} \in \mathcal{I}_{\hat{C}_4^z \hat{T}}=\{(0, 0, 0), (\pi, \pi, 0), (0, 0, \pi), (\pi, \pi, \pi)\}$.
Since $(\hat{C}_4 \hat{I})^4 = -1$, the eigenvalues of $S(\bs{k})$ are the fourth roots of $-1$. Due to $[\hat{C}_4 \hat{I}, \hat{I} \hat{T}] = 0$, and $\hat{I} \hat{T}$ being anti-unitary, they have to come in complex-conjugated pairs $\{\xi_{\vec{k}} e^{\mathrm{i}\pi/4}, \xi_{\vec{k}}e^{-\mathrm{i}\pi/4}\}$ with $\xi_{\vec{k}}=+1$ or $\xi_{\vec{k}}=-1$. The symmetry action of $\hat{C}_4 \hat{I}$ on the filled subspace at a high symmetry point $\vec{k} \in \mathcal{I}_{\hat{C}_4^z \hat{T}}$ is then characterized by the set $\xi_{n,\vec{k}}, n = 1,\cdots, N/2$, where $N$ is the number of filled bands.

We may therefore define the topological invariant in the presence of $\hat{I} \hat{T}$ symmetry as
\begin{matriz} \label{eq: c4iformula}
(-1)^{\nu_{\rm{c}}} = \prod_{n=1}^{N/2} \prod_{\vec{k} \in \mathcal{I}_{\hat{C}_4^z \hat{T}}} \xi_{n, \vec{k}},
\end{matriz}
which resembles the Fu-Kane formula for inversion-symmetric TIs~\cite{Kane07}. This invariant is well defined for the following reason: by the bulk-boundary correspondence of chiral HOTIs, when breaking $\hat{T}$ but preserving $\hat{C}_4^z \hat{T}$, the band inversion which in the first-order TI case led to gapless surfaces now induces gapless channels along the hinges separating the $x$ and $y$ surfaces. To detect this phase transition, Fu and Kane have used a product over the inversion $\hat{I}$ eigenvalues at the time-reversal invariant momenta, which should be $-1$ in the nontrivial case. Here, we do not have $\hat{I}$ symmetry, but only $\hat{C}_4^z \hat{I}$, which we may however use in the very same way, as indicated in Eq.~\eqref{eq: c4iformula}. 
We prove this formula by adiabatically interpolating to the unperturbed case (with $\hat{T}$, $\hat{C}_4^z$, and  $\hat{I}$ symmetry),
which can be done without a bulk gap closing.
In this situation, both the Fu-Kane formula and Eq.~\eqref{eq: c4iformula} evaluate to the same result, i.e., 
the eigenvalues of $\hat{C}_4^z \hat{I}$ show the same inversion as those of $\hat{I}$ for the following reason: 
the $\hat{C}_4^z$ eigenvalues of the occupied bands at $(k_x,k_y)=(0,0)$ and $(k_x,k_y)=(\pi,\pi)$ are well-defined for all $k_z$ and---since the system is insulating---are independent of $k_z$. Thus, each Kramers pair of $\hat{C}_4^z$ eigenvalues enters twice in  
the product in Eq.~\eqref{eq: c4iformula}, once at $k_z=0$ and once at $k_z=\pi$.
Thus, any possible inversion in the $\hat{C}_4^z$ eigenvalues is rendered trivial by the product in Eq.~\eqref{eq: c4iformula}, and only the band inversions of $\hat{I}$ enter. In particular, only the band inversions at the $\bs{k} \in \mathcal{I}_{\hat{C}_4^z \hat{T}}$ enter, as for all other points, $\hat{C}_4$ symmetry makes inversion eigenvalues come in pairs [e.g., $(0, \pi, 0)$ and $(\pi, 0, 0)$ have the same inversion eigenvalues].

\begin{widetext}
More explicitly, if we denote by $\chi_{n,\bs{k}}$ the inversion eigenvalues of band pair $n$ at the time-reversal invariant momenta $\mathcal{I}_{\hat{T}}$, the following identity holds for a $\hat{C}_4^z$ and $\hat{T}$ invariant system:
\begin{matriz}
\begin{split}
 \prod_n \prod_{\bs{k}\in\mathcal{I}_{\hat{T}}} \chi_{n,\bs{k}} 
=&\, \prod_n \chi_{n,(0,0,0)} \chi_{n,(\pi,\pi,0)}\chi_{n,(0,0,\pi)} \chi_{n,(\pi,\pi,\pi)} \\ 
=&\,\prod_n \xi_{n,(0,0,0)}^2\xi_{n,(\pi,\pi,0)}^2 \chi_{n,(0,0,0)} \chi_{n,(\pi,\pi,0)} \chi_{n,(0,0,\pi)} \chi_{n,(\pi,\pi,\pi)} \\ 
=&\, \prod_n \left(\xi_{n,(0,0,0)} \chi_{n,(0,0,0)}\right) \left(\xi_{n,(\pi,\pi,0)}  \chi_{n,(\pi,\pi,0)}\right) \left(\xi_{n,(0,0,\pi)} \chi_{n,(0,0,\pi)}\right) \left(\xi_{n,(\pi,\pi,\pi)} \chi_{n,(\pi,\pi,\pi)} \right).
\label{eq: bla}
\end{split}
\end{matriz}
Here we used that $\xi^2_{n,\bs{k}}=1$, $\prod_n \xi_{n,(0,0,0)}=\prod_n \xi_{n,(0,0,\pi)}$ and $\prod_n \xi_{n,(\pi,\pi,0)}=\prod_n \xi_{n,(\pi,\pi,\pi)}$ as well as $\prod_n \chi_{n,(\pi,0,k_z)}= \prod_n \chi_{n,(0,\pi,k_z)}$ for $k_z=0,\pi$. The left-hand side of Eq.~\eqref{eq: bla} is exactly the Fu-Kand band inversion formula and the right-hand side is Eq.~\eqref{eq: c4iformula}. Since the right-hand side is also well-defined if only $\hat{C}_4^z\hat{T}$ is a symmetry, but not $\hat{C}_4$ and $\hat{T}$ alone, and the two cases are connected without gap closing, it constitutes a well-defined invariant for this case.
In the case of the model given by Eq.~(\ref{eq: H}) in the main text, there is a single inversion $\xi_{\vec{k}}=-1$ at $\vec{k} = (\pi, \pi, \pi)$ or $\vec{k} = (0,0,0)$ for $1<M<3$ or $-1>M>-3$, respectively, confirming $\nu_{\rm{c}} = -1$ in these parameter regimes.

\end{widetext}


\section{Degeneracies in the chiral higher-order TI bulk and Wilson loop spectra}
\label{sec: kramerswilson}

Here we discuss the degeneracies in the chiral HOTI bulk and Wilson loop spectra enforced by $\hat{C}_4^z \hat{T}$.
At the four $\hat{C}_4^z \hat{T}$ invariant momenta $\vec{k}$ taken from $\mathcal{I}_{\hat{C}_4^z \hat{T}}=\{(0, 0, 0), (\pi, \pi, 0), (0, 0, \pi), (\pi, \pi, \pi)\}$ the relation $(\hat{C}_4^z \hat{T})^4 = -1$ enforces a Kramers-like degeneracy: if we assume $C_4 T \ket{\psi} = e^{i \alpha} \ket{\psi}$ for an energy eigenstate $\ket{\psi}$ at the high-symmetry points, applying $(C_4 T)^3$ from the left on this equation leads to the contradiction $\ket{\psi} = - \ket{\psi}$, hence a doublet must exist.

We now show that a similar degeneracy is induced in the band structure of the Wilson loop $W^z(k_x,k_y)$ (which was defined in the methods section entitled ``Wilson loop'') at the $\hat{C}_4\hat{T}$ invariant momenta $(k_x,k_y)=(0,0)$ and $(k_x,k_y)=(\pi,\pi)$.

In general, $\hat{C}_4^z \hat{T}$ invariance implies
\begin{matriz}
C_4^z T \ket{u_n(\vec{k})} = B_{nm}(\vec{k}) \ket{u_m(D_{\hat{C}_4^z \hat{T}}\vec{k})},
\end{matriz}
where $B_{nm}(\vec{k}) = \langle u_m({D_{\hat{C}_4^z \hat{T}} \vec{k}})| C_4^z T | u_n({\vec{k}}) \rangle$ is the unitary sewing matrix that connects states at $\vec{k}$ with those at $D_{\hat{C}_4^z \hat{T}} \vec{k}$ which have the same energy. Summation over repeated indices is implied here as well as below.

Since the filled subspace at all two-dimensional momenta $(k_x, k_y)$ is invariant under $\hat{C}_4^z \hat{T}$, all projectors $P(\vec{k})$ that enter  $W^z(k_x,k_y)$ commute with $C_4^z T$. Therefore, at the $\hat{C}_4^z \hat{T}$-invariant momenta $(k_x,k_y) \in \{(0,0), (\pi,\pi)\}$, $W^z(k_x, k_y)$ transforms under $\hat{C}_4^z \hat{T}$ as
\begin{matriz}
\begin{aligned}
&W_{mn}^z(k_x, k_y) = \\ &\quad B_{mi}(k_z=0) \left[\left[W^{z}(k_x, k_y)\right]^{-1}\right]^{*}_{ij}  \left[B^{-1}(k_z=0)\right]_{jn}.
\end{aligned}
\label{eq: symm of W}
\end{matriz}
When writing $W^z(k_x, k_y) = e^{\mathrm{i} \mathcal{H}_{\mathrm{W}}(k_x,k_y)}$, Eq.~\eqref{eq: symm of W} implies that $\mathcal{H}_{\mathrm{W}} (k_x,k_y)= [B(\vec{k})  \mathit{K}] \mathcal{H}_{\mathrm{W}}(k_x,k_y) [B(\vec{k})  \mathit{K}]^{-1}$, i.e., $[B(\vec{k}) \mathit{K}]$ is an anti-unitary symmetry of $\mathcal{H}_{\mathrm{W}}(k_x,k_y)$.
Due to the relation $[B(\vec{k}) \mathit{K}]^4 = -1$ for each $\vec{k}$ [which can be derived by writing out $B_{mn}(\vec{k})$ as the matrix elements of $C_4^z T$ in the filled subspace], the eigenvalues of $\mathcal{H}_{\mathrm{W}}$ are degenerate for the $\hat{C}_4^z \hat{T}$-invariant values of $(k_x,k_y)$ by the same argument as above.





\begin{figure}[t]
\begin{center}
\includegraphics[width=0.46 \textwidth]{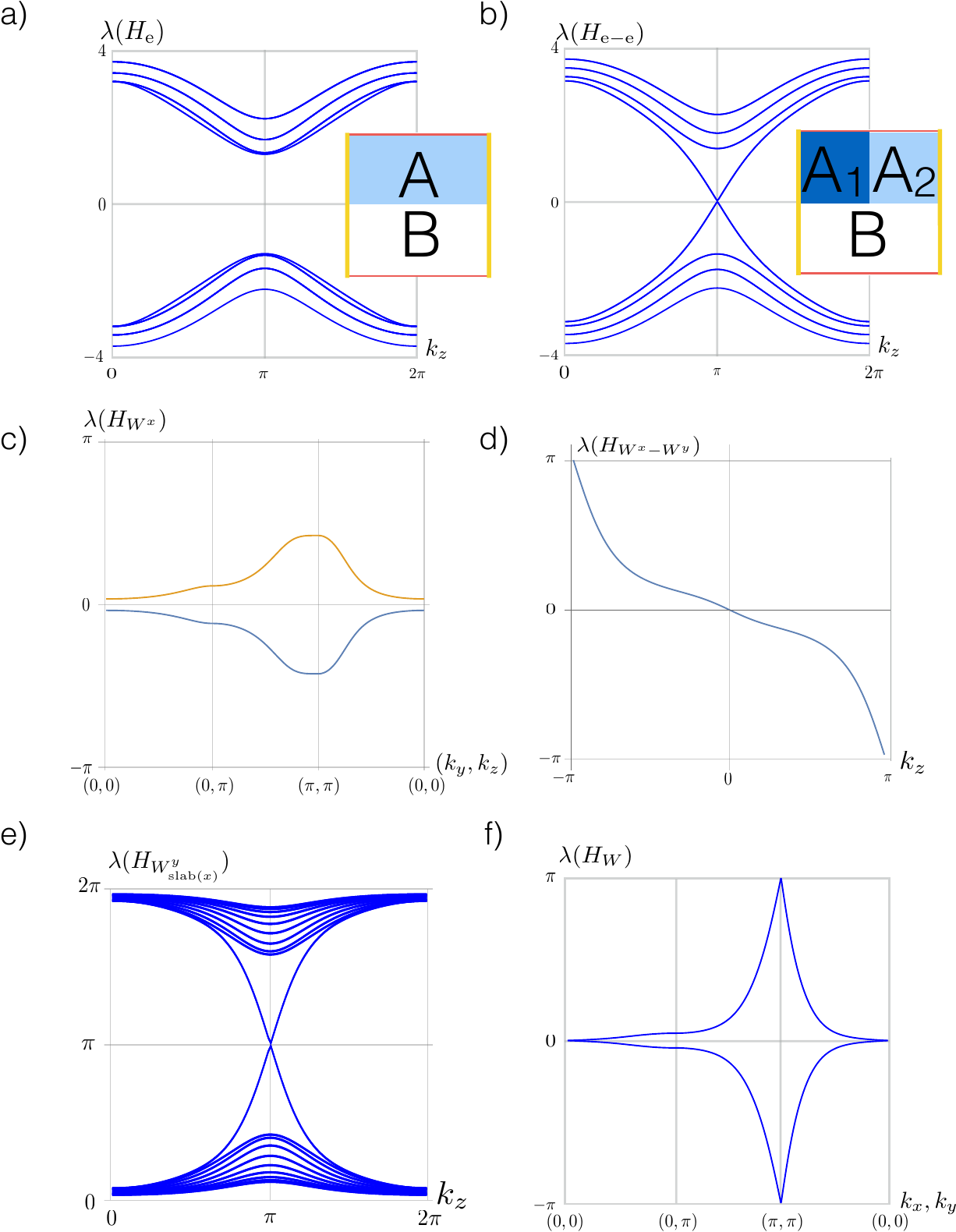}
\caption{
Nested entanglement and Wilson loop spectra for the second-order 3D chiral TI model defined in Eq.~\eqref{eq: H} in the main text with $M/t=2$ and $\Delta_1/t = \Delta_2/t = 1$.
(a) Gapped Entanglement spectrum of $H_{\mathrm{e}}(k_y,k_z)$ for a bipartitioning of the system in the $x$-$y$-plane as shown in the inset.
(b) Nested entanglement spectrum obtained from the ground state of the entanglement Hamiltonian $H_{\mathrm{e}}(k_y,k_z)$ itself by further tracing out the subsystem $A_2$. The chiral gapless modes, localized at the hinges of region $A_1$ reflect the presence of chiral hinge modes.
(c) Eigenvalues of the Wilson loop Hamiltonian $H_{\mathrm{W}}(k_y, k_z)$ in $x$-direction. The spectrum is gapped for all $(k_y, k_z)$, in accordance with the gapped surface spectrum of a second-order 3D TI.
(d) Higher-order Berry phase spectrum defined by diagonalizing the nested, second-order Wilson loop Hamiltonian in the ``filled" subspace of the gapped first-order Wilson loop Hamiltonian $H_{\mathrm{W}}(k_y, k_z)$ along the $k_y$-direction, obtaining a Wilson-of-Wilson loop Hamiltonian that only depends on $k_z$. The nontrivial winding along $k_z$ is in one-to-one correspondence with gapless hinge-excitations in the geometry of Fig.~\ref{fig: physical picture}~(a).
(e) The same nontrivial winding can be obtained when computing the Wilson loop spectrum of the slab Hamiltonian of model~\eqref{eq: H}. Being the Wilson loop of a slab, the number of bands is proportional to the linear system size (the thickness of the slab).
(f) Non-nested $z$-direction Wilson loop spectrum.
}
\label{fig: Wilson spectrum}
\end{center}
\end{figure}

\section{Nested boundary spectra: entanglement spectrum and Brillouin zone Wilson loop}
\label{sec: nestedspectra}
In this section we study the boundary degrees of freedom of HOTIs. We focus on the chiral case as exemplified by $\mathcal{H}_{\mathrm{c}}(\bs{k})$ given in Eq.~\eqref{eq: H} of the main text. The most direct way to determine the boundary spectrum is to perform a \emph{slab calculation}, where open instead of periodic boundary conditions are imposed in one direction (chosen to be $x$ here). The slab Hamiltonian $\mathcal{H}_{\mathrm{slab}}(k_y,k_z)$ has one less good momentum quantum number and its spectrum is gapless if the system is a first-order TI.
In the following we employ two alternative approaches that allow us to infer information about the (topological) boundary spectrum of a bulk gapped Hamiltonian $\mathcal{H}(\bs{k})$ and hence about the topological bulk-boundary correspondence.

\subsection{Nested entanglement spectrum}
The single-particle \emph{entanglement spectrum}~\cite{Li08
,Peschel03
} is the spectrum of the logarithm of the reduced density matrix $\rho_A$ of a system that is obtained by subdividing the single-particle Hilbert space into two parts $A$ and $B$ and tracing out the degrees of freedom of $B$
\begin{matriz}
\rho_A = \Tr_B \ket{\Psi}\bra{\Psi} \equiv \frac{1}{Z_{\mathrm{e}}} e^{-\mathcal{H}_{\mathrm{e}}},
\end{matriz}
where $\ket{\Psi}$ is the gapped many-body ground state of $\mathcal{H}(\bs{k})$.
The last equality then defines the entanglement Hamiltonian $\mathcal{H}_{\mathrm{e}}$, with normalization
$Z_{\mathrm{e}} = \Tr \,e^{-\mathcal{H}_{\mathrm{e}}}$.
Here we are interested in a real-space cut separating regions $A$ and $B$ such that all lattice sites $\bs{r}$ with $x>0$ are in $A$, and $B$ is the complement of $A$.  
In this case, $k_y$ and $k_z$ are good quantum numbers which label the blocks of $\mathcal{H}_{\mathrm{e}}$ as  $\mathcal{H}_{\mathrm{e}}(k_y,k_z)$.
The entanglement spectrum [the spectrum of $\mathcal{H}_{\mathrm{e}}(k_y,k_z)$] of a topological state has been shown to be in direct correspondence with the spectrum of the physical slab Hamiltonian for open boundary conditions in the $x$-direction.~\cite{
Fidkowski10-2
} In fact, for the model given in Eq.~\eqref{eq: H} of the main text it is fully gapped [see Fig.~\ref{fig: Wilson spectrum}~(a)].

To determine the presence of hinge modes for HOTIs, we introduce the \emph{nested entanglement spectrum}. Consider a further subdivision of region $A$ into parts $A_1$ and $A_2$ [right inset of Fig.~\ref{fig: Wilson spectrum}~(b)]. Let $|\Psi_\mathrm{e}\rangle$ be the gapped many-body ground state of $\mathcal{H}_{\mathrm{e}}(k_y,k_z)$, where we consider half-filling. Note that half-filling here itself defines a subspace within the filled energy bands. In fact, the half-filled single particle entanglement spectrum bands physically correspond to the many-body ground state of the density matrix with half of the particles in region $A$ and half of the particles in region $B$. The nested entanglement Hamiltonian $\mathcal{H}_{\mathrm{e-e}} (k_z)$ on $A_1$ is then defined by
\begin{matriz}
\Tr_{A_2} \ket{\Psi_\mathrm{e}}\bra{\Psi_\mathrm{e}} \equiv \frac{1}{Z_{\mathrm{e-e}}} e^{-\mathcal{H}_{\mathrm{e-e}}}.
\end{matriz}
It has one less good momentum quantum number and is in correspondence with the spectrum of the slab with open boundary conditions in two directions, that is, with the spectrum of the physical system in the presence of a surface termination featuring hinges. The nested entanglement spectrum in Fig.~\ref{fig: Wilson spectrum}~(b) shows the gapless chiral hinge modes of $\mathcal{H}_{\mathrm{e-e}}$ (one located at each of the four hinges of region $A_1$).

\subsection{Nested Wilson loop}
\label{sec: nestedwilson}
The \emph{Wilson loop} (along the $k_x$ direction, for example) is an operator on the filled band subspace of  $\mathcal{H}(\bs{k})$ defined as
\begin{matriz}
W_{mn}^x(k_y, k_z) = \bra{u_m(2 \pi,k_y,k_z)} \prod_{k_x}^{2\pi \leftarrow 0} P(\vec{k})\ket{u_n(0, k_y,k_z)},
\end{matriz}
where $\ket{u_{m}(\vec{k})}$ are the Bloch eigenstates of the Hamiltonian $\mathcal{H}(\bs{k})$, the indices $m,n$ run over its filled bands, $P(\vec{k})=\ket{u_m(\bs{k})}\bra{u_m(\bs{k})}$ is the projector on the subspace of filled bands at momentum $\bs{k}$, and summation over repeated indices is implied here as well as below. The spectrum of $W^x(k_y, k_z)$ shares its topological features (such as protected boundary modes) with the physical slab Hamiltonian for open boundary conditions in the $x$-direction~\cite{Yu11, Alexandradinata14}.
Being a unitary operator, $W^x(k_y, k_z)$ can be written as $W^x(k_y, k_z)=e^{\mathrm{i}\mathcal{H}_{\mathrm{W}}(k_y, k_z)}$, where $\mathcal{H}_{\mathrm{W}}(k_y, k_z)$ is a Hermitian operator called the Wilson loop Hamiltonian.~\cite{Benalcazar16}
We observe from direct numerical computation for the chiral HOTI defined in Eq.~\eqref{eq: H} that the spectrum of $W^x(k_y, k_z)$ is fully gapped, reflecting the gapped nature of the surface [see Fig.~\ref{fig: Wilson spectrum}~(c)]. In fact, $\mathcal{H}_{\mathrm{W}}(k_y, k_z)$ can be seen as a Hamiltonian for a 2D insulator. Explicit computation reveals that the half-filled subspace carries a nontrivial Chern number $C=\pm1$.

We can now compute the Wilson loop spectrum of the Wilson loop Hamiltonian $\mathcal{H}_{\mathrm{W}}(k_y, k_z)$, following the concept of nested Wilson loops introduced in Ref.~\onlinecite{Benalcazar16}. To do this, we first diagonalize the gapped Hamiltonian $\mathcal{H}_{\mathrm{W}}(k_y,k_z)$ and evaluate a $y$-directed Wilson loop in its filled subspace. As a result, we obtain an effective one-dimensional system with good momentum quantum number $k_z$ that shows the gapless, symmetry protected spectral flow of a Chern insulator [see Fig.~\ref{fig: Wilson spectrum}~(d)].

Finally, Figure~\ref{fig: Wilson spectrum}~(f) exemplifies the non-trivial $\mathbb{Z}_2$ winding in the spectrum of the first-order (so non-nested) Wilson loop operator $W^z(k_x,k_y)$. The gapless spectrum of the Wilson loop $W^z(k_x,k_y)$ is also in correspondence with the gapless nature of the (001) surface of the model mentioned above. The connectivity of the Wilson loop bands between the the $\hat{C}_4^z \hat{T}$-enforced Kramers pairs at momenta $(k_x, k_y) \in \{(0, 0),(\pi, \pi)\}$ is a $\mathbb{Z}_2$ topological invariant.\cite{Yu11,Alexandradinata14} In contrast to 3D TIs with $\hat{T}$ symmetry, the Wilson loop spectrum has no Kramers degeneracies at momenta $(k_x, k_y) \in \{(0, \pi),(\pi, 0)\}$.

\section{Alternative models for chiral higher-order TI}


\subsection{3D chiral higher-order TI as a perturbation of a 3D first-order TI}

Here we discuss the possibility of realizing a chiral 3D HOTI on the basis of a 3D first-order TI, which is initially invariant under both $\hat{C}_4$ and $\hat{T}$ symmetry, by performing a time-reversal breaking surface manipulation. 
For simplicity, we consider the case of only $\hat{C}_4^z$ rotation symmetry, as found in tetragonal crystal structures (a generalization to cubic crystal structures with three $\hat{C}_4$ symmetries is discussed in the Sec.~\ref{app: isotropicmodel}). We consider such a system in a geometry with open boundary conditions in the $x$ and $y$ directions and periodic boundary conditions in the $z$ direction, as depicted in Fig.~\ref{fig: physical picture} in the main text.

 As its defining feature, the 3D TI has one Dirac cone on each surface. 
Now consider perturbing the 3D TI such that (i) a gap is opened on each surface with a normal lying in the $(x,y)$ plane and (ii) the system remains invariant under the product $\hat{C}^z_4\hat{T}$. Such a perturbation necessarily breaks $\hat{C}^z_4$ and $\hat{T}$ individually. The four hinges are then domain walls at which the magnetization changes from inward to outward pointing. It is well known~\cite{
, Jackiw76} that such a domain wall on the surface of a 3D TI binds a gapless chiral mode, which in the case at hand is reinterpreted as the hinge mode of a HOTI.

Such a $\hat{C}^z_4\hat{T}$ symmetric perturbation could be applied to the surface only by coating the (100) surfaces and the (010) surfaces with ferromagnetic layers whose magnetization is oriented parallel and antiparallel to the surface normal, respectively [see Fig.~\ref{fig: physical picture}~(a)]. 



\subsection{Optical lattice model for chiral higher-order TI}

\label{app: other model}
\begin{figure}[t]
\begin{center}
\includegraphics[width=0.45 \textwidth]{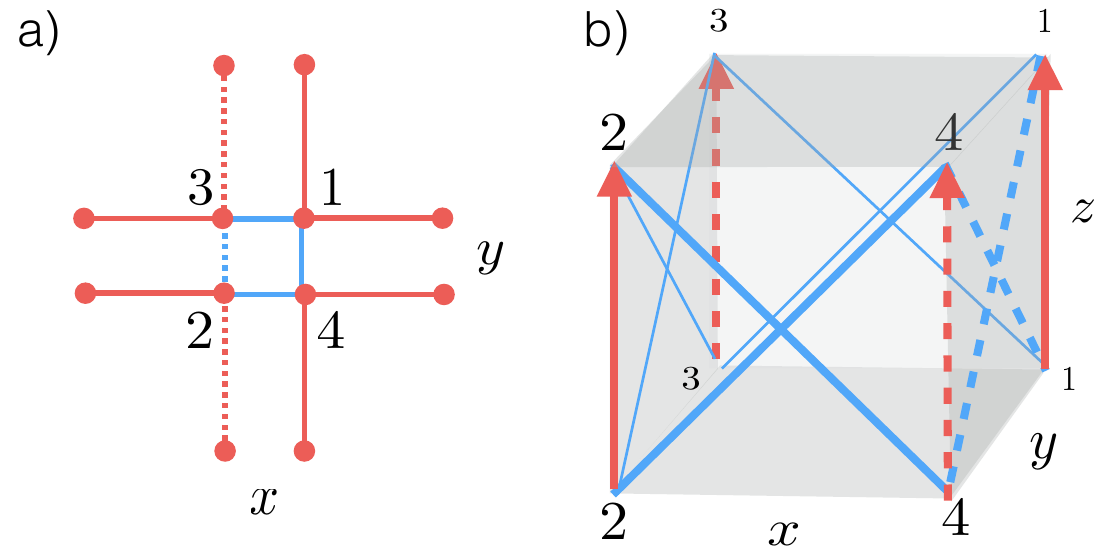}
\caption{
Real-space hopping picture for the optical lattice model with $H_{4} = \sum_{\langle  ij\rangle\in(x,y)} \left(t^{(x,y)}_{ij} c^{\dagger}_{z,i} c_{z,j} + t^{z}_{ij} c^{\dagger}_{z,i} c_{z+1,j}\right)$. (a) In the $(x,y)$-plane, we take over the hopping terms $t^{(x,y)}_{ij}$ from the 2D quadrupole model presented in Ref.~\onlinecite{Benalcazar16}. Here, solid blue lines contribute a hopping amplitude of $M$ to $t^{(x,y)}_{13}=t^{(x,y)}_{24}=t^{(x,y)}_{14}$, dashed blue lines contribute $-M$ to $t^{(x,y)}_{23}$, and the additional red lines, with amplitues $\pm 1$ for solid and dashed lines, respectively, implement a dimerization as long as $M \neq 1$. (b) We then couple the $(x,y)$-planes in $z$-direction with nearest and next-to-nearest neighbor hoppings $t^{z}_{ij}$. Here, solid blue lines stand for a hopping amplitude $t^z_{13}= t^z_{24} = 1$, dashed blue lines for $t^z_{14} = -1$, solid red lines for $t^z_{11}= t^z_{22} = \mathrm{i}$ (for hoppings in the direction of the arrow), and dashed red lines for $t^z_{33}= t^z_{44} = -\mathrm{i}$ (for hoppings in the direction of the arrow).
}
\label{fig: opticalmodel}
\end{center}
\end{figure}

Here, we present an alternative model for a second-order chiral TI with protected hinge states, which naturally lends itself to an interpretation in terms of magnetic fluxes and nearest as well we next-to-nearest neighbor hoppings, and is thus possibly realizable in ultra-cold atomic systems (see Fig.~\ref{fig: opticalmodel} for a real-space picture). This model corresponds to a topological-to-trivial tuning of the second-order 2D TI, or electric quadrupole TI, from Ref.~\onlinecite{Benalcazar16} along the $z$ direction.

In momentum space, the Hamiltonian of the 4-band model has the form
\begin{matriz} \label{eq: Halt_4}
\begin{aligned}
\mathcal{H}(\vec{k}) = \, &
\left[M-\cos(k_z)+\cos(k_x)\right] \, \tau_x \sigma_0 \\&- \,
\left[M-\cos(k_z)+\cos(k_y)\right] \, \tau_y \sigma_y\\
&-\, \Delta_1\left[ \sin(k_x) \, \tau_y \sigma_z + \sin(k_y) \, \tau_y \sigma_x \right. \\ 
&+ \, \left.\sin(k_z) \, \tau_z \sigma_0 \right]  + \Delta_2 \, \tau_z \sigma_y,
\end{aligned}
\end{matriz}
where $\sigma_0$ and $\sigma_i$, $i=x,y,z$, as well as are $\tau_0$ and $\tau_i$, $i=x,y,z$, are the $2\times 2$ identity matrix and the three Pauli matrices. 

The Hamiltonian $\mathcal{H}$ respects $\hat{C}^z_4\hat{T}$, and the parameter $\Delta_2$ can be turned on weakly to break the product of inversion and time-reversal symmetry $\hat{I} \hat{T}$, with $I = \tau_0 \sigma_y$ and $T = \mathit{K}$. The $\hat{C}_4^z$ rotation is represented by the real matrix $C^z_4 = \bigl(\begin{smallmatrix}0&1 \\ -i \sigma_y&0\end{smallmatrix} \bigr)$. Note that for spinless fermions with $\hat{T}^2 = 1$ we usually have $(\hat{C}^z_4)^4 = 1$. However, the present model features a $\pi$-flux piercing the $(x,y)$-plane, which contributes a phase factor $e^{\mathrm{i} \pi}$ to any wavefunction after a fourfold rotation, and therefore implies $(\hat{C}^z_4)^4 = -1$. Thus, \emph{all} of the $\hat{C}_4^z \hat{T}$ eigenvalues are complex, and by the complex conjugation afforded by $\hat{T}$ occur necessarily in degenerate pairs at the $\hat{C}_4 \hat{T}$ invariant momenta $\vec{k} \in \mathcal{I}_{\hat{C}_4^z \hat{T}} = \{(0, 0, 0), (\pi, \pi, 0), (0, 0, \pi), (\pi, \pi, \pi)\}$ (since these are left unchanged by $\hat{C}_4^z$).


\subsection{Isotropic chiral higher-order TI}
\label{app: isotropicmodel}

The anisotropic model Hamiltonian~(\ref{eq: H}) for a chiral 3D HOTI from the main text can be straightforwardly generalized to an isotropic 12-band model with Hamiltonian
\begin{matriz}
\begin{split}
\mathcal{H}_{\mathrm{c},12}(\vec{k}) = &\,
\hat{e}^{(1)} \otimes \mathcal{H}_{\mathrm{c}}(\bs{k}) 
+\hat{e}^{(2)} \otimes R\mathcal{H}_{\mathrm{c}}(D_{\hat{R}}\bs{k})R^{-1}
\\
&
+\hat{e}^{(3)} \otimes R^2\mathcal{H}_{\mathrm{c}}(D_{\hat{R}^2}\bs{k})R^{-2}
,
\end{split}
\label{eq: 12-band model}
\end{matriz}
defined on a simple cubic lattice with three copies of the degrees of freedom found on each lattice site of model~(\ref{eq: H}).  Here, $\hat{e}^{(n)}$ are $(3\times3)$ matrices with elements $\hat{e}^{(n)}_{i,j}=\delta_{n,i}\delta_{n,j}$, for $n,i,j\in \{1,2,3\}$. The rotation operator $R$ about the $\bs{v}=(1,1,1)/\sqrt{3}$ axis acts on spin-orbital space of Hamiltonian~(\ref{eq: H}) as $R=\tau_0e^{-\mathrm{i}\pi \bs{v}\cdot\bs{\sigma}/3}$, with momentum representation $D_{\hat{R}}\bs{k}=(k_y,k_z,k_x)$.
The symmetry representations on $\mathcal{H}_{\mathrm{c}}$ are as before, for the fourfold rotations we choose the isotropic generalization $C_4^i \equiv \tau_0 e^{-\mathrm{i} \frac{\pi}{4} \sigma_i}$.

The Hamiltonian~\eqref{eq: 12-band model} features three $\hat{C}_4 \hat{T}$ symmetries, one for every direction of space, which induce mixings between the three sublattices. These symmetries are represented by the $(12\times12)$ matrices
\begin{matriz}
\begin{gathered}
\tilde{C}_4^x \equiv 
\begin{pmatrix}
0&0&1\\
0&1&0\\
1&0&0
\end{pmatrix}
\otimes C^x_4,
\quad
\tilde{C}_4^y \equiv 
\begin{pmatrix}
0&1&0\\
1&0&0\\
0&0&1
\end{pmatrix}
\otimes C^y_4, 
\\
\tilde{C}_4^z \equiv 
\begin{pmatrix}
1&0&0\\
0&0&1\\
0&1&0
\end{pmatrix}
\otimes C^z_4.
\end{gathered}
\end{matriz}
These symmetries are implemented in the same way as in Eq.~\eqref{eq: C4ZTRep}, with the momentum-space transformations acting as $D_{\hat{C}^{x}_4\hat{T}} \vec{k} = (-k_x, k_z, -k_y)$ and $D_{\hat{C}^{y}_4 \hat{T}} \vec{k} = (-k_z, -k_y, k_x)$, in addition to $D_{\hat{C}^{z}_4\hat{T}} \vec{k} = (k_y, -k_x, -k_z)$ from before.

\begin{figure}[t]
\begin{center}
\includegraphics[width=0.45 \textwidth]{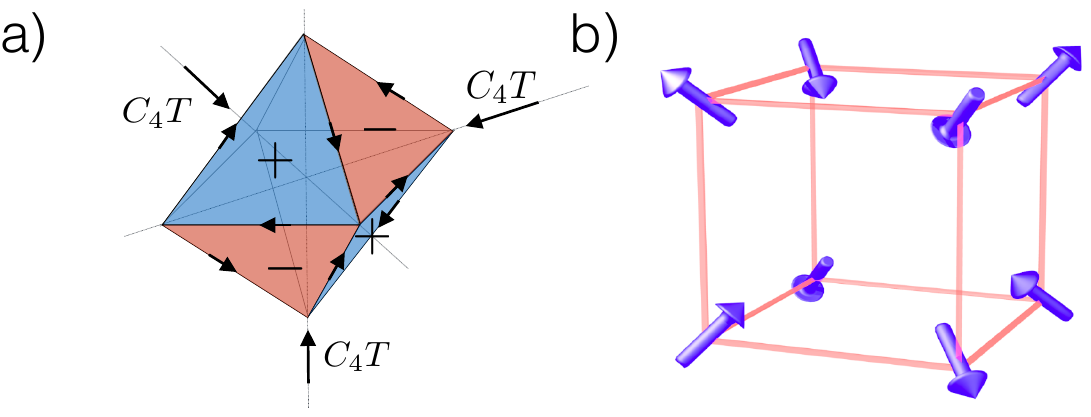}
\caption{
(a) For a $\hat{C}_4\hat{T}$ symmetric termination, second-order 3D TI has gapped surfaces but gapless chiral hinge states (black arrows). The surface magnetization is parallel to the surface normal on blue surfaces (marked by +) and antiparallel to the surface normal on the red surfaces (marked by $-$). 
(b) One unit cell of 3D triple-$q$ magnetic order. When imposed on a system with the band structure of a conventional 3D TI, a second-order 3D TI can be obtained.
}
\label{fig: octahedron}
\end{center}
\end{figure}

For the anisotropic model Hamiltonian~(\ref{eq: H}) from the main text, we found
that its (001) surface termination is gapless, a property that was protected by $\hat{C}_4^z\hat{T}$ symmetry. 
Therefore, by construction, the Hamiltonian~\eqref{eq: 12-band model} has protected gapless states on the (100), (010), and (001) surfaces. However, other surfaces, such as (111), are generically gapped. As a consequence, if we consider a geometry where all surfaces are obtained by the application of arbitrary powers of $\hat{C}_4^i\hat{T}$, ($i=x,y,z$) to the (111) surface, the hinges will carry chiral modes. The required geometry for all surface states to be gapped in the isotropic model is therefore the dual polyhedron of a cube, the octahedron, as shown in Fig.~\ref{fig: octahedron}~(a). In particular, coating the surfaces of an octahedral first-order TI with ferromagnets whose magnetization is alternating between being parallel [blue in Fig.~\ref{fig: octahedron}~(a)] and antiparallel (red) to the surface normal in a $\hat{C}_4\hat{T}$-invariant fashion results in chiral states on the hinges of the octahedron.~\cite{Jackiw76}

Physical realizations of isotropic HOTIs could be found in 3D TI materials with broken time-reversal symmetry, but preserved $\hat{C}^i_4\hat{T}$ symmetries. 
An example of the type of magnetic order that preserves these symmetries
is the $\hat{C}_4\hat{T}$ invariant triple-$Q$ $\left(\pi, \pi, \pi \right)$ magnetic order depicted in Fig.~\ref{fig: octahedron} (b) and also recently discussed in Ref.~\onlinecite{Hanke16}. We can consider a 3D TI band structure that is weakly perturbed by such magnetic order. If the energy scale of the magnetic order is much smaller than the bulk gap of the unperturbed TI, it will mainly affect the electronic structure of the TI surface states. The exponentially decaying surface states will mainly couple to the layer of magnetic moments closest to the surface. For the octahedral sample shown in Fig.~\ref{fig: octahedron}~(a), the triple-$Q$ $\left(\pi, \pi, \pi \right)$ magnetic order naturally terminates with a ferromagnetic layer of the same alternating orientation as discussed above. We conclude that a 3D TI coupled to (sufficiently weak) $\hat{C}_4\hat{T}$ preserving magnetic order is a bulk realization of a second-order 3D TI. 

\section{Bulk-surface-hinge correspondence for the helical higher-order TI}

\subsection{Constraints on mirror eigenvalues of domain wall modes in two-dimensional systems}

\begin{figure*}[t]
\begin{center}
\includegraphics[width=0.8 \textwidth]{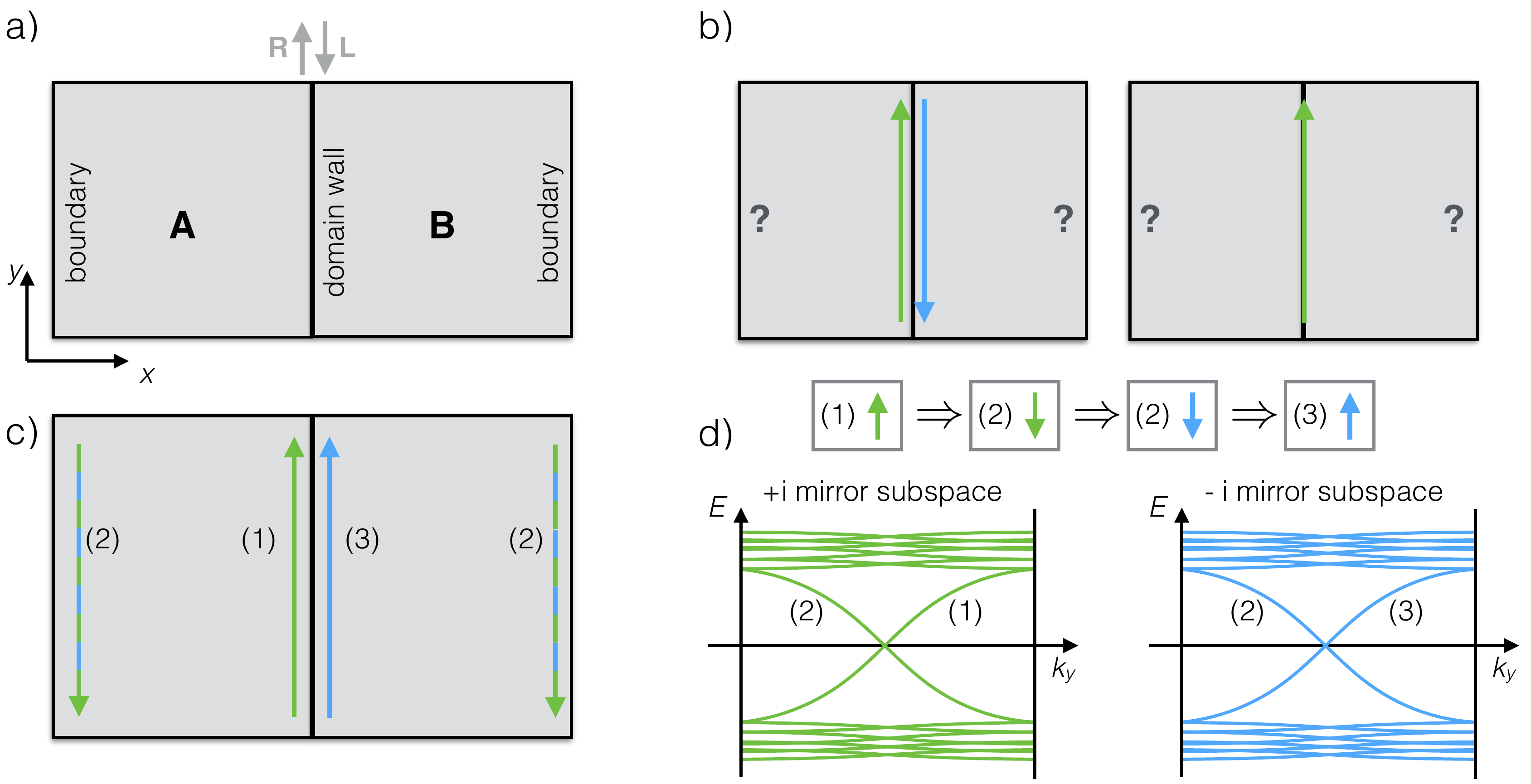}
\caption{
a) Domain wall geometry considered for the proof that there is no excess of mirror eigenvalue $+\mathrm{i}$ chiral states moving chirally in definite direction along the domain wall.
b) These two configurations and any superposition thereof are not allowed in strictly two-dimensional systems. Green and blue arrows stand for modes with mirror eigenvalue $+\mathrm{i}$ and $-\mathrm{i}$, respectively.
c) The minimal allowed configuration with domain wall modes.
d) Logical chain of implications why the existence of mode (1) implies the existence of mode (3). Each implication is explained in the text. The first and third implication follow from the spectral continuity of the system as a function of $k_y$ when periodic boundary conditions are imposed in the $y$ direction: the number of bands  of a given mirror eigenvalue below the Fermi level cannot change as $k_y$ is advanced by $2\pi$. Hence chiral modes at the domain wall must be compensated by antichiral modes at the boundary.  
}
\label{fig: domain wall}
\end{center}
\end{figure*}

In this subsection we discuss constraints on domain wall modes in strictly two-dimensional systems, that are imposed by mirror symmetry. 
Consider a two-dimensional system with a mirror symmetry that maps one half of the system ($A$) to the other ($B$). 
For concreteness, let us consider a spinful system with mirror eigenvalues $\pm \mathrm{i}$, while noting that the proof also applies to spinless systems with mirror eigenvalues $\pm 1$. 
Let the system be insulating in either half and consider a situation, where along the line that is left invariant under the mirror symmetry, gapless bound states may propagate [see Fig.~\ref{fig: domain wall}~(a)]. We will denote the mirror invariant line as \emph{domain wall} and the two edges, which are mapped into each other by the mirror symmetry, as the \emph{boundary} of the system.
\emph{We want to show that necessarily the number of right-movers (R) along the domain wall with mirror eigenvalues $+\mathrm{i}$ equals the number of right-movers with mirror eigenvalue $-\mathrm{i}$.} The same is true for the left-movers (L).
Domain wall modes as shown in Fig.~\ref{fig: domain wall}~(b) are then disallowed, where green and blue arrows denote mirror eigenvalues $+\mathrm{i}$ and $-\mathrm{i}$, respectively. 
 These statements hold independent of whether or not the system is time-reversal symmetric. If present, time-reversal symmetry simply enforces the additional constraint that the number of right-movers with mirror eigenvalue $\pm\mathrm{i}$ equals the number of left-movers with 
mirror eigenvalue $\mp\mathrm{i}$.

We will first give a simple intuitive argument and then a more detailed technical proof. 
For concreteness, let the mirror symmetry $\hat{M}_x$ send $(x,y)\to (-x,y)$. Furthermore, let the system be translationally invariant with periodic boundary conditions along the $y$ direction for simplicity [see Fig.~\ref{fig: domain wall}~(a)].
All eigenstates of the mirror symmetric system, independent of their localization, can be labeled by mirror eigenvalues $\pm \mathrm{i}$. We consider the spectrum in the $+\mathrm{i}$ subspace and of the $-\mathrm{i}$ subspace separately. 
Suppose now, we have one R mode with eigenvalue $+\mathrm{i}$ at the domain wall, represented by the green arrow in Fig.~\ref{fig: domain wall}~(c). As $k_y$ is advanced by $2\pi$, this chiral mode connects the valence and conduction band of the two-dimensional bulk, as shown in Fig.~\ref{fig: domain wall}~(d). However, the total number of bands below the bulk gap has to be equal at $k_y=0$ and $k_y=2\pi$ for the spectrum to be periodic. This implies that there is one antichiral L mode (2) with the same mirror eigenvalue $+\mathrm{i}$ in the system. Since by assumption this mode is not localized at the domain wall and the bulk is gapped, it needs to be localized at the system boundary. To be an eigenstate of mirror symmetry, it needs to be localized on both boundaries at the same time. However, since we cannot place `half' of a mode on each boundary, there must be another gapless L mode (2), which is also localized on both boundaries and also a mirror eigenstate. Since we have already completed the band structure of the $+\mathrm{i}$ mirror subspace with one R and one L gapless mode, this second L boundary mode must have $-\mathrm{i}$ mirror eigenvalue. (Appropriate equal amplitude superpositions of the L modes with $+\mathrm{i}$ and $-\mathrm{i}$ mirror eigenvalues will give the L modes fully localized on only one side of the boundary.)
By the argument of spectral continuity, the $-\mathrm{i}$ mirror subspace cannot support a single L  mode either. Rather, we conclude that there exists also a R mode (3) with mirror eigenvalue $-\mathrm{i}$. We cannot split this mode up to be localized on both sides on the boundary (without also introducing an additional 
$+\mathrm{i}$ R mode). Hence, the R mode (3) must be localized at the domain wall. Thus, via the chain of implications shown in Fig.~\ref{fig: domain wall}~(d), we have argued that the existence of exactly one R  mode at the domain wall with $+\mathrm{i}$ mirror eigenvalue  implies the existence of exactly one R  mode at the domain wall with $-\mathrm{i}$ mirror eigenvalue. This line of arguments carries over to any integer number of modes.


Let us now turn to a more technical elaboration on one aspect of the above argument:
we want to show that each mode that is localized on the system boundary (not at the domain wall) with a specific mirror eigenvalue has a degenerate partner with the opposite mirror eigenvalue that is also localized at the boundary.

Boundary-localized modes with definite mirror eigenvalue (i.e., mirror eigenstates) have to have equal weight on both boundaries. Their nature can be understood as follows: boundary modes have support, up to exponentially small corrections, only near the two boundaries. Denote such an eigenstate of mirror eigenvalue $+\mathrm{i}$ by
\begin{matriz}
\psi_+(k_y)=[\psi_A(k_y),\psi_B(k_y)]^{\mathsf{T}},
\end{matriz}
where $\psi_A(k_y)$ [$\psi_B(k_y)$] is the part of the wave function that lives in the single particle Hilbert space of part $A$ [$B$] of the system at given $k_y$. 
The mirror symmetry exchanges $A$ and $B$ and therefore has a representation of the form
\begin{matriz}
M_x=\mathrm{i}
\begin{pmatrix}
0 & m_x \\
m^{-1}_x & 0
\end{pmatrix}
\label{eq: mirror rep}
\end{matriz}
with $m_x$ unitary in order for $M_x$ to be unitary. 

Further, $\psi_+(k_y)$ is eigenstate of the Bloch Hamiltonian of the system
\begin{matriz}
\mathcal{H}(k_y)=
\begin{pmatrix}
\mathcal{H}_{AA}(k_y)&\mathcal{H}_{AB}(k_y)\\
\mathcal{H}_{AB}(k_y)^\dagger&\mathcal{H}_{BB}(k_y)
\end{pmatrix}.
\end{matriz}
Here, $\mathcal{H}_{AB}(k_y)$ are terms in the Hamiltonian that couple parts $A$ and $B$, i.e., local terms near the domain wall. Due to the localization properties of the boundary mode, 
\begin{matriz}
\mathcal{H}_{AB}(k_y)\psi_B(k_y), \qquad 
\mathcal{H}_{AB}(k_y)^\dagger \psi_A(k_y),
\end{matriz}
are exponentially small in the width of the system over the correlation length in the bulk of either part, compared to 
$\mathcal{H}_{BB}(k_y)\psi_B(k_y)$ and 
$\mathcal{H}_{AA}(k_y) \psi_A(k_y)$.
As a consequence, $\psi_+(k_y)$ is (up to these exponential corrections) also an eigenstate of 
\begin{matriz}
\tilde{\mathcal{H}}(k_y)=
\begin{pmatrix}
\mathcal{H}_{AA}(k_y)&0\\
0&\mathcal{H}_{BB}(k_y)
\end{pmatrix}.
\end{matriz}
This in turn implies that $\psi_A(k_y)$ is an eigenstate of $\mathcal{H}_{AA}(k_y)$ and $\psi_B(k_y)$ is an eigenstate of $\mathcal{H}_{BB}(k_y)$, both with the same eigenvalue. 
Notice also that the mirror symmetry $M_x$ in Eq.~\eqref{eq: mirror rep} commutes with both $\mathcal{H}(k_y)$ and $\tilde{\mathcal{H}}(k_y)$.

Now, $\psi_+(k_y)$ is by assumption a mirror eigenstate with eigenvalue $+\mathrm{i}$, i.e.,
\begin{matriz}
\psi_A(k_y)= m_x \psi_B(k_y).
\end{matriz}
Then, the state 
\begin{matriz}
\psi_-(k_y)=\left(\psi_A(k_y),-\psi_B(k_y)\right)^{\mathsf{T}},
\end{matriz}
is an eigenstate of $M_x$ with mirror eigenvalue $-\mathrm{i}$ because
\begin{matriz}
M_x\psi_-(k_y)
=-
\begin{pmatrix}
m_x\psi_B(k_y)
\\
-m^{-1}_x \psi_A(k_y)
\end{pmatrix}
=-\psi_-(k_y).
\end{matriz}
Furthermore, by the above arguments $\psi_-(k_y)$ is  an eigenstate of $\tilde{\mathcal{H}}(k_y)$ and thus also of $\mathcal{H}(k_y)$, with the same energy and chirality (R/L) as $\psi_+(k_y)$.

We have thus shown that for any mirror eigenstate localized at the system boundaries and of definite chirality R/L  that has mirror eigenvalue $+\mathrm{i}$ we can construct a degenerate state with the same chirality, but the opposite mirror eigenvalue. 
Then also the domain wall has to bind an equal number of chiral modes with mirror eigenvalue $+\mathrm{i}$ and $-\mathrm{i}$ that propagate in the R direction, for example. We denote this number by $N_{\mathrm{R}}$. The same is true for the L direction, with an equal number of $N_{\mathrm{L}}$ modes in each mirror subspace. In a time-reversal breaking system, these two numbers $N_{\mathrm{R}}$ and $N_{\mathrm{L}}$ can in general differ and their difference is the Hall conductivity (in units of $e^2/h$) on the A side of the system, which differs by a minus sign from the Hall conductivity on the B side.

If in addition we consider a time-reversal symmetric system, the Hall conductivities have to vanish, and the number of $-\mathrm{i}$ R modes equals the number of $+\mathrm{i}$ L modes at the domain wall. Thus, $N_{\mathrm{L}}=N_{\mathrm{R}}$, i.e., the number of R modes equals the number of L modes in the $+\mathrm{i}$ sector. However, counter-propagating modes in a given sector can be gapped pairwise by local perturbations at the domain wall. Thus, all potentially existing modes in the $+\mathrm{i}$ sector can generically gap. The same holds for the $-\mathrm{i}$ sector. We conclude that the domain wall in a mirror and time-reversal symmetric two-dimensional system does not host any protected modes.


\subsection{Correspondence between mirror Chern number and hinge modes}

Here,  we show that a non-vanishing mirror Chern number $C_{\mathrm{m}}$ pertaining to the mirror symmetry $\hat{M}_{xy}$ implies the presence of $C_{\mathrm{m}}/2$ Kramers pairs of hinge modes in a geometry terminated by the (100) and (010) surfaces.

We begin with a Hamiltonian that models the combination of $\mathcal{H}_{\mathrm{D},+}$ and $\mathcal{H}_{\mathrm{D},-}$ (which are defined in Eq.~\eqref{eq: low-energy massive surface Dirac} in the main text) on the surfaces with normals $\bs{n}_+$ and $\bs{n}_-$, respectively, which meet at a hinge as depicted in Fig.~\ref{fig: bulk-boundary}~(c) in the main text. Denote by $k_1$ the in-plane momentum of the (110) surface that is perpendicular to $k_z$. The hinge is then modeled as a domain wall in the mass term $m(x_1)$, with $x_1$ the position conjugate to $k_1$. We consider the effective Dirac Hamiltonian
\begin{matriz}
\mH(k_1,k_z) = v_1 k_1 \sigma_z + v_z k_z \sigma_x + m(x_1) \sigma_y,
\end{matriz}
where we have chosen $k_1^{(0)} = k_z^{(0)} = 0$ as an expansion point.
Since we want to find topologically protected hinge modes, we can choose $m(x_1) = \bar{m} \tanh \left(x_1/\lambda \right)$, as a smooth interpolation to represent the domain wall.
Solving for a zero-energy state at $k_z = 0$, we find two solutions, 
\begin{matriz}
\ket{\pm} = f_\pm(x_1) (1,\pm1)^{\mathsf{T}}, \quad f_\pm(x_1) = \mathcal{N} \big[ \cosh \left( x\right) \big]^{\mp \bar{m}\lambda / v_1}  ,
\end{matriz}
where $\mathcal{N}$ is a normalization constant.
The solutions $\ket{\pm}$ have  eigenvalues $\pm \mathrm{i}$, respectively, under the mirror symmetry $\hat{M}_{xy} = \mathrm{i} \sigma_x$ that sends $x_1\to -x_1$. For a given solution to be normalizable, we require $\pm \frac{\bar{m}\lambda}{v_1} > 0$. Thus, for either choice of the sign of $\frac{\bar{m}\lambda}{v_1}$, exactly one solution is normalizable. To determine the chiral dispersion of the solution, we reinstate $k_z$ and consider the energy shift to first order
\begin{matriz}
\Delta E_\pm = \bra{\pm}v_z k_z \sigma_x \ket{\pm} = \pm v_z k_z.
\end{matriz}
From this we deduce that the domain wall either binds a R moving mode with mirror eigenvalue $\mathrm{i} \,\mathrm{sgn}(v_z)$, or an L moving mode with mirror eigenvalue $-\mathrm{i}\,\mathrm{sgn}(v_z)$, as claimed in the main text.

\section{Alternative model for helical higher-order TI}
\label{app: other TRS model}
To define a natural helical, i.e., time-reversal symmetric, generalization of the chiral model considered in the main text, we consider spinful electrons hopping on a lattice with two sets of orbitals labeled as $(d^\mu_{x^2-y^2}, d^\mu_{xy})_\alpha=v^\mu_\alpha$  on each site (so that $\alpha, \mu=0,1$). The model Hamiltonian reads in real-space
\begin{matriz}
\begin{aligned}
H_{\mathrm{h}} = &\hphantom{+} \frac{M}{2} \sum_{\bs{r},\alpha,\mu} (-1)^\alpha \, c^\dagger_{\bs{r},\alpha,\mu} c_{\bs{r},\alpha,\mu} \\
&+ \frac{t}{2} \sum_{\bs{r},\alpha,\mu} \sum_{i=x,y,z}
 (-1)^\alpha \, c^\dagger_{\bs{r}+\bs{\hat{e}}_i,\alpha,\mu} c_{\bs{r},\alpha,\mu} 
\\
&+ \frac{\Delta_1}{2} \sum_{\bs{r},\alpha,\mu} \sum_{i=x,y,z} \, 
c^\dagger_{\bs{r}+\bs{\hat{e}}_i,\alpha+1,\mu} \, \sigma_i \, c_{\bs{r},\alpha,\mu}
 \\
&+ \frac{\Delta_2}{2} \sum_{\bs{r},\alpha,\mu} \sum_{i=x,y,z} (-1)^{\alpha} \, n_i \, c^\dagger_{\bs{r}+\bs{\hat{e}}_i,\alpha + 1,\mu+1} c_{\bs{r},\alpha,\mu}
\\&
+\mathrm{h.c.},
\end{aligned}
\end{matriz}
where $\alpha$ and $\mu$ are defined modulo $2$, $\hat{\bs{n}} = (1,-1,0)$, and $c^\dagger_{\bs{r},\alpha,\mu}$ creates a spinor $c^\dagger_{\bs{r},\alpha,\mu}=(c^\dagger_{\bs{r},\alpha,\mu,\uparrow},c^\dagger_{\bs{r},\alpha,\mu,\downarrow})$ in orbital $(\alpha,\mu)$ at lattice site $\bs{r}$. We denote by $\sigma_0$ and $\sigma_i$, $i=x,y,z$, respectively, the $2\times 2$ identity matrix and the three Pauli matrices acting on the spin 1/2 degree of freedom.

In momentum space, the corresponding Bloch Hamiltonian takes the form
\begin{matriz} \label{eq: trsH}
\begin{aligned}
\mathcal{H}_{\mathrm{h}}(\vec{k}) = &\Bigl(M+t \sum_i \cos k_i\Bigr) \, \tau_z \rho_0 \sigma_0 \\&+ \Delta_1 \sum_i \sin k_i \, \tau_x  \rho_0 \sigma_i \\&+\Delta_2(\cos k_x - \cos k_y) \, \tau_y \rho_y \sigma_0.
\end{aligned}
\end{matriz}
Here, the Pauli matrices $\tau_i$ and $\rho_i$ act on the $\alpha$ and $\mu$ index of the $v^\mu_\alpha$ orbital vector, respectively, $\sigma_i$ acts on spin as before, and zero components such as $\tau_0$ stand for an identity matrix. We now describe how the helical HOTI phase of this Hamiltonian in the parameter range $1<|M|<3$ can be protected by either mirror or fourfold rotational symmetries.

\subsection{Protection by mirror symmetries}
Consider the symmetry representations
\begin{matriz}
\begin{aligned}
&T \equiv \tau_0 \rho_0 \sigma_y \mathit{K}, \quad D_{\hat{T}} \, \vec{k} = -\vec{k},\\ 
&M_x \equiv \mathrm{i} \, \tau_z \rho_0 \sigma_x, \quad D_{\hat{M}_x} \, \vec{k} = (-k_x,k_y,k_z),\\
&M_y \equiv \mathrm{i}\, \tau_z \rho_0 \sigma_y, \quad D_{\hat{M}_y} \, \vec{k} = (k_x,-k_y,k_z),\\
&C_4^z \equiv \tau_0 \rho_0 e^{-\mathrm{i} \frac{\pi}{4} \sigma_z}, \quad D_{\hat{C}^{z}_4} \vec{k} = (-k_y,k_x,k_z).
\end{aligned}
\label{eq: symmetries higher-order}
\end{matriz}
For $\Delta_2=0$, the model preserves all of these symmetries individually. Note that $\hat{C}_4^z$ together with the $\hat{M}_{x/y}$ mirror symmetries implies the diagonal mirror symmetries $\hat{C}_4^z \hat{M}_y = \hat{M}_{xy}$ and $\hat{C}_4^z \hat{M}_x = \hat{M}_{x\bar{y}}$. All mirror symmetries allow the definition of mirror Chern numbers on mirror-invariant subspaces of the Brillouin zone: For the $\hat{M}_x$ and $\hat{M}_y$ symmetries, these are the $k_x = 0, \pi$ and $k_y = 0, \pi$ planes, respectively, for the $\hat{M}_{xy}$ and $\hat{M}_{x\bar{y}}$ symmetries they are the $k_x = k_y$ and $k_x = -k_y$ planes, respectively. 

While the mirror Chern numbers of model~\eqref{eq: trsH} with $1<M<3$ and $\Delta_2=0$ on the $k_{x/y} = 0$ planes are $0$, on the $k_{x/y} = \pi$ and $k_x = \pm k_y$ planes they are equal to $2$, and therefore protect two surface Dirac cones each on bulk terminations that preserve the respective mirror symmetries. Turning on $\Delta_2$ weakly gaps out the Dirac cones on the surfaces obtained by terminating the system in $x$ and $y$ direction with opposite masses. This breaks $\hat{C}_4^z$ and $\hat{M}_{x/y}$ individually, rendering the respective mirror Chern numbers ill-defined, but preserves $\hat{M}_{xy}$ and $\hat{M}_{x\bar{y}}$, which retain their mirror Chern number of $2$ on the $k_x = \pm k_y$ planes, and still enforce two Dirac cones each on the (110), (1\=10), and (001) surfaces. From Eq.~\eqref{eq: bulk-hinge helical} in the main text we therefore deduce the presence of one Kramers pair of helical modes located at the hinges separating the (100) and (010) surface terminations.

\begin{figure}[t]
\begin{center}
\includegraphics[width=0.48 \textwidth]{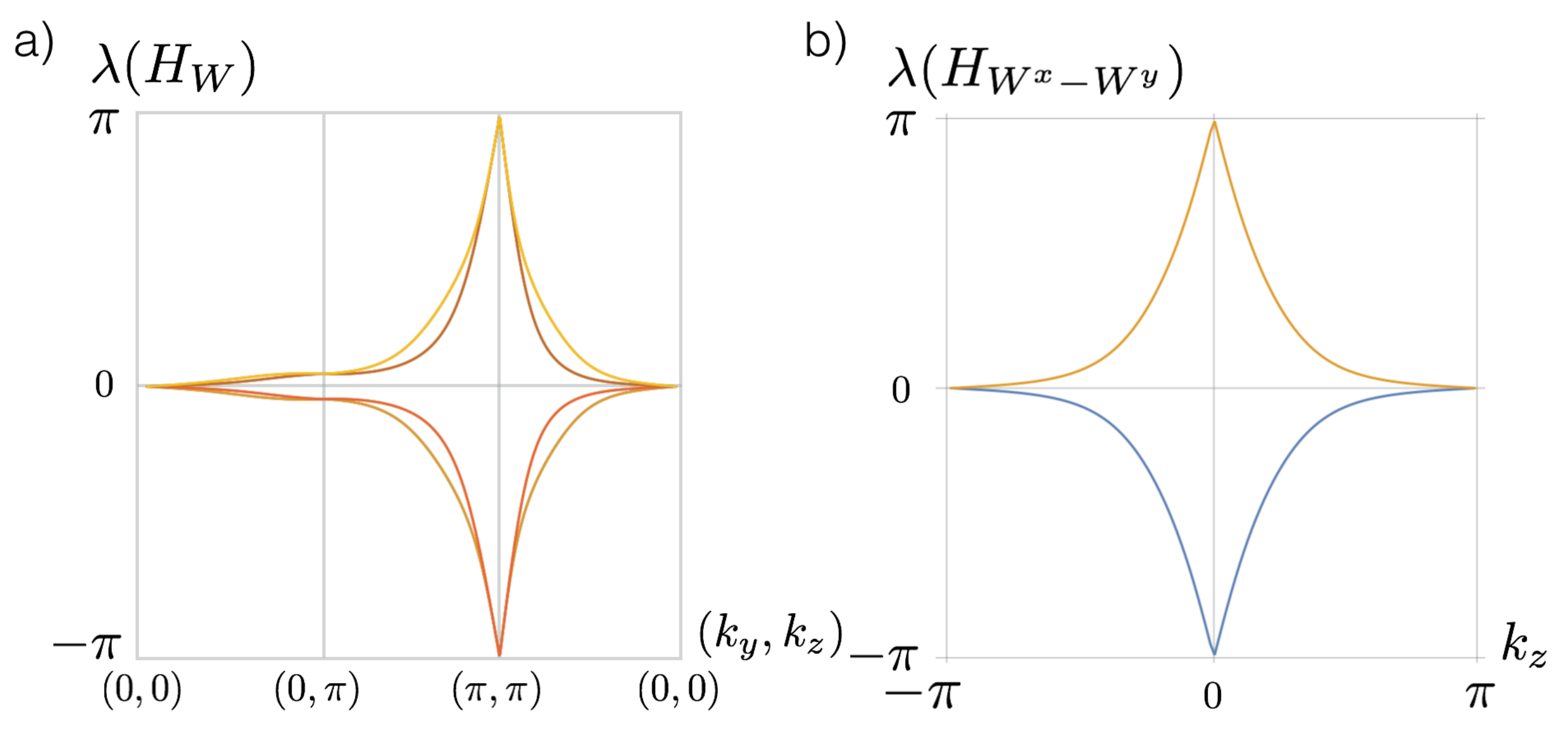}
\caption{
(a) Spectra of the $\hat{C}^z_4$-graded Wilson loops $W_{\xi=+}^{\hat{C}_4^z} (k_x, k_y)$ and $W_{\xi=-}^{\hat{C}_4^z} (k_x, k_y)$ defined in Eq.~\eqref{eq: C4 graded wilsonloop}, evaluated for model~\eqref{eq: trsH} with $M/t=2$ and $\Delta_1/t = \Delta_2/t = 1$. Both spectra are exactly equal to one another. 
All states in each spectrum have an extra double degeneracy. Thus, the degeneracies at $(k_x,k_y)=(0,0)$ and $(k_x,k_y)=(\pi,\pi)$ correspond to a superposition of four Dirac points. This corresponds to the minimal nontrivial degeneracy, as all states at these $\hat{C}^z_4$-invariant momenta are four-fold repeated by the construction of the $\hat{C}^z_4$-graded Wilson loops in Eq.~\eqref{eq: C4 graded wilsonloop}.
(b) Nested Wilson loop spectrum for model~\eqref{eq: trsH} as defined in Supplementary Note~\ref{sec: nestedspectra}~B. The nontrivial winding along $k_z$ is in one-to-one correspondence with gapless hinge-excitations in the geometry of Fig.~\ref{fig: physical picture}~(a).}
\label{fig: alternativeHelical}
\end{center}
\end{figure}

\subsection{Protection by $\hat{C}_4$ symmetry}
Consider now an alternative but equally valid set of symmetry representations for Hamiltonian~\eqref{eq: trsH}, which feature a $\hat{C}_4^z$ symmetry preserved by the term proportional to $\Delta_2$
\begin{matriz}
\begin{aligned}
&T \equiv \tau_0 \rho_0 \sigma_y \mathit{K}, \, D_{\hat{T}} \, \vec{k} = -\vec{k},\\ 
&C_4^z \equiv \tau_0 \rho_1 e^{-\mathrm{i} \frac{\pi}{4} \sigma_z}, \, D_{\hat{C}^{z}_4} \vec{k} = (-k_y,k_x,k_z).
\end{aligned}
\end{matriz}
Since $(\hat{C}_4^z)^4 = -1$, all the $\hat{C}_4^z$ eigenvalues are complex, and due to time reversal symmetry necessarily occur in complex conjugated pairs $\{\xi_{\vec{k}} e^{\mathrm{i}\pi/4},\xi_{\vec{k}} e^{-\mathrm{i}\pi/4}\}$. Along the $\hat{C}_4^z$ invariant lines $(k_x, k_y) \in \{(0,0),(\pi,\pi)\}$, this allows us to group the bands of~\eqref{eq: trsH} into two subspaces with $\xi_{\vec{k}} = +1$ and $\xi_{\vec{k}} = -1$, respectively. To protect a helical HOTI by $\hat{C}_4$ symmetry, we require that the first-order $\mathbb{Z}_2$ invariant, defined through the CS term in Eq.~\eqref{eq: theta invariant}, is nontrivial in \emph{each} of the two subspaces with $\xi_{\vec{k}} = +1$ and $\xi_{\vec{k}} = -1$ individually. 
(If only one subspace had a nontrivial $\mathbb{Z}_2$ invariant, the system would be a conventional 3D TI with gapless surfaces protected by $\hat{T}$ symmetry.)

To measure this $\hat{C}_4$-graded index, we employ a generalized Brillouin zone Wilson loop. In the main text, it was argued that for the chiral HOTI with $\hat{C}_4^z \hat{T}$ symmetry, a nontrivial winding of the $z$-direction Wilson loop $W^z(k_x,k_y)$ between the points $(0,0)$ and $(\pi,\pi)$ is in one-to-one correspondence with a nontrivial $\mathbb{Z}_2$ index. To generalize this concept to the present case of $\hat{C}_4^z$ \emph{and} $\hat{T}$ symmetry, we introduce the states
\begin{matriz}
\begin{aligned}
\ket{\Psi_{\alpha} (\bs{k})} = &\bigl(\ket{u(\vec{k})}, e^{-\mathrm{i} \alpha} C_4^z \ket{u(\vec{k})}, \\& e^{-2 \mathrm{i} \alpha} (C_4^z)^2 \ket{u(\vec{k})}, e^{-3 \mathrm{i} \alpha} (C_4^z)^3 \ket{u(\vec{k})} \bigr),
\end{aligned}
\end{matriz}
where we denote by $\ket{u(\vec{k})} = (\ket{u_m(\vec{k})})_{m=1,\cdots, N}$ the vector of eigenstates of the $N$ filled bands at a given momentum $\bs{k}$ and $\alpha=\pi/4,-\pi/4,3\pi/4,-3\pi/4$. The states $\ket{\Psi_{\alpha} (\bs{k})} $ are eigenstates of the operator
\begin{matriz}
R_4 = \begin{pmatrix} 
0 & 0 & 0 & C_4^z \\
C_4^z & 0 & 0 & 0 \\
0 & C_4^z & 0 & 0 \\
0 & 0 & C_4^z & 0
\end{pmatrix}
\end{matriz}
with eigenvalue $e^{\mathrm{i} \alpha}$, and are defined everywhere in the Brillouin zone.
Here, $\alpha = \pm \pi/4$ and $\alpha = \pm 3\pi/4$ correspond to $\xi = 1$ and $\xi = -1$, respectively. 
\begin{widetext}
By introducing the projector 
\begin{matriz}
\hat{P}_{\xi} (\vec{k})=
\ket{\Psi_{\frac{\pi}{4}(2-\xi)} (\bs{k})} \bra{\Psi_{\frac{\pi}{4}(2-\xi)} (\bs{k})}
+\ket{\Psi_{-\frac{\pi}{4}(2-\xi)} (\bs{k})} \bra{\Psi_{-\frac{\pi}{4}(2-\xi)} (\bs{k})},\qquad \xi=\pm1,
\end{matriz}
onto a time-reversal invariant subspace, we may define the $\hat{C}_4^z$-graded Wilson loop
\begin{matriz}
\begin{aligned}
W_{\xi,(m\lambda)(m'\lambda')}^{\hat{C}_4^z} (k_x, k_y) =
\bra{\Psi^{m}_{\lambda \frac{\pi}{4}(2-\xi)} (k_x,k_y,2 \pi)} 
\prod_{k_z}^{2\pi \leftarrow 0} \hat{P}_{\xi}(\vec{k}) 
\ket{\Psi^{m'}_{\lambda' \frac{\pi}{4}(2-\xi)} (k_x,k_y,0)},
\end{aligned}
\label{eq: C4 graded wilsonloop}
\end{matriz}
where $m,m' = 1...4$ and $\lambda,\lambda' = \pm1$ so that $W^{\hat{C}_4^z}_\xi$ is an $8 \times 8$ matrix.
\end{widetext}
In the two time reversal invariant subspaces, corresponding to $\xi = 1$ ($\alpha = \pm \pi/4$) and $\xi = -1$ ($\alpha = \pm 3\pi/4$), the winding of the eigenvalues of $W_{\xi}^{\hat{C}_4^z} (k_x, k_y)$ between the points $(k_x,k_y)=(0,0)$ and $(\pi,\pi)$ defines a $\mathbb{Z}_2$ topological invariant [see Fig.~\ref{fig: alternativeHelical}~(a)].

Alternatively, we may calculate the nested Wilson loop as described in section~\ref{sec: nestedwilson}: We first evaluate the Wilson loop along the $x$ direction, which is gapped. We then take the nested Wilson loop in its filled subspace along the $y$ direction, retaining only $k_z$ as a good quantum number. Figure~\ref{fig: alternativeHelical}~(b) shows that its spectrum as a function of $k_z$ consists of a pair of helical modes, confirming the gapless nature of the hinge.

\section{$\text{SnTe}$ as a higher-order topological insulator}

\subsection{Tight-binding model with (110) strain}

We consider a modification of a model describing the topological crystalline insulator\cite{Fu11} SnTe~\cite{Hsieh12
}.
SnTe forms a rocksalt lattice structure on top of which we consider a distortion due to uniaxial stress [see Fig.~\ref{fig: TRS TI}~(a)] along the (110) direction modeled by a crystal field splitting parameter $\Delta$.
For pristine SnTe, we have $\Delta=0$.
Strained SnTe with $\Delta\neq0$ has lower octahedral symmetry. 
The Hamiltonian
\begin{matriz}
\begin{split}
H_{\mathrm{h}}=
&\,m\sum_{j}(-1)^j\sum_{\bs{r},s}\bs{c}^\dagger_{j,s,\bs{r}}\cdot \bs{c}_{j,s,\bs{r}}
\\
&+\,
\sum_{j,j'} t_{jj'}\sum_{(\bs{r},\bs{r}'),s}
\bs{c}^\dagger_{j,s,\bs{r}}\cdot \bs{d}_{\bs{r},\bs{r}'}\, \bs{d}_{\bs{r},\bs{r}'} \cdot \bs{c}_{j,s,\bs{r}'}
+\mathrm{h.c.}
\\
&
+\mathrm{i}\lambda\sum_j\sum_{\bs{r},s,s'}
\bs{c}^\dagger_{j,s,\bs{r}}\times \bs{c}_{j,s',\bs{r}}\cdot \bs{\sigma}_{ss'}
\\
&
+\Delta
\sum_j\sum_{\bs{r},s} 
\left(
c^\dagger_{j,s,\bs{r},p_x}c_{j,s,\bs{r},p_y}
+
c^\dagger_{j,s,\bs{r},p_y}c_{j,s,\bs{r},p_x}
\right)
\end{split}
\label{eq: SnTe Hamiltionian}
\end{matriz}
acts on the degrees of freedom on a rocksalt lattice structure [see Fig.~\ref{fig: TRS TI}~(a)], of which Sn atoms ($j=1$) and Te atoms ($j=2$) each form one sublattice. The operators $c^\dagger_{j,s,\bs{r},p_i}$, $i=x,y,z$, create an electron at lattice site $\bs{r}$, sublattice $j$, with spin $s=\uparrow, \downarrow$ in one of the three $p$ orbitals $p_x$, $p_y$, $p_z$, and are combined into the spinor $\bs{c}^\dagger_{j,s,\bs{r}}=(c^\dagger_{j,s,\bs{r},p_x},c^\dagger_{j,s,\bs{r},p_y},c^\dagger_{j,s,\bs{r},p_z})$. The summation $(\bs{r},\bs{r}')$ includes both nearest and next-nearest neighbor hopping and $\bs{d}_{\bs{r},\bs{r}'}$ is the unit vector in the direction of the hopping process. We choose the parameters~\cite{fulga2016} $m=1.65$ for the staggered sublattice potential, $t_{12}=t_{21}=0.9$, $t_{11}=-t_{22}=0.5$, for the nearest and next-nearest neighbor hopping amplitude, and $\lambda=0.7$ for the strength of spin-orbit coupling. To disentangle the hinge electronic structure obtained with $\Delta=-0.4$, we furthermore apply an on-site chemical potential $|\mu_{\mathrm{hinge}}|=0.2$, which is nonvanishing only on the hinge sites, and has positive sign on the pair of sites invariant under the $\hat{M}_{xy}$ mirror symmetry, and negative sign on the pair invariant under $\hat{M}_{x \bar{y}}$. This creates a difference in value between the $xy$ hinge state eigenenergies and the $x \bar{y}$ hinge state eigenenergies.

\begin{figure*}[t]
\begin{center}
\includegraphics[width=0.9 \textwidth]{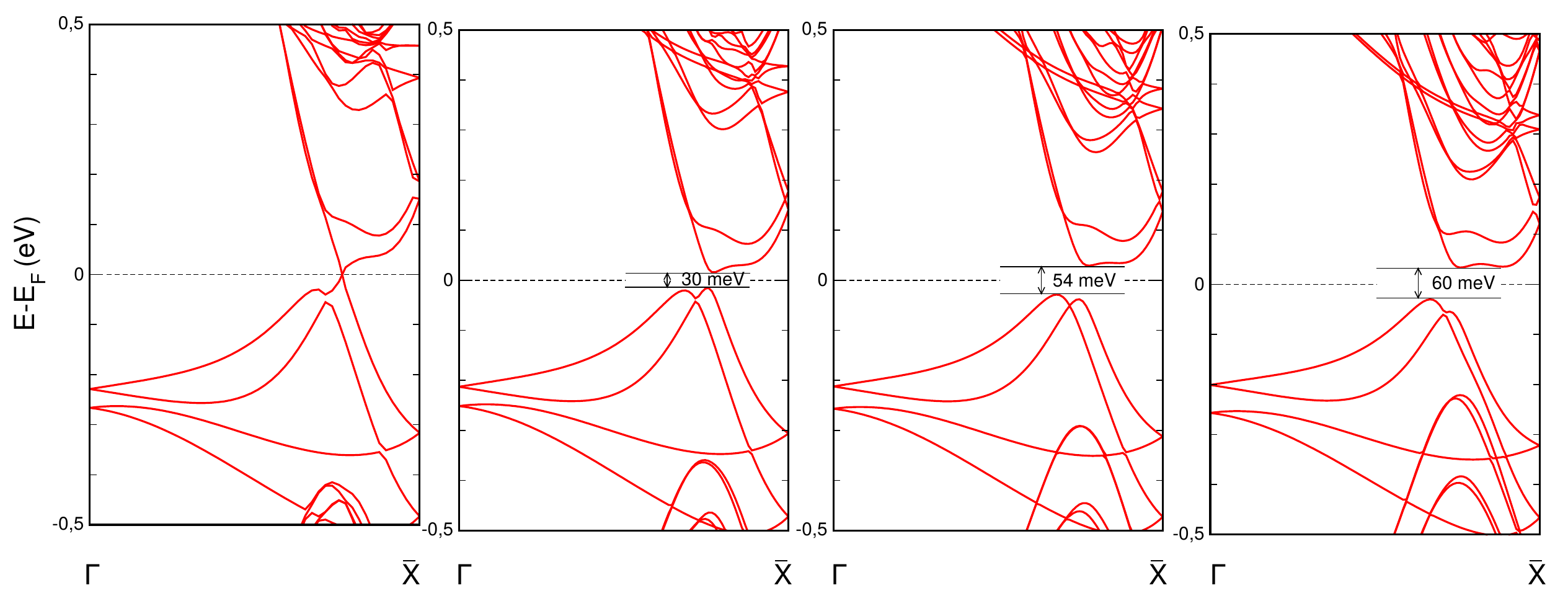}
\caption{
Band structure of the surface Dirac cones of the topological crystalline insulator SnTe 
calculated in a slab geometry, from left to right: undistorted cubic phase with gapless Dirac cones, with 1\%, 3\%, and 5\% strain along the (110) direction. The strain breaks the mirror symmetry that protects the Dirac cones and hence can open a gap.
}
\label{fig: surf-gap}
\end{center}
\end{figure*}

In order to write down the corresponding Bloch Hamiltonian, we choose the lattice spacing $a = 1$ and define $m_{\mathrm{Sn}} = - m_{\mathrm{Te}} = 1.65$ to automatically incorporate the staggered sublattice potential. We also introduce the orbital angular momentum operators
\begin{matriz}
\begin{aligned}
L_x = \mathrm{i} \begin{pmatrix} 0 & 0 & 0\\ 0 & 0 & -1 \\ 0 & 1 & 0 \end{pmatrix}, & \quad L_y = \mathrm{i} \begin{pmatrix} 0 & 0 & 1\\ 0 & 0 & 0 \\ -1 & 0 & 0 \end{pmatrix},\\
L_z = \mathrm{i} & \begin{pmatrix} 0 & -1 & 0\\ 1 & 0 & 0 \\ 0 & 0 & 0 \end{pmatrix},
\end{aligned}
\end{matriz}
as well as the hopping matrices $T_i = \bs{d}_{i}\otimes \bs{d}_{i}$, $i \in \{x,y,z,xy,x\bar{y},xz,x\bar{z},yz,y\bar{z}\}$, where
\begin{matriz}
\begin{aligned}
\bs{d}_{x} = (1,0,0), \quad \bs{d}_{y} = &(0,1,0), \quad \bs{d}_{z} = (0,0,1), \\
\bs{d}_{xy} = (1,1,0)/2, &\quad \bs{d}_{x\bar{y}} = (1,-1,0)/2, \\
\bs{d}_{xz} = (1,0,1)/2, &\quad \bs{d}_{x\bar{z}} = (1,0,-1)/2, \\
\bs{d}_{yz} = (0,1,1)/2, &\quad \bs{d}_{y\bar{z}} = (0,1,-1)/2.
\end{aligned}
\end{matriz}
The on-site Hamiltonian is then given by
\begin{matriz}
\mathcal{H}_{j}^{\text{os}} = m_j + \lambda \sum_{\alpha=1}^3 L_\alpha \otimes \sigma_\alpha, \quad j \in \{\mathrm{Sn},\mathrm{Te}\},
\end{matriz}
and the $\bs{k}$-independent part of the full Hamiltonian reads
\begin{matriz}
\mathcal{H}_0 = \begin{pmatrix} \mathcal{H}_{\mathrm{Sn}}^{\text{os}} & t_{12} T_x^\dagger \\t_{21} T_x & \mathcal{H}_{\mathrm{Te}}^{\text{os}} \end{pmatrix}.
\end{matriz}
We may then set up the hopping matrices $(j \in \{1,2\})$
\begin{matriz}
\begin{aligned}
\mathcal{H}_{jj}^{xy} (\bs{k}) &= t_{jj} \, (e^{\mathrm{i} k_x} T_{xy} + e^{\mathrm{i} (k_z - k_y)} T_{x\bar{y}}),�\\
\mathcal{H}_{jj}^{xz} (\bs{k}) &= t_{jj} \, (e^{\mathrm{i} k_z} T_{xz} + e^{\mathrm{i} (k_x - k_y)} T_{x\bar{z}}),�\\
\mathcal{H}_{jj}^{yz} (\bs{k}) &= t_{jj} \, (e^{\mathrm{i} k_y} T_{yz} + e^{\mathrm{i} (k_x - k_z)} T_{y\bar{z}}),�\\
\mathcal{H}_{12}^{xy} (\bs{k}) &= t_{12} \, (e^{\mathrm{i} (k_x + k_z - k_y)} T_x + e^{\mathrm{i} k_x} T_y), \\
\mathcal{H}_{21}^{xy} (\bs{k}) &= t_{21} \, e^{\mathrm{i} (k_y - k_z)} T_y,\\
\mathcal{H}_{12}^{z} (\bs{k}) &= t_{12} \, e^{\mathrm{i} (k_z)} T_z,\\
\mathcal{H}_{21}^{z} (\bs{k}) &= t_{21} \, e^{\mathrm{i} (k_y - k_x)} T_z,\\
\end{aligned}
\end{matriz}
in order to define 
\begin{widetext}
\begin{matriz}
\mathcal{H}_1 (\bs{k}) = \begin{pmatrix} \mathcal{H}_{11}^{xy} (\bs{k}) + \mathcal{H}_{11}^{xz} (\bs{k}) + \mathcal{H}_{11}^{yz} (\bs{k})�
& \mathcal{H}_{12}^{xy} (\bs{k}) + \mathcal{H}_{12}^{z} (\bs{k}) \\�
\mathcal{H}_{21}^{xy} (\bs{k}) + \mathcal{H}_{21}^{z} (\bs{k}) & 
\mathcal{H}_{22}^{xy} (\bs{k}) + \mathcal{H}_{22}^{xz} (\bs{k}) + \mathcal{H}_{22}^{yz} (\bs{k}) \end{pmatrix}.
\end{matriz}
\end{widetext}
The full Bloch Hamiltonian is then given by
\begin{matriz}
\mathcal{H} (\bs{k}) = \mathcal{H}_0 + \mathcal{H}_1 (\bs{k})+ \mathcal{H}_1 (\bs{k})^\dagger.
\end{matriz}

Diagonalizing the Hamiltonian in the geometry described above with $\Delta\neq0$ reveals a fully gapped electronic structure, except at the hinges [see Fig.~\ref{fig: TRS TI}~(c)], where one Kramers pair of gapless modes is localized at each hinge.
It is worth noting that there are `flat-band' hinge states connecting the $k_z$ projections of the surface Dirac cones already for $\Delta=0$ [see inset of Fig.~\ref{fig: TRS TI}~(c)], reminiscent of the `flat-band' states at the zig-zag edge of graphene 
and states bound to the surface step edges of topological crystalline insulators\cite{Sessi2016}. With these `flat-band' states, undistorted SnTe may be viewed as a higher-order topological semimetal.

\subsection{Surface gaps from DFT}

Figure~\ref{fig: surf-gap} shows the evolution of the surface Dirac cones under uniaxial strain along the (110) direction. 1\% strain is sufficient to open a gap of 30~meV which might be enough to detect hinge states in this gap. All four cones on the (100) and (010) surfaces are symmetry-equivalent with respect to this strain direction. For this reason it is sufficient to study one of them. 

\subsection{Dirac picture of the SnTe surface states} \label{sec: diracpic}
Here we demonstrate the presence of a single helical pair of eigenmodes localized on the hinge between the (100) and (010) surface terminations of a 3D helical HOTI on the basis of an effective continuum Dirac description. 
We discuss both the case of the alternative model~\eqref{eq: trsH} for 3D helical HOTIs and a model pertaining to SnTe. 
The apparent difference between the two is the dimension of the representation of the Dirac equation, i.e., the number of (massive) Dirac cones in the bulk and on the surface. The bulk low-energy physics of model~\eqref{eq: trsH} near the topological phase transition is governed by an $8\times 8$ matrix representation of the Dirac equation, and its surface by a $4\times 4$ representation. The bulk of SnTe, in contrast, is described by four $4\times4$ Dirac equations, i.e., a $16\times 16$ Dirac equation\cite{Hsieh12}, yielding an $8\times 8$ Dirac equation on the surface, i.e., four Dirac cones\cite{Liu13,Wang13}.    
We will discuss that, in order to obtain a single Kramers pair of hinge modes in both cases, it is useful to think of the  hinge as a vortex in a two-component mass term in the Dirac equation.

The bulk low energy physics of Hamiltonian~\eqref{eq: trsH} at the topological transition $M=3$ is that of a massless Dirac equation
\begin{matriz}
\mH(\bs{k}) = \sum_{i=x,y,z} \tau_x \rho_0 \sigma_i \, k_i,
\end{matriz}
where $\tau_0$, $\rho_0$, $\sigma_0$ are $2\times2 $ unit matrices and $\tau_i$, $\rho_i$, $\sigma_i$, $i=x,y,z$, are the corresponding Pauli matrices, acting on the same degrees of freedom as in Eq.~\eqref{eq: trsH}.
We choose the same symmetry representations as in Eq.~\eqref{eq: symmetries higher-order}, i.e.,
\begin{matriz}
\begin{aligned}
&T \equiv \tau_0 \rho_0  \sigma_y \mathit{K}, 
\quad C_4^i \equiv \tau_0 \rho_0 e^{-\mathrm{i} \frac{\pi}{4} \sigma_i}, \\
&M_{x} \equiv \mathrm{i} \, \tau_z \rho_0 \sigma_x, 
\quad M_{y} \equiv \mathrm{i}\, \tau_z \rho_0 \sigma_y, \\
&M_{xy} \equiv \mathrm{i} \, (\tau_z \rho_0 \sigma_y - \tau_z \rho_0 \sigma_x)/\sqrt{2}, \\
&M_{x\bar{y}} \equiv \mathrm{i} \, (\tau_z \rho_0 \sigma_x + \tau_z \rho_0 \sigma_y)/\sqrt{2}.
\end{aligned}\label{eq: symmetries higher-order Dirac}
\end{matriz}

There are four time-reversal symmetric masses,
\begin{matriz}
\begin{aligned}
m_1 =& \tau_z \rho_0 \sigma_0, \quad 
m_2 = \tau_z \rho_x \sigma_0, \\ 
m_3 =& \tau_z \rho_z\sigma_0, \quad 
m_4 = \tau_y \rho_y \sigma_0.
\end{aligned}
\end{matriz}
Of these, $m_1$, $m_2$ and $m_3$ respect all mirror symmetries and may therefore be used as bulk masses. We will choose only $m_1$ to be nonzero, since this is the mass occurring in a low energy expansion of the model in Eq.~\eqref{eq: trsH}.

We now consider a geometry where the topological material is terminated in the $x$ and $y$ directions with a hinge symmetric under $\hat{M}_{xy}$. We take the system to lie in the lower left quadrant of the $(x,y)$ plane, i.e., 
take the prefactor of  the $m_1$ mass term positive for $x<0$, $y<0$ and negative otherwise. 
Since $m_4$ is the only mass breaking $\hat{M}_{x}$ and $\hat{M}_{y}$, it is the only possible mass that can gap out the surface. 
Furthermore, because $m_4$ anticommutes with $\hat{M}_{xy}$, we may add it to the Hamiltonian as a surface term with opposite sign on the (100) and (010) terminations. In $(x,y,k_z)$ hybrid space, the Hamiltonian then reads
\begin{matriz}
\label{eq: hybridDiracH}
\begin{aligned}
\mathcal{H}(k_z) =& 
 \tau_x \rho_0 \sigma_x (-\mathrm{i} \partial_x) + \tau_x \rho_0 \sigma_y (-\mathrm{i} \partial_y)+\tau_x \rho_0 \sigma_z \, k_z  \\&
+ m_1 \, \delta_1 (x,y) + m_4 \, \delta_2 (x,y),
\end{aligned}
\end{matriz}
with real space dependencies
\begin{matriz}
\begin{aligned}
\delta_1 (x,y) =& \begin{cases}
+ 1 & x<0, y<0, \\
-1 & \text{else},
\end{cases} \\
\delta_2 (x,y) =& \begin{cases}
+ 1 & x \approx 0, \\
-1 & y \approx 0, \\
0 & x = y \text{ and else}.
\end{cases}
\end{aligned}
\label{eq: vortex}
\end{matriz}
Since $m_1$ and $m_4$ are anticommuting mass matrices, the spatial dependence of \eqref{eq: vortex} corresponds to a vortex with winding  number 1 in a two-component mass parameter for the Dirac electrons. 
We will now argue that this vortex supports a (Kramers) pair of topologically protected bound states at zero energy for $k_z=0$. These bound states are part of the Kramers pair of propagating hinge states, which are dispersing with $k_z$. 
At $k_z=0$, the Dirac Hamiltonian~\eqref{eq: hybridDiracH} has a unique chiral symmetry $\mathcal{C}=\tau_x\rho_y\sigma_z$ that commutes with time-reversal symmetry, i.e., $\mathcal{C}\mathcal{H}(k_z=0)\mathcal{C}^{-1}=-\mathcal{H}(k_z=0)$, $T\mathcal{C}T^{-1}=\mathcal{C}$. By time-reversal, all eigenstates of Hamiltonian~\eqref{eq: hybridDiracH} at $k_z=0$ have to come in degenerate Kramers pairs. Due to $\mathcal{C}$, a Kramers pair at energy $E$ has to have a partner Kramers pair at energy $-E$. Only at $E=0$ a single Kramers pair can appear as spatially isolated eigenstates of $\mathcal{H}(k_z=0)$ and $\mathcal{C}$, i.e., as topological vortex zero modes. Due to this symmetry protection, such bound states are robust against deformations in the vortex profile [i.e., smooth changes in the spatial dependence of $\delta_1(x,y)$ and $\delta_2(x,y)$], as long as the spectral gap far away from the vortex remains intact. 
We use this freedom to smoothly deform the vortex profile of the hinge geometry~\eqref{eq: vortex} to 
 \begin{matriz}
 \delta_1 (x,y)=y,
 \qquad
 \delta_2 (x,y)=x
 \end{matriz}
 and argue that a Kramers pair of vortex bound states exists in this case.
 Upon conjugation with the matrix $R=\mathrm{exp}(\mathrm{i}\pi/4 \, \tau_y\rho_y\sigma_0)$ the Hamiltonian 
  $\mathcal{H}(k_z=0)$ can be cast in the block off-diagonal form
\begin{matriz}
\begin{aligned}
R\, \mathcal{H}(k_z=0)R^{-1}
=
\begin{pmatrix}
0&0&0& h_1
\\
0&0& h_2&0
\\
0& h_2^\dagger&0&0
\\
h_1^\dagger&0&0&0
\end{pmatrix}
\end{aligned}
\label{eq: vortex block off-diagonal structure}
\end{matriz}
with blocks
\begin{matriz}
\begin{aligned}
h_1
=
\begin{pmatrix}
-\partial_x+\mathrm{i}\partial_y &\mathrm{i} \delta_1(x,y)-\delta_2(x,y)\\
\mathrm{i} \delta_1(x,y)+\delta_2(x,y)&\partial_x+\mathrm{i}\partial_y
\end{pmatrix}
\end{aligned}
\end{matriz}
and
\begin{matriz}
\begin{aligned}
h_2
=
\begin{pmatrix}
-\partial_x-\mathrm{i}\partial_y &\mathrm{i} \delta_1(x,y)-\delta_2(x,y)\\
\mathrm{i} \delta_1(x,y)+\delta_2(x,y)&\partial_x+\mathrm{i}\partial_y
\end{pmatrix}.
\end{aligned}
\end{matriz}
The problem of finding zero modes of $h_1$, $h_2$, $h_1^\dagger$, and $h_2^\dagger$ 
goes back to the vortex solution found by Jackiw and Rossi in Ref.~\onlinecite{Jackiw81}. It was shown that either $h_1$ or $h_1^\dagger$, and either $h_2$ or $h_2^\dagger$ support exactly one normalizable zero-energy state that is localized at the vortex. 
Explicitly, we can rewrite $\partial_z=(\partial_x-\mathrm{i}\partial_y)/2$  and $\partial_{\bar{z}}=(\partial_x+\mathrm{i}\partial_y)/2$, where $z=x+\mathrm{i} y$ and $\bar{z}=x-\mathrm{i} y$. In terms of these variables
\begin{matriz}
\begin{aligned}
h_1
=
\begin{pmatrix}
-\partial_z &-\bar{z}\\
z&\partial_{\bar{z}}
\end{pmatrix},
\quad
h_1^\dagger
=
\begin{pmatrix}
\partial_{\bar{z}} &\bar{z}\\
-z&-\partial_{z}
\end{pmatrix}
\end{aligned}
\end{matriz}
and
\begin{matriz}
\begin{aligned}
h_2
=
\begin{pmatrix}
-\partial_{\bar{z}} &-\bar{z}\\
z&\partial_{\bar{z}}
\end{pmatrix},
\quad
h_2^\dagger
=
\begin{pmatrix}
\partial_{z} &\bar{z}\\
-z&-\partial_{z}
\end{pmatrix}.
\end{aligned}
\end{matriz}
One verifies that 
$\psi_1(z,\bar{z})\equiv\mathcal{N}\, e^{-z\bar{z}}(0,0,0,0,0,0,1,1)^{\mathsf{T}}$ and 
$\psi_2(z,\bar{z})\equiv\mathcal{N}\, e^{-z\bar{z}}(0,0,1,1,0,0,0,0)^{\mathsf{T}}$, with $\mathcal{N}$ being the appropriate normalization factor, are zero modes of Eq.~\eqref{eq: vortex block off-diagonal structure} with nontrivial action in the blocks $h_1$ and $h_2^\dagger$, while $h_2$ and $h_1^\dagger$ do not support zero modes.
Their $k_z$ dispersion can be inferred from the matrix elements
\begin{matriz}
\begin{pmatrix}
\left\langle\psi_1|\mathcal{H}(k_z)|\psi_1\right\rangle&
\left\langle\psi_1|\mathcal{H}(k_z)|\psi_2\right\rangle\\
\left\langle\psi_2|\mathcal{H}(k_z)|\psi_1\right\rangle&
\left\langle\psi_2|\mathcal{H}(k_z)|\psi_2\right\rangle
\end{pmatrix}
=
\begin{pmatrix}
0&-\mathrm{i} k_z\\
\mathrm{i} k_z&0
\end{pmatrix},
\end{matriz}
which yields two counter-propagating Kramers paired eigenmodes $|\pm\rangle=|\psi_1\rangle\pm\mathrm{i}|\psi_2\rangle$ at the hinge with dispersions $\epsilon_\pm=\pm k_z$.

\begin{figure}[t]
\begin{center}
\includegraphics[width=0.48 \textwidth]{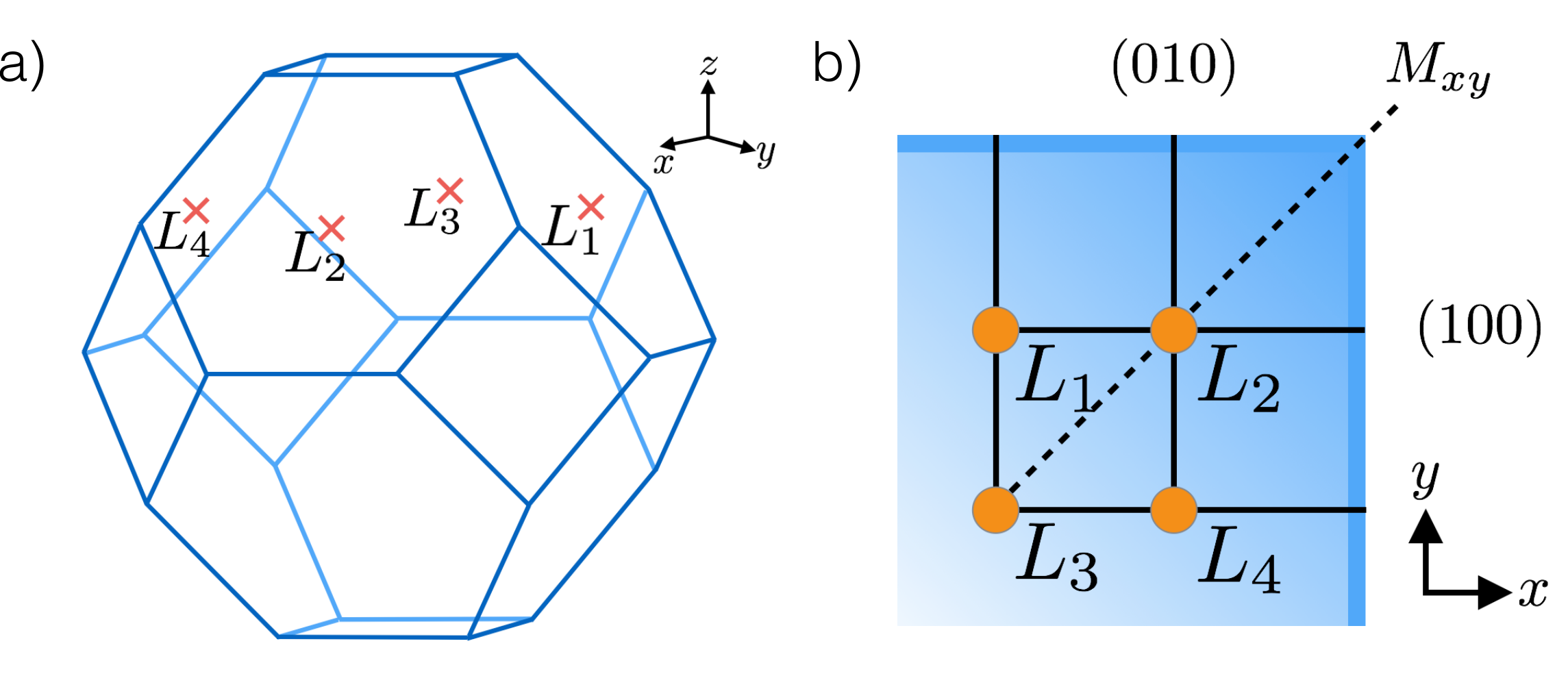}
\caption{
(a) Bulk brillouin zone of SnTe. The low-energy electronic structure is made up of four gapped Dirac cones at the time-reversal invariant L points. (b) Hinge geometry considered in Supplementary Note~\ref{sec: diracpic} for the Dirac equation modeling SnTe. Periodic boundary conditions in $z$ direction are assumed. The four L points in the bulk Brillouin zone project onto two time-reversal symmetric points in each of the surface Brillouin zones of the (100) and (010) surfaces.}
\label{fig: hingegeometrydirac}
\end{center}
\end{figure}

To make contact with the low energy electronic structure of SnTe, we consider a model that contains four Dirac cones in the bulk which are located at four time-reversal invariant points in the 3D Brillouin zone, the L points shown in Fig.~\ref{fig: hingegeometrydirac}. We take them to be described by the bulk Hamiltonian\cite{Hsieh12}
\begin{matriz}
\mH_{\mathrm{b}}(\bs{k}) = \sum_{i=x,y,z} X_{00xi} \, k_i,
\label{eq: Dirac Ham SnTe}
\end{matriz}
where $X_{ijkl} = \rho_{1,i} \otimes \rho_{2,j} \otimes \tau_{k} \otimes \sigma_{l}$ and $\{\rho_1, \rho_2, \tau, \sigma\}$ are four different sets of Pauli matrices.
We furthermore choose the symmetry representations
\begin{matriz}
\begin{aligned}
&T \equiv -\mathrm{i} \, X_{000y} \mathit{K}, \\
&M_{x} \equiv \mathrm{i} \, X_{x0zx}, \quad M_{y} \equiv \mathrm{i} \, X_{0xzy}, \\
&M_{xy} \equiv \mathrm{i} \, \sum_{\mu=0,x,y,z} (X_{\mu\mu zx} - X_{\mu\mu zy})/\sqrt{8}, \\
&M_z \equiv \mathrm{i} X_{xxzz}.
\end{aligned}
\end{matriz}
Comparing to Fig.~\ref{fig: hingegeometrydirac}, this choice corresponds to the set of matrices $\rho_{1,\mu},\, \mu=0,x,y,z$, acting on the $k_x$ coordinate of the L points (i.e.,  $ \rho_{1,x}$ in $M_{x}$ exchanges $\text{L}_1$ and $\text{L}_2$ as well as $\text{L}_3$ and $\text{L}_4$), while the set of matrices $ \rho_{2,\mu},\, \mu=0,x,y,z$, acts on the  $k_y$ coordinate of the L points (i.e.,  $ \rho_{2,x}$ in $M_{y}$ exchanges $\text{L}_1$ and $\text{L}_3$ as well as $\text{L}_2$ and $\text{L}_4$). Hamiltonian~\eqref{eq: Dirac Ham SnTe} is then interpreted as a set of four identical $4\times 4$ Dirac equations $\sum_i \tau_{x} \otimes \sigma_{i} \, k_i$, one at each of the four time-reversal invariant L points.
There is one bulk mass, $m_{\mathrm{b}} = X_{00z0}$, which commutes with time-reversal and all mirror symmetries, and furthermore respects the bulk translational symmetry in the sense that it does not couple different L points.

Next, possible surface masses have to be determined. They should (i) anticommute with $\mH_{\mathrm{b}}(\bs{k})$ and $m_{\mathrm{b}}$, (ii) be time-reversal symmetric, (iii) respect the translation symmetry on the (100) and (010) surfaces, respectively.
These conditions are obeyed by two pairs of masses,
\begin{matriz}
m_{\mathrm{s}1} = X_{0yy0}, \qquad
m_{\mathrm{s}2} = X_{y0y0},
\end{matriz}
and
\begin{matriz}
m_{\mathrm{s}1}' = X_{zyy0}, \qquad
m_{\mathrm{s}2}' = X_{yzy0},
\end{matriz}
which are mapped into each other under $M_{xy}$, i.e., $M_{xy}m_{\mathrm{s}1} M_{xy}^{-1}=-m_{\mathrm{s}2}$ and $M_{xy}m_{\mathrm{s}1}' M_{xy}^{-1}=-m_{\mathrm{s}2}'$. The difference between the pairs $(m_{\mathrm{s}1},m_{\mathrm{s}2})$ and $(m_{\mathrm{s}1}',m_{\mathrm{s}2}')$ is that the former preserves $\hat{M}_z$ symmetry, while the latter breaks $\hat{M}_z$. With the goal of finding an effective model for (110) uniaxial strain, which preserves $\hat{M}_z$, we focus on $(m_{\mathrm{s}1},m_{\mathrm{s}2})$. Note that $m_{\mathrm{s}1}$ exchanges $\text{L}_1$ and $\text{L}_3$ as well as $\text{L}_2$ and $\text{L}_4$ and thus preserves translation symmetry on the (010) surface only and the opposite is true for  $m_{\mathrm{s}2}$.
We thus add $m_{\mathrm{s}1}$ on the (010) surface and $m_{\mathrm{s}2}$ on the (100) surface with a prefactor of same magnitude and opposite sign to obey  $\hat{M}_{xy}$ symmetry.

Having identified the masses we now consider a geometry where the angle between the two planes forming the hinge is increased from $\pi/2$ to $\pi$. (Rather than removing the hinge, this transformation should be thought of as a coordinate transformation.) We expect possible topological zero-energy Kramers pairs at $k_z=0$ to be unaffected as long as time-reversal symmetry is not broken. Then, as before, this smooth deformation cannot remove or add a single Kramers pair of modes to the hinge when the gap at large distances is preserved during the transformation. We thus choose
\begin{matriz}
\begin{aligned}
\mathcal{H}_{\text{hinge}}(k_z) = &\mathcal{H}_{\mathrm{b}}(k_z) + m_{\mathrm{b}}\,r \cos(\phi) \\&+ m_{\mathrm{s}1} r\,(1+\sin \phi) - m_{\mathrm{s}2} r\, (1-\sin \phi),
\end{aligned}
\end{matriz}
written in polar coordinates defined through $x=r\cos\,(\phi+\pi/4)$ and $y=r\sin\,(\phi+\pi/4)$, such that $\hat{M}_{xy}$ maps $\phi\to -\phi$. Similar to Eq.~\eqref{eq: vortex block off-diagonal structure}, this has a two-component mass vortex with winding number 1, as we will now show. 

Note that except for the surface masses, all terms in $\mathcal{H}_{\text{hinge}}$ act trivially on the degrees of freedom acted on by $\rho_{1}$ and $\rho_{2}$. We can thus block-diagonalize $\mathcal{H}_{\text{hinge}}(k_z)$ in this subspace first. The only part of $\mathcal{H}_{\text{hinge}}(k_z)$ that is not proportional to the identity in this subspace  is proportional to the operator 
\begin{matriz}
\rho_{1,0} \rho_{2,y} (1+\sin \phi) - \rho_{1,y} \rho_{2,0} (1-\sin \phi),
\label{eq: op subsapce snTe}
\end{matriz}
which has the spectrum $\{\pm 2, \pm 2 \sin \phi \}$. The eigenvectors with eigenvalues $\pm2$ are independent of $\phi$.
We now argue that no topological zero-energy bound states (at $k_z=0$) can arise in the subspace with eigenvalues $\pm2$, which lack a $\phi$ dependence. Starting from a limit in which the prefactors of the surface masses are much larger than the prefactor of the bulk mass $m_{\mathrm{b}}$, we can smoothly tune the prefactor of the bulk mass  $m_{\mathrm{b}}$ to 0 in the $\pm2$ subspace without closing the gap far away from the vortex. The resulting Hamiltonian has no $\phi$-dependence in the $\pm2$ subspace and the node in the surface mass term at the hinge can be smoothly removed. This results in a fully gapped spectrum with no bound states at zero energy in the $\pm2$ subspace.

To find zero-energy bound states at the hinge, it is thus sufficient to consider the action of $\mathcal{H}_{\text{hinge}}(k_z)$ in the subspace with eigenvalues  $\pm 2 \sin \phi$ of the operator~\eqref{eq: op subsapce snTe}.
This subspace is equal to the eigenspace with eigenvalue $+1$ of the operator $\rho_{1,y} \rho_{2,y}$. Representing operators acting on this subspace by the set of matrices $\rho_\mu$, $\mu = 0,x,y,z$, we end up with the reduced $8 \times 8$ Dirac Hamiltonian
\begin{matriz}
\begin{split}
\mathcal{H}_{\text{hinge}}^{\mathrm{red}}(k_z) = 
&
+\rho_0\tau_x\sigma_x \, (-\mathrm{i}\partial_x)+
\rho_0\tau_x\sigma_y \, (-\mathrm{i}\partial_y)+
\rho_0\tau_x\sigma_z \, k_z
\\& + \rho_0\tau_z\sigma_0\,\delta_1(x,y)+ 
 \rho_z\tau_y\sigma_0\,\delta_2(x,y),
\end{split}
\end{matriz}
where $\delta_1(x,y) = \frac{x+y}{\sqrt{2}}$ and $\delta_2(x,y)=\frac{-x+y}{\sqrt{2}}$ again form a vortex with winding number $1$, as was the case for Hamiltonian~\eqref{eq: vortex block off-diagonal structure}. We conclude that it supports a single Kramers pair of modes propagating along the hinge.  

In summary, we have shown that the model~\eqref{eq: trsH} for 3D helical HOTIs, which corresponds to an $8\times 8 $ Dirac equation in the bulk, and SnTe, which corresponds to a $16\times 16$ Dirac equation in the bulk, each have a single topologically protected Kramers pair of hinge modes. 

\end{document}